\documentclass[10pt]{myiopart}
\usepackage{citesort}
\usepackage{color}

\usepackage{iopams}

\usepackage{graphicx}
\usepackage{ae}
\usepackage{bm}

\usepackage{tikz}

\newcommand{\text}{\mathrm}

\newcommand{\Rg}{\ensuremath{R_\mathrm{gyration}}}
\newcommand{\uw}{\ensuremath{u^\mathrm{w}}}

\newcommand{\sex}{\ensuremath{s^\mathrm{ex}}}
\newcommand{\sid}{\ensuremath{s^\mathrm{id}}}

\newcommand{\scf}{\ensuremath{s_\mathrm{c}}}
\newcommand{\kB}{\ensuremath{k_\mathrm{B}}}
\newcommand{\taurelax}{\ensuremath{\tau_{\mathrm{relax}}}}
\newcommand{\tauHS}{\ensuremath{\tau_{\mathrm{HS}}}}
\newcommand{\xiwall}{\ensuremath{\xi_{\mathrm{wall}}}}
\newcommand{\ew}{\ensuremath{\epsilon_{\mathrm{w}}}}
\newcommand{\diff}{\ensuremath{{\mathrm{d}}}}
\newcommand{\varphic}{\ensuremath{\varphi_{\text{c}}}}
\newcommand{\varphiavg}{\ensuremath{\varphi_{\text{avg}}}}
\newcommand{\Dbulkx}{\ensuremath{D^\text{x}_\text{bulk}}}
\newcommand{\Lbox}{\ensuremath{L_\text{box}}}
\newcommand{\amin}{\ensuremath{a_\text{min}}}

\newcommand\delete[1]{\textcolor{red}{}}

\newcommand\revision[1]{\textcolor{black}{#1}}

\newcommand{\beq}{\begin{equation}}
\newcommand{\eeq}{\end{equation}}
\newcommand{\beqn}{\begin{equation*}}
\newcommand{\eeqn}{\end{equation*}}
\newcommand{\bmult}{\begin{multline}}
\newcommand{\emult}{\end{multline}}

\newcommand{\eqref}[1]{Eq.~(\ref{#1})}

\newcommand{\bfig}{\begin{figure}}
\newcommand{\efig}{\end{figure}}

\renewcommand{\figurename}{Figure}

\usepackage[normalem]{ulem}
\renewcommand{\vec}[1]{\bi{#1}}

\def\be{\begin{equation}}       \def\ee{\end{equation}}
\def\bea{\begin{eqnarray}}      \def\eea{\end{eqnarray}}

\begin{document}

% -------------------------------------
% Header
% -------------------------------------

\topical[Non-monotonic effect of confinement---Draft \today]{Non-monotonic effect of confinement on the glass transition}

\author{Fathollah Varnik$^1$ and Thomas Franosch$^2$}

\address{ 
$^1$Interdisciplinary Centre for Advanced Materials Simulation (ICAMS), Ruhr-Universit\"at Bochum, Universit\"atsstra{\ss}e 150, D-44780 Bochum, Germany}
\address{$^2$Institut f\"ur Theoretische Physik, Leopold-Franzens-Universit\"at Innsbruck, Technikerstr. 21A, A-6020 Innsbruck, Austria}

\eads{\mailto{fathollah.varnik@rub.de}, \mailto{Thomas.Franosch@uibk.ac.at}}

% -------------------------
% Abstract
% --------------------------

\begin{abstract}
	
The relaxation dynamics of glass forming liquids and their structure are influenced in the vicinity of confining walls. This effect has mostly been observed to be a monotonic function of the slit width. Recently, a qualitatively new behaviour has been uncovered by Mittal and coworkers, who reported that the single particle dynamics in a hard-sphere fluid confined in a planar slit varies in a non-monotonic way as the slit width is decreased from five to roughly two particle diametres [Mittal et al., Phys.\ Rev.\ Lett.\ {\bf 100}, 145901 (2008)]. In view of the great potential of this effect for applications in those fields of science and industry, where liquids occur under strong confinement (e.g., nano-technology), the number of researchers studying various aspects and consequences of this non-monotonic behaviour has been rapidly growing. This review aims at providing an overview of the research activity in this newly emerging field. We first briefly discuss how competing mechanisms such as packing effects and short-range attraction may lead to a non-monotonic glass transition scenario in the bulk. We then analyse confinement effects on the dynamics of fluids using a thermodynamic route which relates the single particle dynamics to the excess entropy. Moreover, relating the diffusive dynamics to the Widom's insertion probability, the oscillations of the local dynamics with density at moderate densities are fairly well described. At high densities belonging to the supercooled regime, however, this approach breaks down signaling the onset of strongly collective effects. Indeed, confinement introduces a new length scale which in the limit of high densities and small pore sizes competes with the short-range local order of the fluid. This gives rise to a non-monotonic dependence of the packing structure on confinement, with a corresponding effect on the dynamics of structural relaxation. This non-monotonic effect occurs also in the case of a cone-plate type channel, where the degree of confinement varies with distance from the apex. This is a very promising issue for future research with the possibility of uncovering the existence of alternating glassy and liquid-like domains.
\end{abstract}

% Uncomment for PACS numbers title message
\pacs{61.20.Ja, 61.25.Hq, 64.70.Pf}

%\submitto{\JPCM}

% \maketitle

% -------------------------------------
% Table of contents
% -------------------------------------
\newpage
\tableofcontents

% -------------------------------------
% Text
% -------------------------------------
%%%%%%%%%%%%%%%%%%%%%%%%%%%%%%%%%%%%%%%%%%%%%%%%%%%%%%%%%%

\newpage
\clearpage

%%%%%%%%%%%%%%%%%%%%%%%%%%%%%%%%%%%%%%%%%%%%%%%%%%%%%%%%%%%%%%%%%%%%%%%%%
%%%%%%%%%%%%%%%%%%%%%%%%%%%%%%%%%%%%%%%%%%%%%%%%%%%%%%%%%%%%%%%%%%%%%%%%%
%%%%%%%%%%%%%%%%%%%%%%%%%%%%%%%%%%%%%%%%%%%%%%%%%%%%%%%%%%%%%%%%%%%%%%%%%
%%%%%%%%%%%%%%%%%%%%%%%%%%%%%%%%%%%%%%%%%%%%%%%%%%%%%%%%%%%%%%%%%%%%%%%%%
\section{Introduction}
\label{sec:introduction}

Understanding the glassy state of matter and the associated transition from the supercooled liquid state to the amorphous solid is considered a great challenge for theoretical condensed matter physics~\cite{Anderson1995,Stillinger1995,Angell1995,Binder1999b,Pham2002}. Intense experimental, theoretical and  computer simulation studies in the past decades have addressed various aspects of this phenomenon and have paved the way for a variety of applications, ranging from nano-technology to large scale structural systems \cite{Tant1999,Loeffler2003,Telford2004,LeBourhis2007,Greer2009}. A new research direction in this quickly evolving field has been to introduce competing mechanisms which may lead to glass transition scenarios with non-monotonic behaviour.

\begin{figure}
\begin{center}
a)\includegraphics[height=4.5cm]{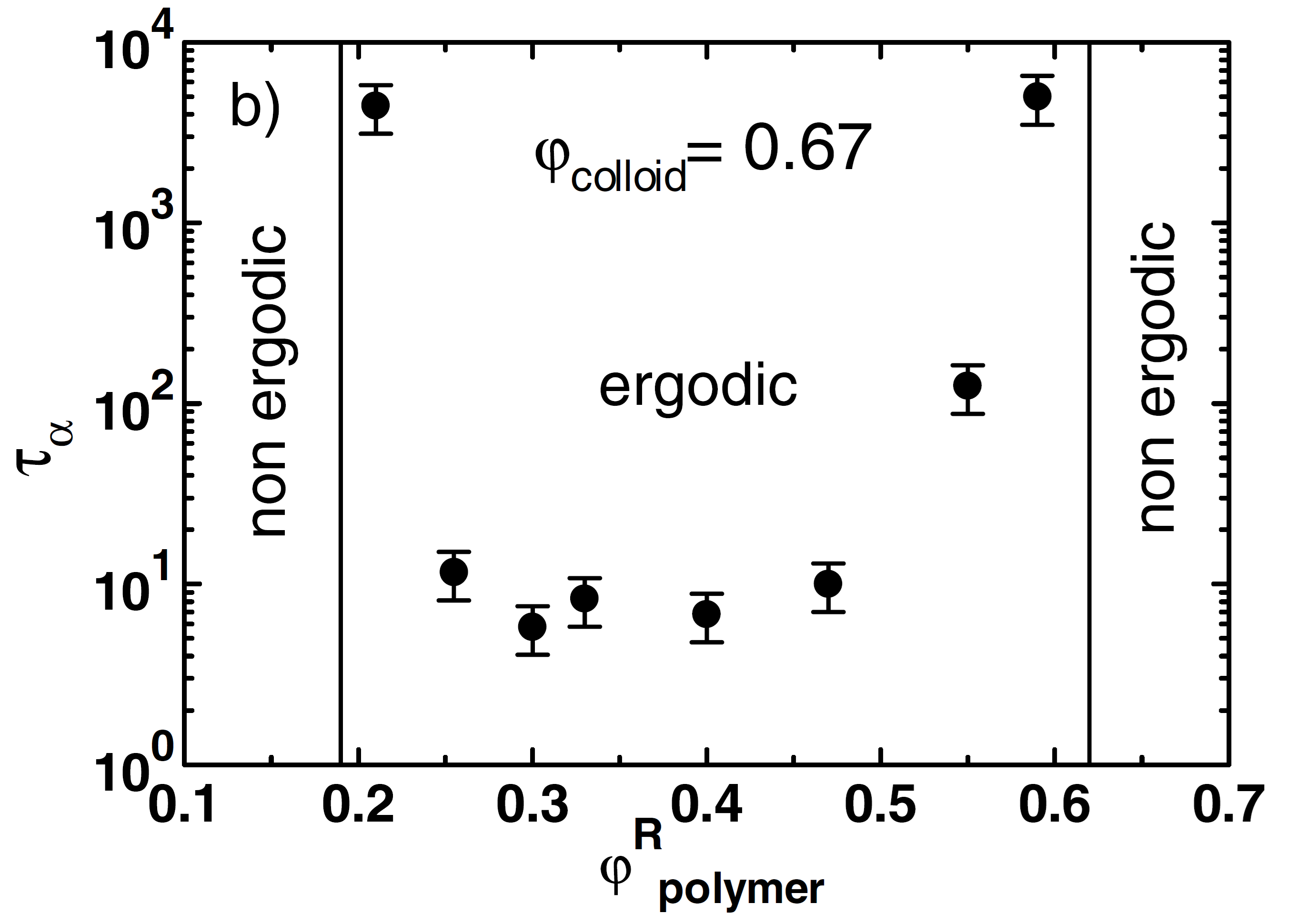}
(b)\includegraphics[height=4.5cm]{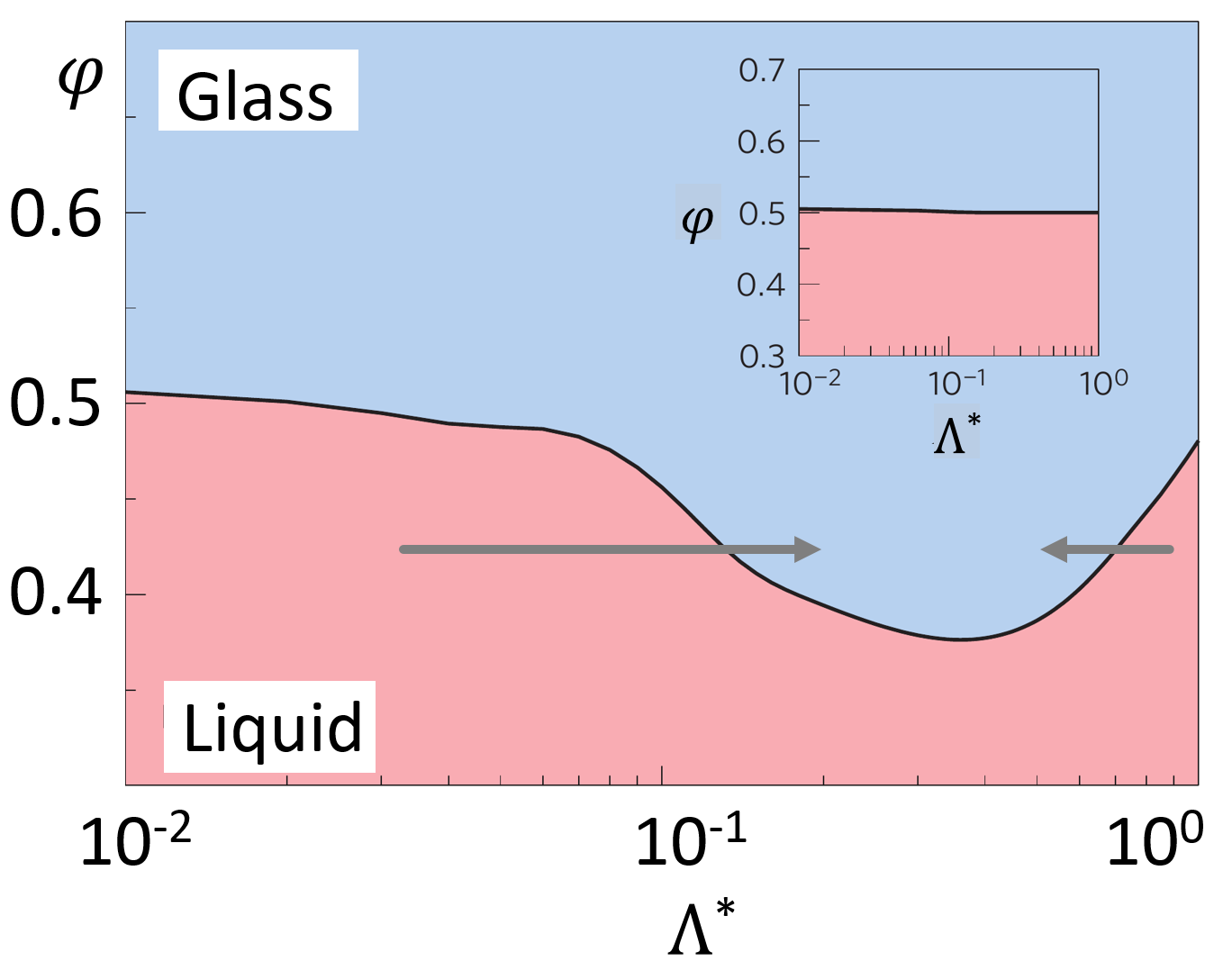}
\end{center}
\caption[]{(colour online).
(a) A classical reentrant scenario in a polymer-colloid mixture. $\tau_{\alpha}$ is the structural relaxation time. The non-ergodic state is reached as $\tau_{\alpha}$ exceeds some empirical value beyond the experimental time window. At low polymer concentrations,  
 $\varphi^{\text{R}}_{\text{polymer}}\lesssim  0.2$,the system is in a non-ergodic state. The relaxation time decreases upon increasing $\varphi^{\text{R}}_{\text{polymer}}$ such that a transition from an arrested to a liquid 
state occurs. At concentrations above $\varphi^{\text{R}}_{\text{polymer}}\gtrsim  0.6$, on the other hand, 
the system dynamics is arrested again, leading to a reentrance in the 
non-ergodic state. (b) Quantum mode-coupling theory (QMCT) prediction on 
the dynamic phase diagram of a hard-sphere fluid. $\varphi$ is the  packing fraction and $\Lambda^*=\Lambda/\sigma$, is the ratio of the thermal de Broglie wavelength to the particle size. High (low) values of $\Lambda^*$ indicate strong (weak) quantum effects. The inset shows the non-equilibrium state diagram obtained from classical mode-coupling theory, where the necessary input have been delivered by quantum mechanical calculations of the structure factor.  This emphasizes the importance of QMCT to predict the reentrant glass behaviour. Images (modified) from (a)~\cite{Eckert2002} and (b)~\cite{Markland2011} (reproduced with permission).
}
\label{fig:reentrant-glass-bulk}
\end{figure}

Historically, the concept of a reentrant glass has been introduced long ago in the spin glass community to describe a transition from a spin glass state to a ferromagnetic phase and back to the glassy state~\cite{Sherrington1975,Binder1986,Mezard1987,Nisoli2013}.  It has  been shown later that a 
reentrant behaviour may also arise in structural glasses from the competition of entropic and energetic effects as, e.g., in polymer-colloid mixtures. In the 1990s, it has been conjectured that the colloid-colloid short-range attraction due to the presence of polymers (depletion interaction) may give rise to interesting new phase behaviour~\cite{Pusey1991}. Roughly a decade later, theoretical work based on the mode-coupling theory (MCT) of the glass transition led to the prediction of a reentrant glass transition and revealed the existence of a new arrested state, dominated by the short-range attraction, the so-called attractive glass as opposed to the well-known repulsive glassy state in hard-sphere (HS) colloids~\cite{Dawson2000}. These predictions have been confirmed by experiments on colloid-polymer mixtures~\cite{Eckert2002,Poon_TOP2002,Poon2003,Pham2002,Pham2004,Charbonneau2007} and computer simulations~\cite{Pham2002,Zaccarelli2009}
(Fig.~\ref{fig:reentrant-glass-bulk}a).

Introducing size disparity has also been demonstrated to provide an alternative route for the emergence of a reentrant glass-transition scenario in colloidal hard-sphere mixtures~\cite{Foffi2003,Zaccarelli2004}. This leads to competing near-ordering in colloidal hard-sphere mixtures and a dependence of the relaxation dynamics on the concentration of the additive component (e.g., smaller particles). The ratio of the short-time dynamics  of the smaller particles relative to the dynamics of the larger ones plays an important role here~\cite{Zaccarelli2004}. 

Inserting the liquid in a frozen disordered host structure has been identified to be another way to induce non-monotonic effects \cite{Krakoviack2005,Krakoviack2007,Krakoviack2011,Kurzidim2009,Kim2009}. 
\revision{A genuinely new type of non-monotonic glass behaviour has been recently predicted via the quantum mode-coupling theory, developed by  Reichman and co-workers and found in path integral-based molecular dynamics simulations~\cite{Markland2011}, applied to a quantum version of the well-known Kob-Andersen binary Lennard-Jones mixture~\cite{Kob1994}. These studies ---which were motivated by the discovery of the 'superglass' state in $^4$He, a state characterized at the same time by superfluidity and a frozen amorphous structure~\cite{Boninsegni2006,Biroli2008,Hunt2009}--- reveal that quantum fluctuations may both enhance and hinder the system dynamics with respect to the classical limit~\cite{Markland2011}. As a quantitative measure for the degree of quantumness, the ratio of the thermal de Broglie wavelengths, $\Lambda=h/\sqrt{2\pi m \kB T}$, to the particle diametre, $\sigma$, is used, $\Lambda^*=\Lambda/\sigma$. Here, $h$ is  Planck's constant, $\kB$ is the Boltzmann constant, $m$ denotes the particle mass and $T$ is temperature. More specifically, as indicated by the arrows in Fig.~\ref{fig:reentrant-glass-bulk}b, the dynamically arrested state can be reached either by starting from the classical limit (small $\Lambda^*$) and increasing $\Lambda^*$ gradually, or via decreasing it from the quantum-dominated regime ($\Lambda^*\approx 10^{0}$) towards lower values~\cite{Markland2011}. Since the above defined ratio depends on temperature, particle mass and size, the mechanism based on quantum fluctuations provides a qualitatively new access to the reentrant glass-transition phenomenon (Fig.~\ref{fig:reentrant-glass-bulk}b).}

All the above mentioned cases of the reentrant glass do have in common that they occur in  bulk, i.e., in the absence of interfaces or confinement. On the other hand, recent works provide evidence that strong geometric confinement may also lead to a reentrant glass-transition behaviour. 
Walls introduce a competition between the oscillations of the density profile (layering) and the local packing structure, 
thereby bringing about the issue of commensurability~\cite{Lang2010,Mandal2014}. Recent 
experimental~\cite{Nugent2007}, theoretical~\cite{Lang2010,Lang2012,Lang2013} and computer simulation~\cite{Mittal2008,Mandal2014} studies have revealed that this competition may lead to a non-monotonic dependence of the relaxation dynamics on the wall-to-wall separation and that this effect extends to higher densities and leads to a multiple reentrant glass transition scenario~\cite{Mandal2014}.

This review is organized as follows. We first give in the next section a brief account of reentrant glass-transition phenomenon in  bulk. Section \ref{sec:Reentrant-glass-in-confinement} then provides a survey of a selected set of observations regarding the glass transition upon confinement. Starting with the vastly studied case of what we call 'weak confinement', we elaborate an argument as to why non-monotonic effects could not be observed in these works. This brings us naturally to the domain of strong confinement, where the competition of confinement-induced layering and local packing may and often does give rise to non-monotonic effects and reentrant phenomena. In section \ref{sec:Theory-and-simulation}, a number of ideas are presented aiming at rationalizing these observations. There, we first invoke arguments based on the  excess entropy~\cite{Rosenfeld1977,Rosenfeld1999,Dzugutov1996} and the particle insertion probability~\cite{Widom1963,Mittal2008,Bollinger2015}. The newly developed mode-coupling theory of the glass transition in confinement~\cite{Lang2010,Lang2012,Lang2013,Franosch:2014} is then introduced in section \ref{subsec:mct}. This is followed by a test of the predictions of this theory regarding the non-monotonic effects of confinement on the non-equilibrium state diagram via computer simulations. Section \ref{sec:summary} closes this review with a summary and perspectives for future work.

A last remark is still at order here. We are well aware of the fact that, despite all our efforts, this review cannot be 
exhaustive  with regard to ongoing research in this rapidly growing field. Therefore, we apologize if one or the other researcher misses his/her relevant contribution in this review. It is also inevitable that the present article is biased with regard to the authors' own work.
This is, after all, what we know the best and can describe most adequately. Given these restrictions, we, nevertheless, hope that the present review will serve as a reliable guide for those scientists who start their work in this research area, helping them to identify unresolved questions and locate the potentials for new effects.

%%%%%%%%%%%%%%%%%%%%%%%%%%%%%%%%%%%%%%%%%%%%%%%%%%%%%%%%%%%%%%%%%%%%%%%%%
%%%%%%%%%%%%%%%%%%%%%%%%%%%%%%%%%%%%%%%%%%%%%%%%%%%%%%%%%%%%%%%%%%%%%%%%%
%%%%%%%%%%%%%%%%%%%%%%%%%%%%%%%%%%%%%%%%%%%%%%%%%%%%%%%%%%%%%%%%%%%%%%%%%
%%%%%%%%%%%%%%%%%%%%%%%%%%%%%%%%%%%%%%%%%%%%%%%%%%%%%%%%%%%%%%%%%%%%%%%%%
\section{Reentrant glass in the bulk}
\label{sec:Reentrant-glass-in-the-bulk}

There are only few experiments providing evidence for the reentrant phenomenon in glass forming systems. Among these, experiments on colloid-polymer mixtures seem to provide the most direct information (see, e.g., Fig.~\ref{fig:reentrant-glass-bulk}a) and thus deserve some attention here. 

\begin{figure}
\begin{center}
\hspace*{10mm}(a)\includegraphics[height=2.5cm]{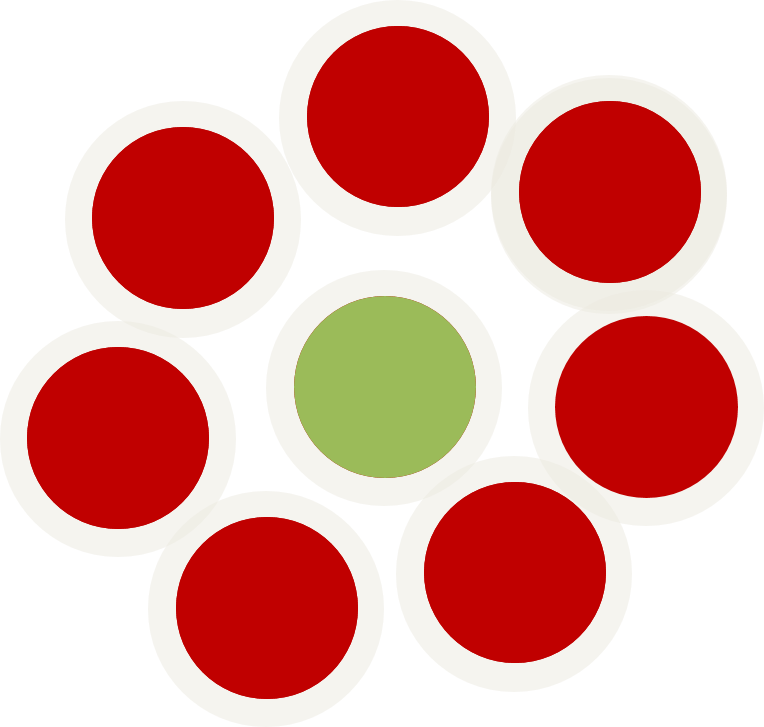}\hspace*{2mm}
(b)\includegraphics[height=2.5cm]{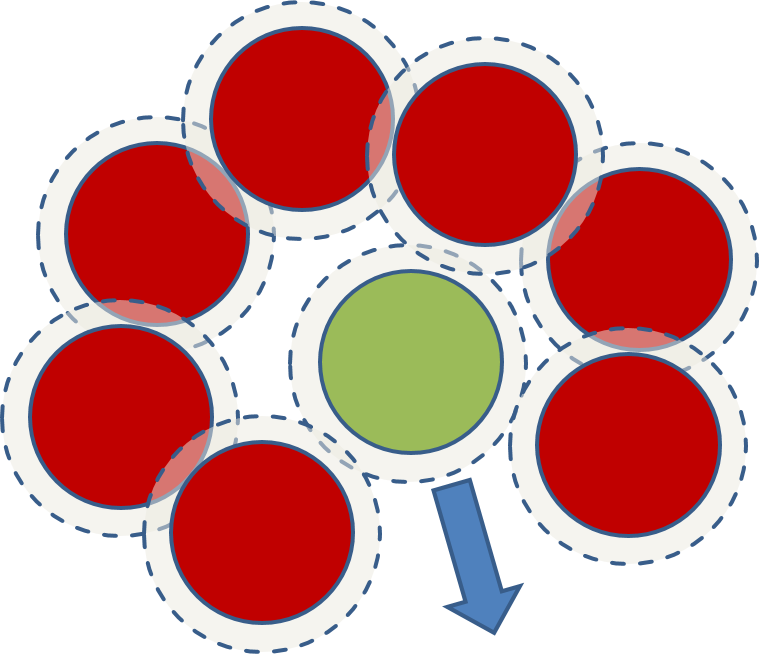}\hspace*{2mm}
(c)\includegraphics[height=2.5cm]{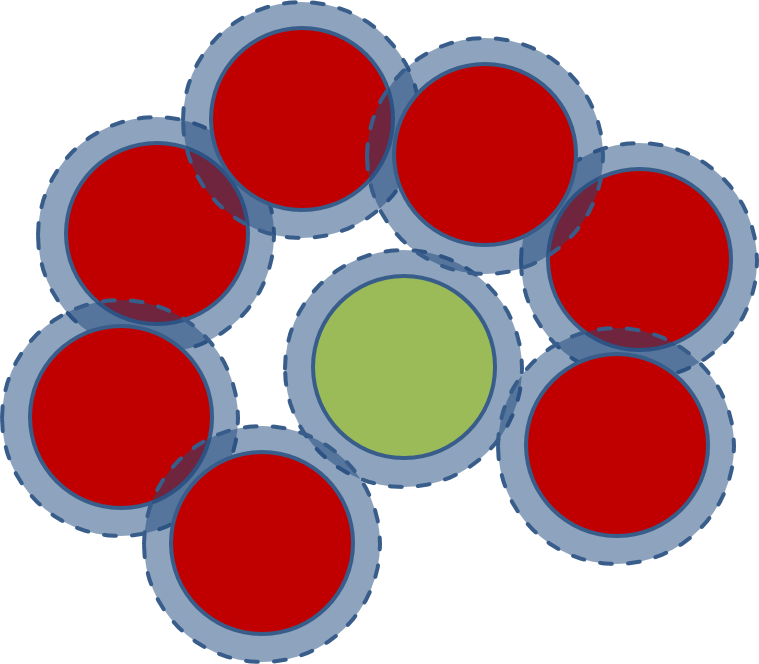}\\
\unitlength=1mm
\begin{picture}(200,40)
\put(17,0){(d)}
\put(30,5){\includegraphics[height=2.5cm]{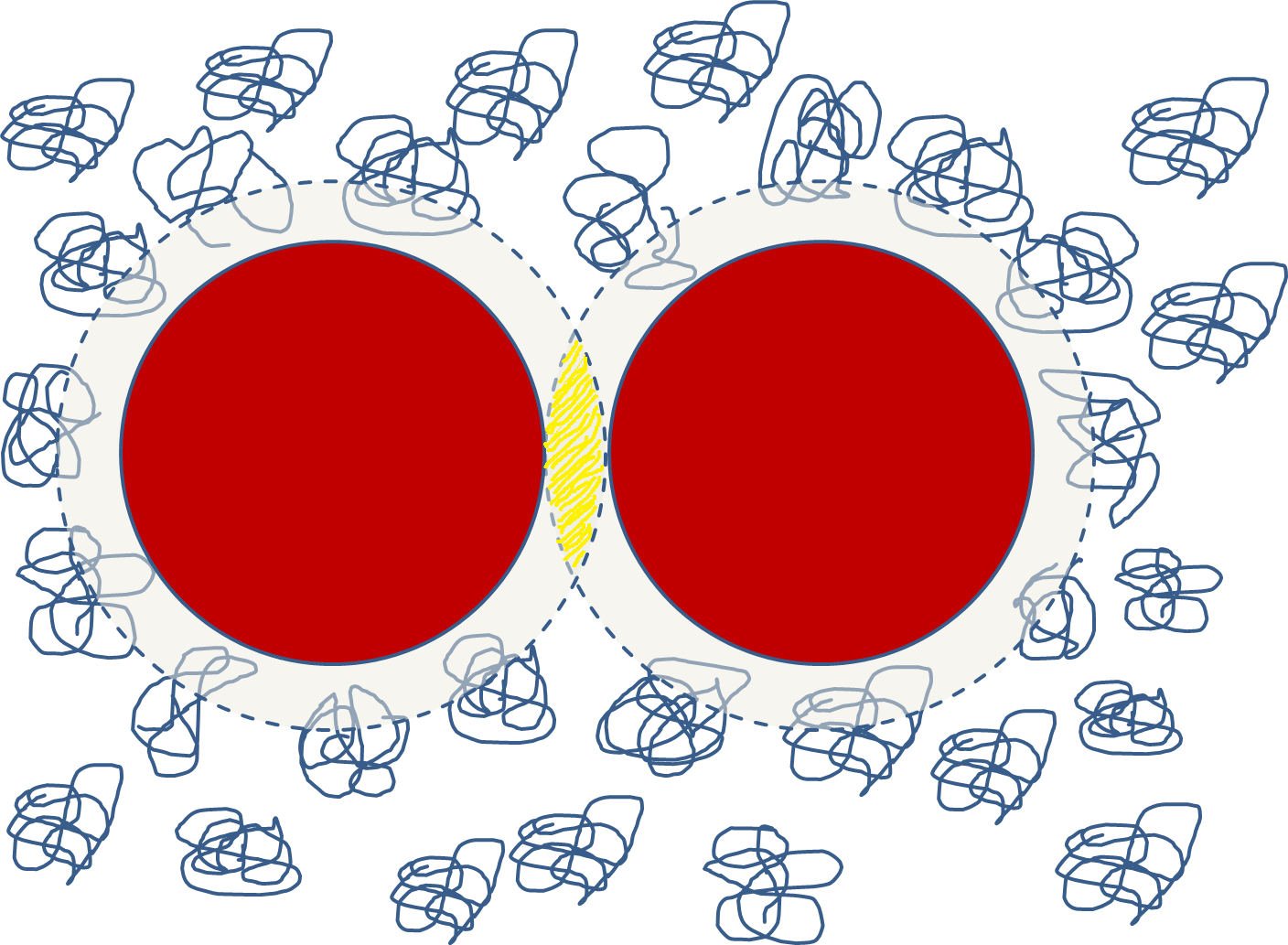}}
\put(80,0){(e)\includegraphics[height=4cm]{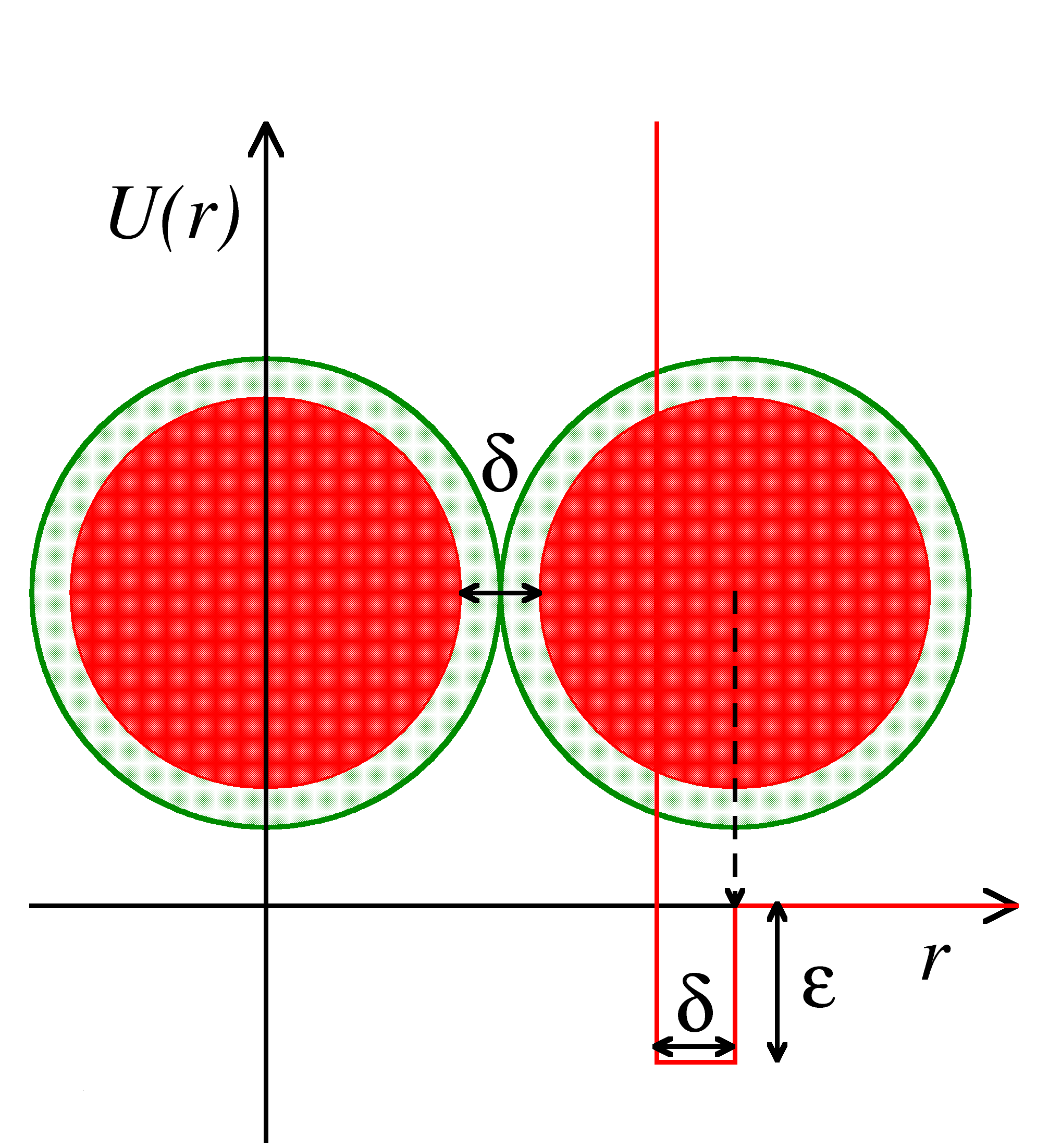}}
\end{picture}
\caption[]{ (colour online).
(a-c) An interpretation of the reentrant glass transition---as proposed by Wilson Poon in his topical review article~\cite{Poon_TOP2002}---in a colloidal glass, induced by adding short-range attraction. (a) For sufficiently weak, or in the absence of, attraction (low polymer concentration in the solvent), the green particle in the centre is caged by its neighbours. This is 
the case of a repulsive glass. (b) Introducing short-range attraction via 
adding (more) polymers causes the particles forming the cage to come closer together upon formation of intermittent bounds, thereby opening a path for the central particle to escape. This eventually melts the glass and marks the polymer-induced glass-liquid transition. (c) At sufficiently high attraction strength (high polymer concentration), long-lived bonds emerge between all touching particles and the whole system arrests into another glassy state, an attractive glass.
(d) A schematic representation of the depletion interaction. The entropic force associated with restricted polymer conformation close to a hard surface~\cite{Varnik2002e} hinders the centre of a polymer molecule from coming closer than a certain distance, approximately its own radius of gyration, $\Rg$, to the surface of a  colloid. (e) A simple step potential-well to mimic the depletion interaction. The energy is reduced by an 
amount of $\epsilon$, as soon as the distance between the centers of two colloids becomes smaller than the sum of their hard-core 
radii plus twice the thickness of the depletion zone, $\delta$. For distances beyond this, 
the polymers do not 'see' one another. For distances below the sum of the radii, on the other hand, the hard-core repulsion pushes the colloids apart. 
The range of the depletion interaction is of the order of the polymers' radius of gyration, $\delta \sim \Rg$. The depth, $\epsilon$, of the depletion interaction is controlled by the polymer concentration~\cite{Fuchs2000,Fuchs2001}.}
\end{center}
\label{fig:poon+varnik}
\end{figure}

Following Wilson Poon~\cite{Poon_TOP2002}, a qualitative interpretation of the reentrant glass phenomenon in colloid-polymer mixtures may be given 
as follows. Consider a suspension containing hard-sphere colloids at a relatively high packing fraction, where each colloid particle is arrested in the cage formed by its neighbours and cannot leave it within the observation time window (green circle in Fig.~\ref{fig:poon+varnik}a). This is an example of a repulsive glass. Due to entropy, however, each colloid explores a slightly larger volume than its exact size. 
Adding polymers to the suspension introduces a short-range attraction which gives rise to rather weak and 
intermittent ``bonds'' between colloids. As a result, some particles come closer together, forming 
an (intermittent) cluster and leaving space to others, which may use the thus opened 'door' to escape from 
the cage (Fig.~\ref{fig:poon+varnik}b). The glass thus melts under the action of a moderate short-range attraction. As the polymer concentration 
increases further, the strength and hence the life time of the bonds grows, eventually leading to long-living bonds and 
a new, kinetically arrested, state (Fig.~\ref{fig:poon+varnik}c). Since this behaviour is 
dominated by attractive forces, it is called the attractive glass~\cite{Zaccarelli2004}.

It is remarkable that these investigations have largely been motivated by theoretical predictions, 
based on the mode-coupling theory, that adding short-range attraction to a hard-sphere colloidal glass may give rise 
to a non-monotonic non-equilibrium state diagram with the possibility of a reentrant glass~\cite{Dawson2000}. A test of these predictions became possible with the advent of colloid-polymer mixtures. Indeed, a well-controlled way 
to introduce a short-range attraction in a suspension of hard-sphere particles is to add polymers. 
In a solvent, a polymer chain explores a certain spatial domain with a linear dimension of its  radius of gyration, $\Rg$. If the centre of mass of a polymer coil comes to a distance significantly below $\Rg$ from the surface of a colloid, it cannot explore the full conformation space but must stretch along the direction parallel to the colloid's surface~\cite{Varnik2002e}. This conformational distortion prevents polymers from being too close to the surface of a colloid. Each colloid is, therefore, surrounded by an essentially polymer-free layer (depletion zone) of thickness, $\delta \sim \Rg$  (Fig.~\ref{fig:poon+varnik}d). The presence of this depletion layer reduces the effective volume accessible to the polymers' centers of mass and thus leads to a decrease of the associated configurational entropy. If, however, the surfaces of two colloids come closer than $\delta$, then the depletion zones overlap and the total excluded volume diminishes by the size of the overlap region. The corresponding raise in the entropy manifests itself as a net force which pushes the particles together. This is the very origin of the so-called depletion attraction~\cite{Asakura1954,Tuinier2000,Fuchs2000,Fuchs2001}.

Regarding the reentrant glass transition in binary mixtures, it turns out that an interpretation of this phenomenon is not 
as straightforward as in the case of colloidal hard spheres with short-range attraction. In particular, one finds that 
the occurrence of a reentrant behaviour crucially depends on the ratio of the short-time dynamics  of the smaller (additive) particles relative to the dynamics of the larger colloids~\cite{Zaccarelli2004}.

%%%%%%%%%%%%%%%%%%%%%%%%%%%%%%%%%%%%%%%%%%%%%%%%%%%%%%%%%%%%%%%%%%%%%%%%%
%%%%%%%%%%%%%%%%%%%%%%%%%%%%%%%%%%%%%%%%%%%%%%%%%%%%%%%%%%%%%%%%%%%%%%%%%
%%%%%%%%%%%%%%%%%%%%%%%%%%%%%%%%%%%%%%%%%%%%%%%%%%%%%%%%%%%%%%%%%%%%%%%%%
%%%%%%%%%%%%%%%%%%%%%%%%%%%%%%%%%%%%%%%%%%%%%%%%%%%%%%%%%%%%%%%%%%%%%%%%%
\section{Reentrant glass in confinement}
\label{sec:Reentrant-glass-in-confinement}

In this section, we collect some important facts and observations regarding confinement effects on the glass transition. In the first part, we discuss how, to our understanding, studies of confined systems became interesting to the glass community. 
The second part of this section presents some of the first observations of non-monotonic effects on the dynamics of fluids, 
arising from geometric confinement. Here, the set of physical systems is rather diverse and the underlying mechanisms are 
by part qualitatively distinct. The guiding idea in this section is not to provide a simple universal picture. Rather, we wish to first give here a flavour of the range of possibilities. A discussion, focused on identifying the generic effects of confinement on the glass transition, will be presented in section \ref{sec:Theory-and-simulation}.

%%%%%%%%%%%%%%%%%%%%%%%%%%%%%%%%%%%%%%%%%%%%%%%%%%%%%%%%%%%%%%%%%%%%%%%%%
%%%%%%%%%%%%%%%%%%%%%%%%%%%%%%%%%%%%%%%%%%%%%%%%%%%%%%%%%%%%%%%%%%%%%%%%%
\subsection{Monotonic effects of weak confinement}
\label{subsec:waek-confinement}

For more than two decades, the study of confinement effects on the glass transition was mainly motivated by the desire to shed light onto the possible existence of a growing length scale associated with the glass transition~\cite{Fischer1992,Donth2001,Biroli2006}. Half a century ago, Adam and Gibbs introduced the concept of cooperatively rearranging regions, whose characteristic size is supposed to diverge upon approaching the glass transition~\cite{Adam1965}. Consequently, within the model proposed by Adam and Gibbs, the structural relaxation time would diverge via a power law , $\taurelax \sim l^z$, with $l$ being a correlation length (the characteristic size of a cooperatively rearranging region) and $z$ the corresponding critical exponent. 

In this context, the fact that confinement might set a limit to the growth of the long sought diverging length scale appeared 
as a promising route: If there really is a correlation length which grows on approaching the glass transition, 
it should stop growing further along the confined direction as soon as it reaches the linear dimension of the confinement. 
In this case, one expects a faster dynamics under confinement as compared to the reference bulk system.

However, intense experiments \cite{Fryer2001,MorineauEtal:JCP2003,AlbaEtal:EPJE2003,JacksonMcKenna1996,WendtRichert1999,PissisEtal:JPCM1998,ArndtEtal:PRL1997,DemirelGranick:JCP2001,KeddieEtal:EPL1994,Jones:Current1999,ForrestDalnoki:AdvColSci2001,Forrest:EPJE2002,Herminghaus:EPJE2002,HerminghausEtal:PRL2004,Johannsmann:EPJE2002,Reiter:EPJE2002,KajiyamaEtal:Polymer1998,FukaoEtal:PRE2000,SchoenhalsStauga:JCP1998,EllisonTorkelson:NatureMaterials2003,SinghEtal:SolidFilms2004,Pye2011,Yin2013,Yin2013b}, computer simulations \cite{VarnikEtal:PRE2002,VarnikEtal:EPJE2002,VarnikEtal:EPJE2003,VarnikBinder:JCP2002,XuMattice:Macro2003,StarrEtal:Macro2002,YoshimotoEtal:JCP2005,JaindePablo:PRL2004,Torres2000,MansfieldEtal:Macro1991,ManiasEtal:ColSurf2001,BaljonEtal:Macro2005,Roder:JCP2001,Scheidler2000,Scheidler2002,Scheidler2004,GalloEtal:EPL2002,FehrLoewen:PRE1995,BoddekerTeichler:PRE1999,Xia2013,Krishnan2012,Ingebrigtsen2013,Williams2013,Rodriguez-Fris2014}, and theoretical work \cite{PGG:EPJE2000,McCoyCurro:JCP2002,Truskett2003,Mittal2004,Chow:JPCM2002,LongLequeux:EPJE2001,MerabiaEtal:EPJE2004,Mayes:Macro1994,Ngai:EPJE2003,Mirigian2014}, revealed a far richer phenomenology than originally expected. 
In particular, it was found that the confinement effects largely depend on the particle scale structure 
of the confining walls and the energetics of the wall-liquid interactions. 
For example, strongly attractive wall-liquid interactions may hinder the dynamics of 
liquid particles by, e.g., trapping them in a potential minimum~\cite{Torres2000}. A special role here is also 
played by particle scale wall corrugations~\cite{Scheidler2000b,Baschnagel2005} and the specific arrangement of the wall 
particles. Crystalline walls, for example, are known to enhance layering of fluid particles, while the effect of 
amorphous walls is relatively weak in this respect~\cite{Varnik2009}. Similarly, crystalline walls usually 
lead to a stronger decrease of particle dynamics as compared to amorphous walls~\cite{VarnikEtal:PRE2002}.

Despite the diversity of the observed confinement effects, all these studies show a \emph{monotonic} dependence of \revision{structural relaxation time, diffusion coefficient and shear viscosity} on the linear dimension of the confinement, e.g., the plate separation in a planar geometry. This behaviour can be rationalized as follows. A careful survey reveals that, in all the above studies, the range of 
confinement allows for the existence of a bulk-like region sufficiently far from the wall. In other words, there is a separation 
between the length scale of the confinement, given by the plate separation, $H$, and the characteristic range, $\xiwall$, over 
which the wall may induce layering and influence system dynamics. Thus, as long as the wall-to-wall separation is larger 
than $H_{\mathrm{c}} \simeq 2\xiwall$, an increase of $H$ will only increase the extension of the bulk-like 
region away from the walls, thereby weakening the wall effects in a monotonic manner. In order to observe non-monotonic effects, the 
plate separation shall be smaller than this characteristic size, $H<H_\mathrm{c}$. In the next sections of this review, we are going to explore this interesting range.

%%%%%%%%%%%%%%%%%%%%%%%%%%%%%%%%%%%%%%%%%%%%%%%%%%%%%%%%%%%%%%%%%%%%%%%%%
%%%%%%%%%%%%%%%%%%%%%%%%%%%%%%%%%%%%%%%%%%%%%%%%%%%%%%%%%%%%%%%%%%%%%%%%%
\subsection{Reentrance in strong confinement}
\label{subsec:strong-confinement}

A question arising from the above consideration is what happens if the dimension of the confinement falls \emph{below} 
the range of the wall effects, i.e., if $H<2\xiwall$? In this case, the entire system is composed of close-wall regions only, 
whose properties (packing structure, density, dynamics) may vary significantly upon a change of the wall-to-wall separation.

This limit of strong confinement has been the focus of a number of recent studies and the interest in this field is growing rapidly at  present~\cite{Mittal2006,Mittal2008,Satapathy2008,Satapathy2009,Krishnan2012,Lang2010,Lang2012,Bordin2012,Ingebrigtsen2013,Nygard2014}.

\begin{figure}
\begin{center}
\includegraphics[height=5cm]{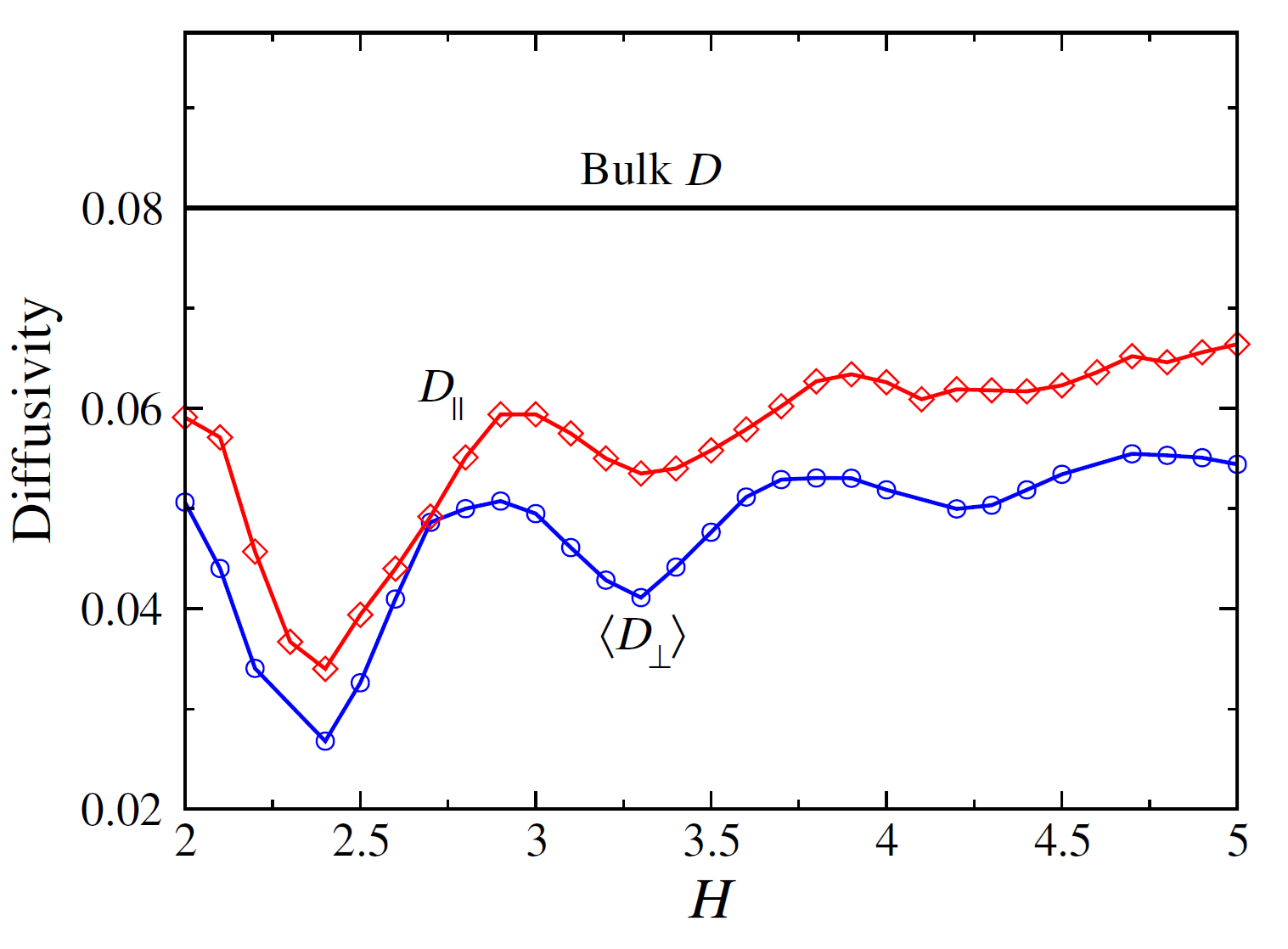}
\end{center}
\caption[]{
(colour online).
Molecular dynamics results on the single particle dynamics of a HS fluid, confined between two planar smooth hard walls, obtained by Mittal and coworkers. The plot shows the slit-averaged diffusion coefficient in directions parallel \revision{($D_\parallel$)} and perpendicular  \revision{($D_\perp$)} to the walls. The packing fraction (averaged over the entire slit) is $\phi=0.40$. The horizontal line marks the diffusion coefficient of the corresponding bulk system at the same packing fraction. Figure from~\cite{Mittal2008} (reproduced with permission).}
\label{fig:Mittal}
\end{figure}

Mittal and co-workers were among the first groups to report on a non-monotonic effect of confinement on particle dynamics~\cite{Mittal2008}. Via event driven molecular dynamics (MD) simulations of a hard-sphere fluid confined between two planar and parallel hard walls, 
these authors revealed an oscillatory behaviour of diffusive dynamics as function of slit width both 
for the motion parallel to the wall surface as well as along the normal direction (Fig.~\ref{fig:Mittal}). 
The packing fraction in these simulations was well below the fluid-crystal coexistence regime so that crystallization issues did not show up.

\begin{figure}
\begin{center}
(a)\includegraphics[height=5cm]{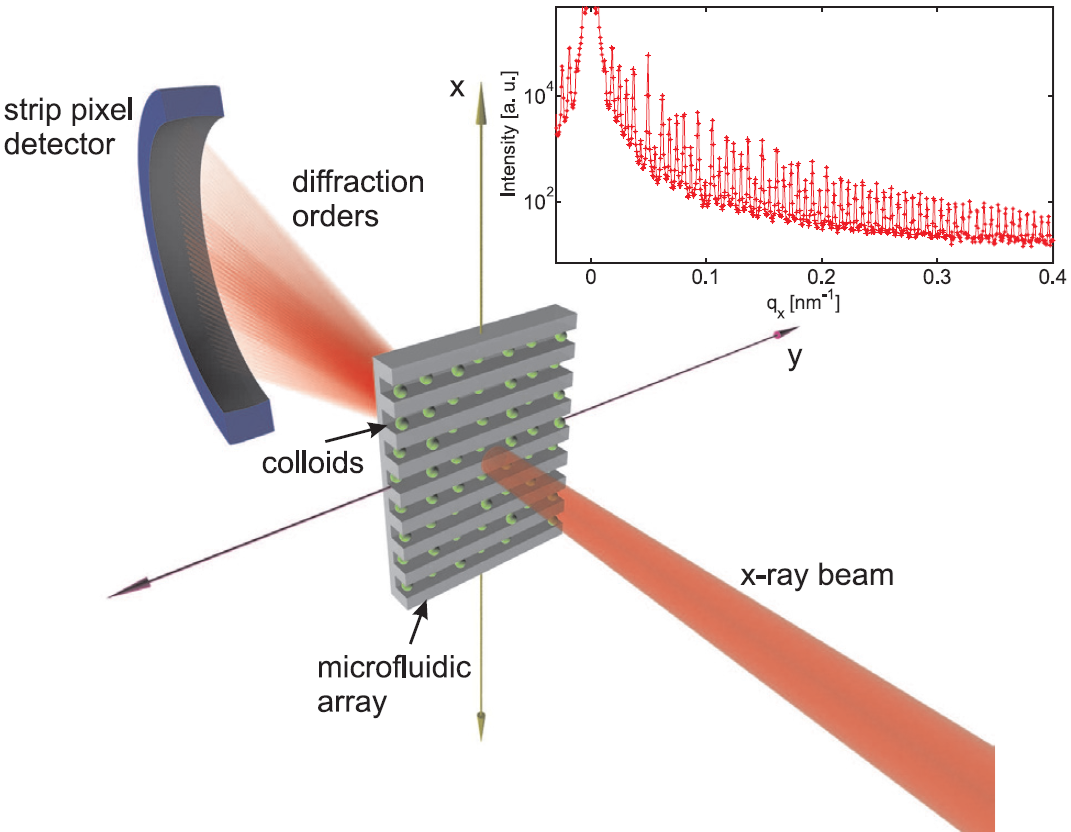}\hspace*{5mm}
(b)\includegraphics[height=5cm]{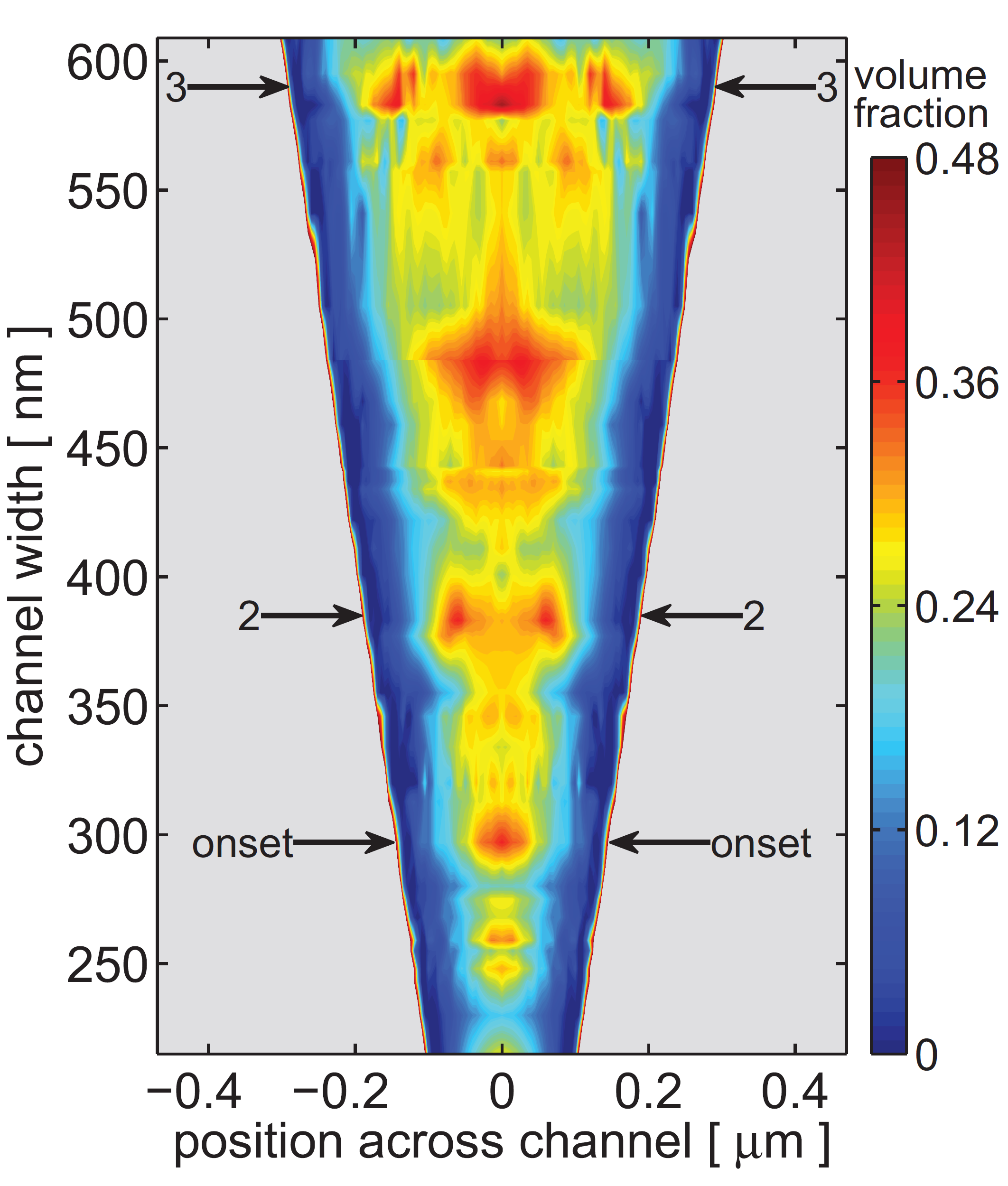}
\end{center}
\caption[]{
(colour online).
(a) Schematic illustration of the setup used by Satapathy and coworkers in their X-ray diffraction experiments on charged colloidal suspensions. The periodic arrays of rectangular channels act as fluid containers and are directly connected to a reservoir of colloidal suspension which allows unhindered transfer of particles and ions. The inset shows a typical diffraction pattern measured from such a microfluidic channel array filled with a colloidal dispersion.
The colloidal diametre  is 120nm. (b) Experimental contour plots of the local packing fraction versus the width of the channel.  The channel walls are like-charged thus pushing the colloids towards the centre of channel. The arrows mark the onset of buckling and the cross sections with two- and three-layer structures. Figure from~\cite{Satapathy2009} (reproduced with permission).}
\label{fig:Satapathy}
\end{figure}

Another example for the occurrence of non-monotonic effects of confinement is provided by experiments of Satapathy and co-workers~\cite{Satapathy2008,Satapathy2009} who resolved, via X-ray scattering, the spatial variations of the colloidal packing fraction across an array of planar slits of 
various widths, embedded in a solvent which served as the reservoir for colloidal particles. By analysing the thus obtained data, they could demonstrate oscillations of the local packing 
fraction across the slit with quasi-perfect layering for integer values of the slit 
width (Fig.~\ref{fig:Satapathy}). These authors, however, did not address the effect of this layering transition on the system dynamics.

Even though the presence of electric charges does not allow a direct comparison to uncharged hard-sphere colloids, it is interesting to note that the observed layering is qualitatively similar to the results of computer simulations of hard spheres in a wedge-shaped channel by Mandal and coworkers~\cite{Mandal2014}. As will be discussed in section~\ref{subsec:simulations}, these simulation results are supported 
by accurate theoretical calculations and suggest that the observed density variations are a generic feature induced by confinement. To the best of our knowledge, there is yet no experimental 
study of such an effect for the case of charge neutral hard-sphere colloids.

\begin{figure}
\begin{center}
(a)\hspace*{5mm} \includegraphics[width=0.5\linewidth]{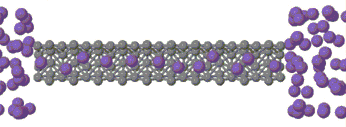}
(b)\includegraphics[width=0.55\linewidth]{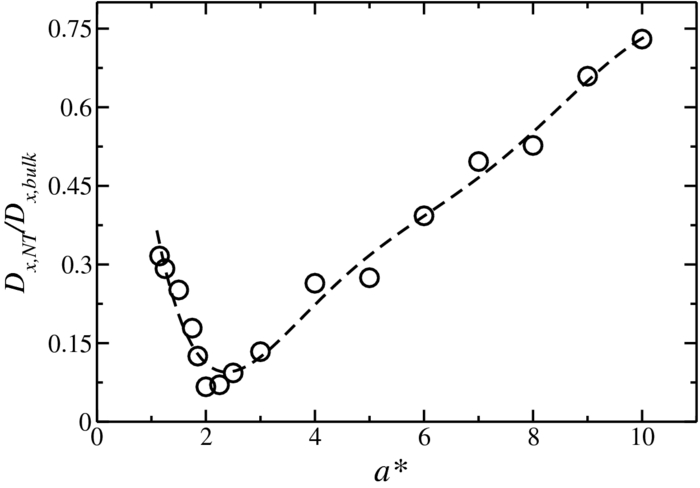}
(c)\includegraphics[width=0.55\linewidth]{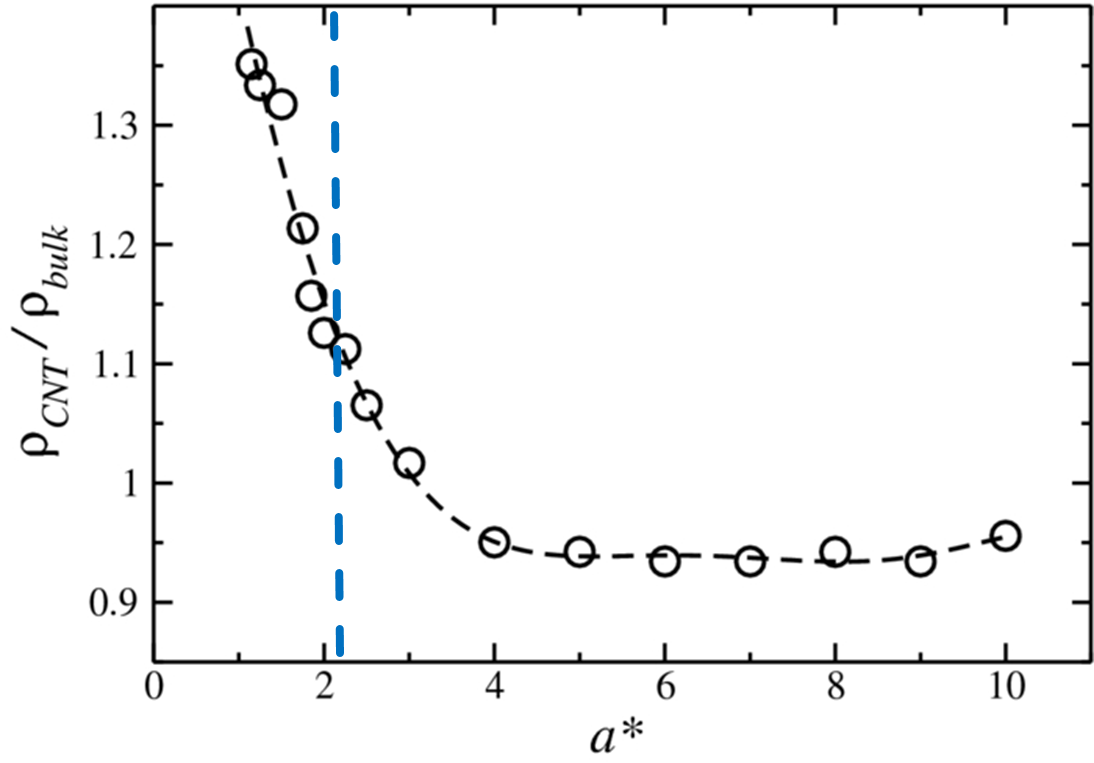}
\end{center}
\caption[]{(colour online). 
(a) The tube geometry studied by Bordin et al.~\cite{Bordin2012}. The tube is coupled to a particle reservoir, the latter being kept at constant pressure via movable side walls (laying outside the visible range here). (b) Fickian diffusion coefficient versus pore radius, $a^*=a/\sigma$. (c) Monotonic dependence of average density on pore size. The vertical dashed line marks the tube radius, $\amin^*\approx 2$, for which the self diffusion coefficient attains a minimum. It serves to highlight the qualitatively different 
dependence of density and diffusion on confinement. The figure (slightly adapted) 
is taken from~\cite{Bordin2012} (reproduced with permission).}
\label{fig:Bordin-tube}
\end{figure}

Among other groups focusing on strong confinement, Bordin et al.\ recently report on a non-monotonic effect in a 
simulated core-softened model fluid (mimicking water) in a cylindrical nano-pore (Fig.~\ref{fig:Bordin-tube}a). Starting 
at a large tube radius, $a$, and decreasing it gradually, the effect of confinement is found to reduce the single particle diffusion coefficient (Fig.~\ref{fig:Bordin-tube}b). This trend is, however reversed as the tube radius falls roughly  below twice the core-diametre , $a=\amin \approx 2.1 \sigma$. For $a<\amin$, the diffusion coefficient increases upon a further decrease of $a$. It is, noteworthy that these simulations are performed at constant pressure and that the pore-averaged density varies by roughly 50\% in the studied range of pore radii. A survey of the average density, however, reveals that it is a monotonic function of the pore radius~(Fig.~\ref{fig:Bordin-tube}c). Thus, the non-monotonic dependence of the single particle dynamics on pore radius in Fig.~\ref{fig:Bordin-tube}b is not a mere consequence of density variations. Seeking for other causes of this non-monotonic behaviour, Bordin and coworkers report that the minimum in the diffusion coefficient at $\amin$ coincides with the onset of a depletion of hydrogen bonds. Following Bordin and coworkers, the confinement-induced reduction of particle mobility competes with the enhancement of the dynamics resulting from a decrease in number of hydrogen bonds per particle. The latter, being also a consequence of the presence of confining walls, gains over the former for $a<\amin$~\cite{Bordin2012}.

We also mention the recent work by Krishnan and Ayappa~\cite{Krishnan2012}, who performed molecular dynamics simulations of a 
strongly confined single-component Lennard-Jones fluid containing 2---4 particle layers. Via an analysis of the temperature-dependence of the relaxation times and the diffusion coefficient, these authors concluded that the confined fluid can be classified as a fragile glass. It was also shown that the critical mode-coupling  temperature decreases with increasing confinement. Probably due to the restricted sampling of the wall-to-wall separation (only three values of $H$ have been considered), a non-monotonic behaviour has not been detected in this work.

%%%%%%%%%%%%%%%%%%%%%%%%%%%%%%%%%%%%%%%%%%%%%%%%%%%%%%%%%%%%%%%%%%%%%%%%%
%%%%%%%%%%%%%%%%%%%%%%%%%%%%%%%%%%%%%%%%%%%%%%%%%%%%%%%%%%%%%%%%%%%%%%%%%
%%%%%%%%%%%%%%%%%%%%%%%%%%%%%%%%%%%%%%%%%%%%%%%%%%%%%%%%%%%%%%%%%%%%%%%%%
%%%%%%%%%%%%%%%%%%%%%%%%%%%%%%%%%%%%%%%%%%%%%%%%%%%%%%%%%%%%%%%%%%%%%%%%%
\section{Theory and simulation}
\label{sec:Theory-and-simulation}
In this section, we first discuss how confinement effects on the dynamics of fluids can be rationalized via approaches which relate the structural relaxation dynamics to the number of available configurations, i.e., the excess entropy. A test of these ideas via computer simulations shows remarkable agreement between theory and simulation. In the same spirit, a linear relation between the local dynamics and the local density profile is worked out using Widom's particle insertion method. This relation turns out to provide a good approximation for intermediate densities but fails at high packing fractions, indicating the need for more elaborates theories which account for strong coupling effects at high densities. One of the promising candidates here is the mode-coupling theory of confined fluids. After introducing the basic formalism of this theory, we then present its predictions about the confinement effects on the non-equilibrium state diagram, and compare them to the results obtained from computer simulations. In the last part of this section, these predictions are transferred to a wedge geometry,  and are complemented by direct simulations. This is reminiscent of the cone-plate channel, often used in experiments on colloidal suspensions. Evidence is provided that the non-monotonic effects persist in this case and may even give rise to the occurrence of multiple liquid-glass layers along the wedge.

%%%%%%%%%%%%%%%%%%%%%%%%%%%%%%%%%%%%%%%%%%%%%%%%%%%%%%%%%%%%%%%%%%%%%%%%%
%%%%%%%%%%%%%%%%%%%%%%%%%%%%%%%%%%%%%%%%%%%%%%%%%%%%%%%%%%%%%%%%%%%%%%%%%
\subsection{Excess entropy approaches}
\label{subsec:excess-entropy}

One of the appealing routes to rationalize confinement effects on the dynamics 
of supercooled liquids builds upon ideas relating the dynamics of these 
 systems to their static thermodynamic properties. Already in 1965, Adam and Gibbs 
postulated that the relaxation dynamics in dense liquids is due to 
cooperative rearrangement events in small domains which only weakly interact 
with one another and can essentially be regarded as statistically independent. 
Assuming that the 
probability of a relaxation event is proportional to the logarithm of the configurational 
entropy, they obtained for the average transition probability or the inverse 
relaxation time, $W = 1/\taurelax = A \exp(-C/(T\scf))$, where $A$ is a weakly 
temperature-dependent prefactor and $\scf$ is the configurational entropy of the 
entire system divided by the particle number. The factor $C$ in the exponent is 
given by $C=\Delta \mu \scf^*/ \kB$, where $\kB$ is the Boltzmann constant, 
$\Delta \mu$ is the typical change in chemical potential associated with the 
transition/rearrangement event and $\scf^*$ is the configurational entropy, 
associated with the number of available states \emph{within} the smallest 
cooperatively rearranging region~\cite{Adam1965}.

A different approach, first proposed by Rosenfeld~\cite{Rosenfeld1977} is motivated by the desire 
to relate dynamic properties of liquids to the equation of state~\cite{Rosenfeld1977} and leads to a dependence of 
the system dynamics on the so-called excess entropy, $\sex=\scf-\sid$, where $\scf$, 
as above, denotes the thermodynamic entropy per particle of the liquid and 
$\sid$ is that of the equivalent ideal gas at the same temperature and density. 
The relation for diffusion coefficient then reads $D \propto \exp(-C|\sex|)$, 
with $C$ being a weakly varying prefactor. It is noted that $\sex < 0$, since 
interactions between particles gives rise to correlations, which ultimately 
reduce the number of available configurations. A reduction of $\sex$ (i.e., an 
increase of |\sex|) results from a  more efficient trapping of particles and 
thus is associated with a slower dynamics. Roughly twenty years later, 
Rosenfeld proposed a variational thermodynamic perturbation theory for the justification 
of his model and its extension to the dilute-gas limit~\cite{Rosenfeld1999}. This scaling law was 
found to be consistent with molecular dynamics simulations of a variety of model 
systems including hard-sphere and soft-sphere fluids 
\cite{Rosenfeld1977,Rosenfeld1999} and liquid argon \cite{Bastea2003}.

In a further elaboration of the problem, proposed by Dzugutov~\cite{Dzugutov1996}, two factors 
determine the diffusive transport in a dense medium. On the one hand, the rate 
of hard-sphere-like particle collisions, $\Gamma$, gives the attempt rate to 
escape from the cage formed by the neighbours. On the other hand, the success of any of these attempts linearly scales with the number of available 
configurations per particle, which, compared to an ideal gas, is reduced by a 
factor of $\exp(\sex)$, where $\sex$ is the excess entropy introduced above (measured in units of $\kB$). This argument yields, $D \propto \Gamma \exp(\sex)$ and thus $D^*=D/(\sigma^2\Gamma) \propto \exp(\sex)$, where $\sigma$ is the diametre  of 
the  equivalent repulsive hard-core. It is noted that, in contrast to 
Rosenberg's result, there appears no prefactor in the exponent here. Dzugutov 
himself provided evidence for the validity of this scaling relationship via 
computer simulation studies of a solid state ionic conductor ($\alpha-$AgI) and 
quasicrystals~\cite{Dzugutov1996}. Hoyt and co-workers showed that it also 
applies, to a good approximation, to liquid metals and binary metallic 
liquids \cite{Hoyt2000} but that it fails in the case of silicon (SI). For 
hard-sphere fluids, Bretonnet showed that the range of applicability of the 
scaling proposed by Dzugutov is restricted to high packing 
fractions~\cite{Bretonnet2002}. This shortcoming has been circumvented by 
Samanta and co-workers, who used the mode-coupling theory to derive a more 
accurate relation between relaxation dynamics and excess entropy, thus extending 
the theory to significantly lower densities~\cite{Samanta2004}.

\begin{figure*}
\centering
\includegraphics*[height=7cm]{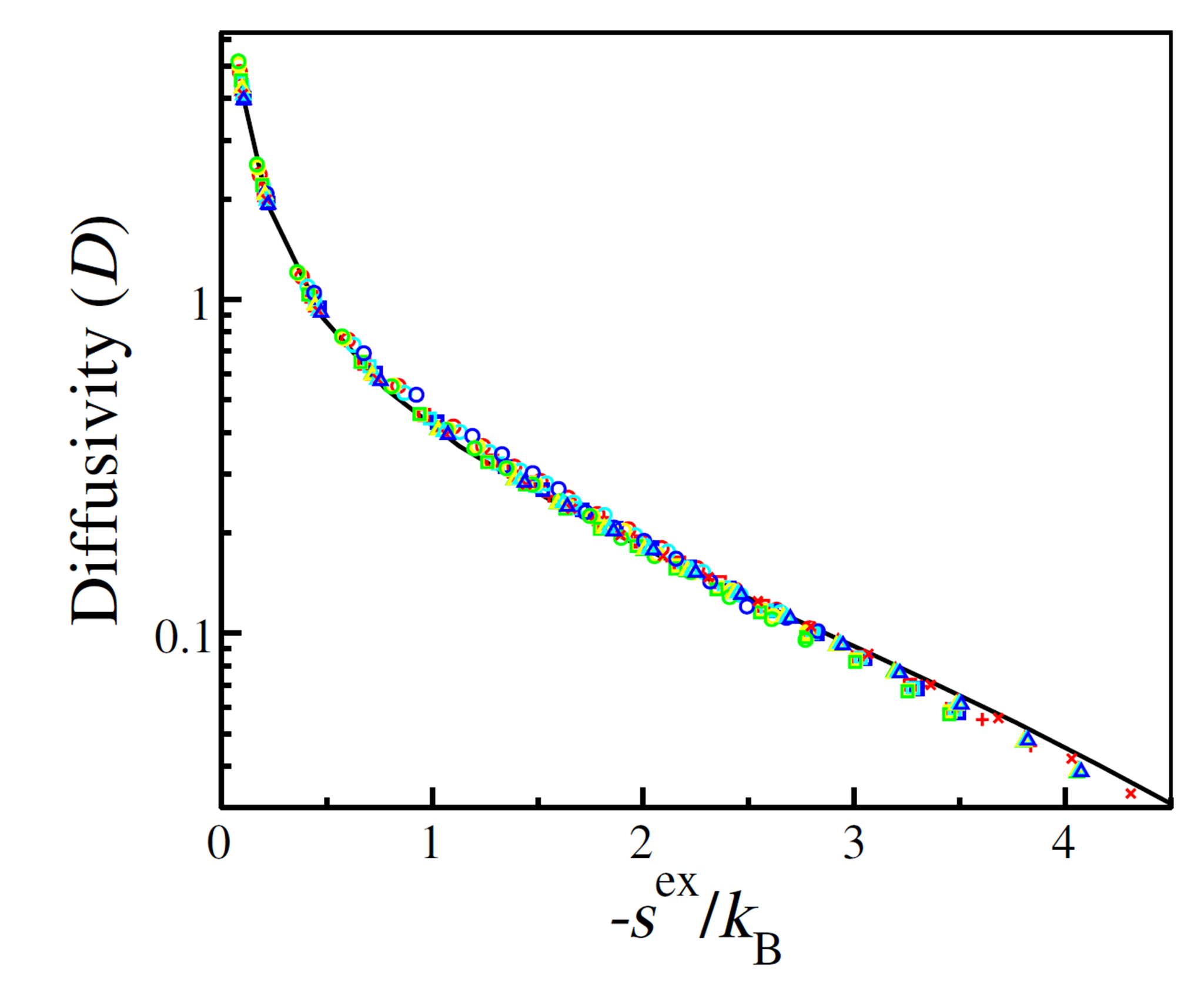}
\caption[]{(colour online). 
Diffusion coefficient, $D$, versus the excess 
entropy per particle, $\sex$, for the bulk HS fluid (solid curve) and for a HS 
fluid confined between two parallel plates (symbols). The symbols correspond to 
plate separations of $H=2.5$, (circle), 5 (square), 7.5 (plus), 10 (triangle 
up), and 15 (cross). The wall-fluid interaction potential is given by $u(r)=\infty$, $\ew$ 
and 0 for $r<1/2$, $1/2<r<1$ and $r>1$, respectively.  The colour codes are 
for the wall-fluid interaction parameters of $\ew=1$ (blue), 0.5 (cyan), 
0 (red), -0.5 (yellow), and -1 (green). Apparently, $D$ is 
directly correlated to the excess entropy regardless of the slit width and the 
wall-fluid interaction, the latter varied here from repulsive ($\ew>0$) through 
neutral ($\ew=0$) to attractive ($\ew<0$). All quantities are in reduced HS units.
Figure (adapted) from~\cite{Mittal2006} (reproduced with permission).
}
\label{fig:slit-D-sex-varying-ew}
\end{figure*}

The significance of the relation between excess entropy and single particle 
dynamics for the present review lies in the possibility to relate the dynamics 
of confined fluids to their static thermodynamic properties. This conjecture has 
been tested intensely in the past years by a number of researchers. In a 
systematic molecular dynamics study, Mittal and coworkers~\cite{Mittal2006,Mittal2008,Goel2008,Goel2009}
have shown that the concept of excess entropy can indeed help to rationalize confinement 
effects on single particle mobility for a number of fluid models and fluid-wall interaction 
parameters. An example is provided in Fig.~\ref{fig:slit-D-sex-varying-ew}, where diffusion coefficient in 
a planar slit is shown to correlate with the excess entropy for a number of 
slit widths and wall-fluid interaction parameters ~\cite{Mittal2006,Mittal2007,Mittal2007b}.

The data shown in Fig.~\ref{fig:slit-D-sex-varying-ew} underline 
the importance of the excess entropy for the dynamics of dense fluids both in the bulk and under confinement.
The non-monotonic effect of confinement is, however, not visible in this plot.
This is not surprising since, as argued above, single particle diffusion is 
expected to be a monotonic function of the excess entropy.

\begin{figure*}
\centering
\includegraphics*[width=0.55\linewidth]{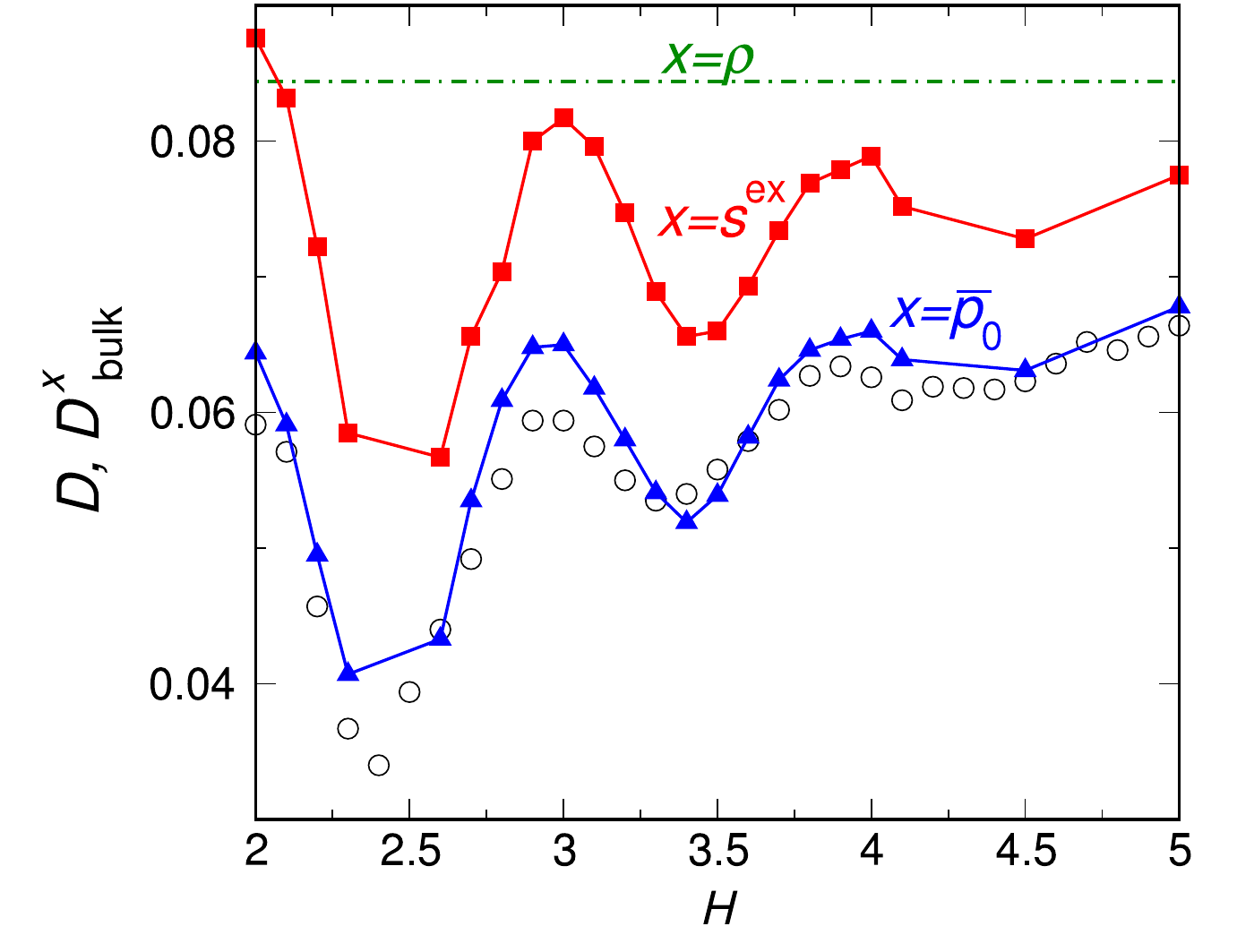}\hspace*{8mm}
\caption[]{(colour online). \revision{Slit-averaged} self diffusion coefficient, $D$, \revision{evaluated using particle displacements parallel to the slit walls, for} a hard-sphere fluid confined between two planar smooth hard plates versus plate separation, $H$ (open circles). The average particle number density is kept constant at $\rho=0.4 \times (6/\pi)$ for all investigated slit widths. This corresponds to a packing fraction of $\varphi=0.4$ (averaged over the entire slit). Data plotted as \revision{dashed line and connected symbols} give the diffusion coefficient of the equivalent bulk system, $\Dbulkx$, where, for each $H$, the value of the parameter x is fixed to that of the slit, x$=\rho=0.4 \times (6/\pi)$ (green dash-dotted line), x$=\sex(H)$ (red square), and x$=\bar{p}_0(H)$ (blue triangle). Figure from \cite{Goel2009} (reproduced with permission).}
\label{fig:Fig3-Goel}
\end{figure*}

The sought-after non-monotonic effect of confinement is revealed in Fig.~\ref{fig:Fig3-Goel}, where the single particle 
diffusion exhibits an oscillatory behaviour as the plate separation 
is varied at a fixed average density~\cite{Goel2009}. This observation provides an interesting opportunity
to examine to which extent the excess entropy determines the single particle dynamics. The idea behind this test is the following. 
If the excess entropy encodes the essential information about the system dynamics, then, one should observe the same 
dynamics both in confinement and in the bulk, provided that the system-averaged 
excess entropy per particle is identical in both cases. This idea has been tested by Goel and coworkers
in a series of carefully designed computer simulations~\cite{Goel2009}. The results obtained 
from these simulations are plotted in Fig.~\ref{fig:Fig3-Goel}. The qualitative agreement of 
the equivalent bulk and slit simulations is quite remarkable. 

Figure \ref{fig:Fig3-Goel} also shows that a still better agreement between the slit and the equivalent bulk simulations is obtained if, instead of the excess entropy, the slit-averaged fractional available volume, $\bar{p}_0$, is used as the relevant control parameter. Here, $\bar{p}_0=\frac{1}{H-\sigma} \int p_0(z) dz$ 
and $ p_0=\rho(z) \exp[-\beta \uw(z)]/\xi$. Further, $\beta=1/(\kB T)$ and $\uw$ is the fluid-wall interaction potential. The parameter $\xi$ is the so-called activity, defined as $\xi \equiv \Lambda^{-3}\exp(\beta\mu)$ with $\mu$ denoting the chemical potential and $\Lambda$ the thermal de Broglie wave length \cite{Hansen2006}. 
Results shown in Fig.~\ref{fig:Fig3-Goel} underline the close connection between the excess entropy and the fractional available volume. However, the shift between the data for x=$\bar{p}_0$ and x=$\sex$ in Fig.~\ref{fig:Fig3-Goel} is not understood yet and calls for further studies of the mutual relation between these two fundamental quantities.

Next we address an interesting---and at the same time intriguing---dependence between the local density and the local diffusion coefficient in a confined system. We start the related considerations by mentioning that, at thermal equilibrium, the chemical potential is spatially uniform. As a consequence, the above defined activity is also spatially constant across the slit. Further, we recall the close connection between activity and the Widom's insertion probability~\cite{Widom1963}, defined as $P_0 \equiv \left< \exp(-\beta u)\right>$, where $u$ is the interaction potential between a single particle with all the other $(N-1)$ particles in the system. For this purpose, we write the thermodynamic definition of the chemical potential and use the relation between the free energy of a system of $N$ particles, $F_N$, and the corresponding canonical partition sum, $Z_N$.
This gives $\mu=\partial F/\partial N = F_N-F_{N-1}= \beta^{-1}\ln(Z_N/Z_{N-1})$, which we rewrite as $\exp(\beta\mu)=Z_N/Z_{N-1}=\frac{\Lambda^3}{N} Q_N/Q_{N-1} $. Here $Q_N$ is the configurational part of the partition function~\cite{Hansen2006}. As first noticed by Widom 
in his seminal article on the theory of fluids~\cite{Widom1963}, one can show 
that $Q_N/Q_{N-1}=V \left< \exp(-\beta u) \right>$
 and thus $\xi \equiv \Lambda^{-3}\exp(\beta\mu) = \frac{V}{N} \left< \exp(-\beta u)\right>$. The particle insertion probability is thus given by $P_0=\rho/\xi$. Since $\xi$ is spatially uniform, it follows that $P_0\propto \rho$. This result is somewhat unexpected as it states that, in an inhomogeneous fluid, high insertion probabilities are associated with high densities.

\label{Widom-insertion}
In the case of a hard-sphere system, the energy, $u$, associated with an attempt to insert a new particle is zero 
if the insertion of that particle does not lead to overlap with any of the existing particles. In this case, 
the insertion probability for this specific move is unity. In the case of an overlap, $u=\infty$, leading to zero probability. 
Thus, the ensemble-averaged insertion probability, $P_0$, is an exact measure of the number of available states. 
In order to link this information to the system dynamics, one assumes that the rate of structural relaxation is proportional 
to the number of available states in the sense of the particle insertion statistics. One then anticipates that the diffusion coefficient scales as $D \sim P_0 \propto \rho$. Similar to the case of the insertion probability, this means that the fastest dynamics occurs at the position with the highest local density.

\begin{figure*}
\centering
\includegraphics*[height=7cm]{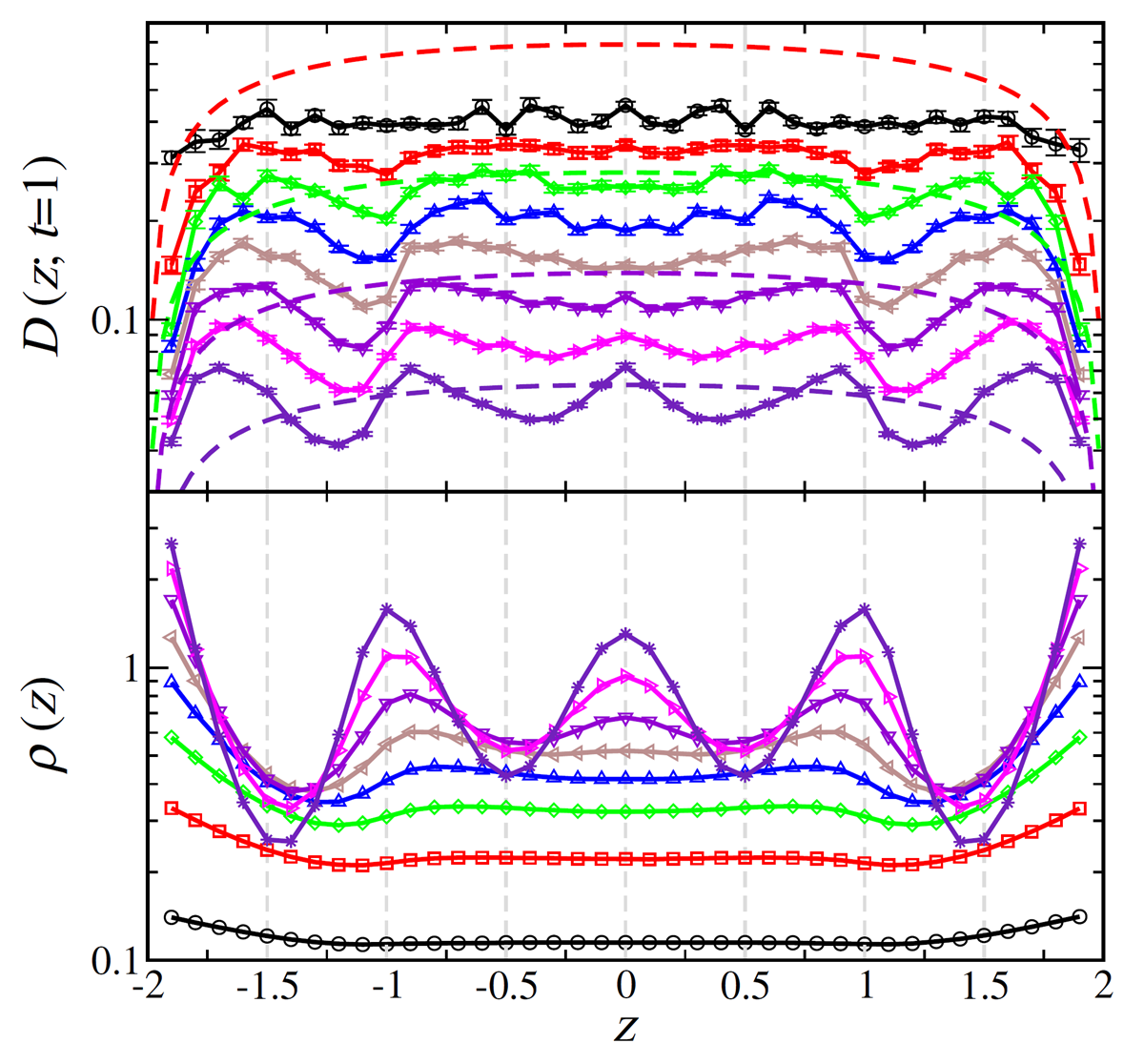}
\caption[]{(colour online). Results of Mittal and co-workers on the correlation between the local diffusion coefficient \revision{in the direction perpendicular to the walls, $D(z,t=1)$, (top) and the local density, $\rho(z)$, (bottom) in a monodisperse HS fluid, confined in a planar slit of width $H=5$ particle diametre . 
The time argument in $D$ is the time interval over which particle dynamics is monitored to determine $D$. It is chosen such that particles move sufficiently far to provide an estimate of $D$ but not too far so that the averaging over $z$, implicitly present in the motion along the perpendicular direction ($z$), does not strongly bias the $z$-dependence of the diffusion coefficient.} The packing fraction is  $\varphi=0.05$, 0.10, 0.15, 0.20, 0.25, 0.30, 0.35, and 0.40 (top to bottom for $D$ and 
bottom to top for $\rho(z)$). Figure from~\cite{Mittal2008} (reproduced with permission).}
\label{fig:D-rho-positive-correlation}
\end{figure*}

This, apparently counter-intuitive, prediction is in qualitative agreement with the results obtained 
from computer simulations (Fig.~\ref{fig:D-rho-positive-correlation}), which unambiguously reveal that peaks of the local diffusion coefficient approximately correlate with the \emph{maxima} of the density profile. Conversely, the minima of the local diffusivity occur 
at those places, where also the density attains the locally smallest value. We also remark that a slight phase shift between $D(z)$ and $\rho(z)$ is visible from the data shown in Fig.~\ref{fig:D-rho-positive-correlation}. 

\begin{figure*}
\centering
\includegraphics*[width=8cm]{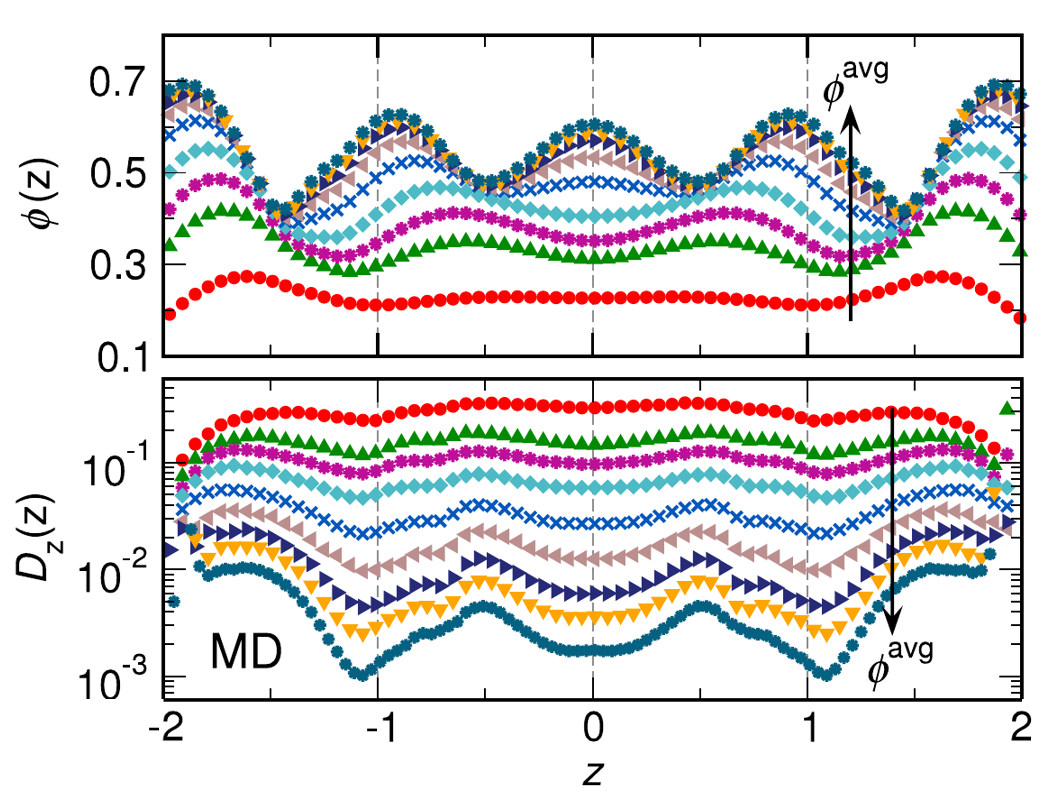}
\caption[]{(colour online). Molecular dynamics simulation results of a binary HS fluid confined between two smooth hard plates exhibiting that the relation between the local density profile and the local single particle dynamics is reversed as the packing fraction approaches 
the supercooled regime $\phi \ge 0.5$. The upper panel shows the
local packing fraction, $\phi(z)$, and the lower panel depicts the local diffusivities in the direction perpendicular to the walls, $D_z(z)$. The width is $H=5$ large particle diametres. The diffusion coefficient is obtained by tracking the motion of the  small particles. 
The slit-averaged packing fraction is $\varphiavg=0.20$, 0.30, 0.35, 0.40, 0.45, 0.48, 0.50, 0.51 and 0.52. 
The figure (adapted) is taken from \cite{Bollinger2015} (reproduced with permission).}
\label{fig:Fig2-Bollinger}
\end{figure*}

In the light of the above discussion, a question arises regarding the generality of the correlation between local particle mobility in an inhomogeneous fluid and  position-dependent static properties. A particularly interesting issue here concerns 
the validity of this conjecture when approaching the glass transition. Recent computer simulations of a 
binary hard-sphere mixture show that the above mentioned phase shift between density and diffusivity 
increases in a systematic way with the packing fraction until a complete reversal of the 
relation between $D(z)$ and $\rho(z)$ is established. At a packing fraction of $\varphi=0.52$, for example, the 
maxima of the local diffusion coefficient no longer correspond to the maxima but to the \emph{minima} of the 
density profile (Fig.~\ref{fig:Fig2-Bollinger}). This emphasizes that the relation between the single particle dynamics and the 
particle insertion probability, used to obtain the relation $D \propto \rho$, becomes inaccurate at 
high packing fractions.

This emphasizes the needed for a microscopic theory which adequately addresses static and dynamic properties of densely packed confined fluids. One of the promising routes in this context is the recent extension of the mode-coupling theory of the glass transition to confined geometry. The next section is devoted to an introduction of the basic concepts of this new theoretical approach. Its most salient predictions are worked out and tested via simulations subsequently.

The remaining of this review essentially compiles the recent contributions made by the authors and their coworkers. The issues discussed below are, however, far from being completely settled. We sometimes only touch upon a new exciting effect but let it largely open to young researchers to explore them amply in more complete studies. A remarkable example is the possible coexistence of multiple liquid-glass domains in a cone-plate type chamber, addressed in the last section. Here, only first evidences are provided. This calls for simulations and experiments at high densities to examine and directly uncover the predicted coexistence.

%%%%%%%%%%%%%%%%%%%%%%%%%%%%%%%%%%%%%%%%%%%%%%%%%%%%%%%%%%%%%%%%%%%%%%%%%
%%%%%%%%%%%%%%%%%%%%%%%%%%%%%%%%%%%%%%%%%%%%%%%%%%%%%%%%%%%%%%%%%%%%%%%%%
\subsection{mode-coupling theory of confined fluids}
\label{subsec:mct}

Many of the phenomena associated with the slowing down of transport upon cooling or compressing a simple liquid have been rationalized within the mode-coupling theory of the glass transition~\cite{Goetze:Complex_Dynamics} developed by Wolfgang G\"otze and collaborators within the last 30 years. The theory makes a series of non-trivial predictions for the directly measurable intermediate scattering functions or, equivalently, the dynamic structure factors both for the collective dynamics as well as for the tagged-particle motion. In particular, derived quantities, such as the mean square displacement and the associated diffusion coefficient, can be calculated within\delete{in} the theory, provided that the static structure factors are used as known input. There are no adjustable parameters, hence  MCT constitutes a microscopic theory.  

Here we review the mode-coupling theory in confinement emphasizing the parallels to the MCT of the glass transition  as well as the necessary modifications to account for the boundaries.

\begin{figure}
\centerline{\includegraphics[height=4.5cm]{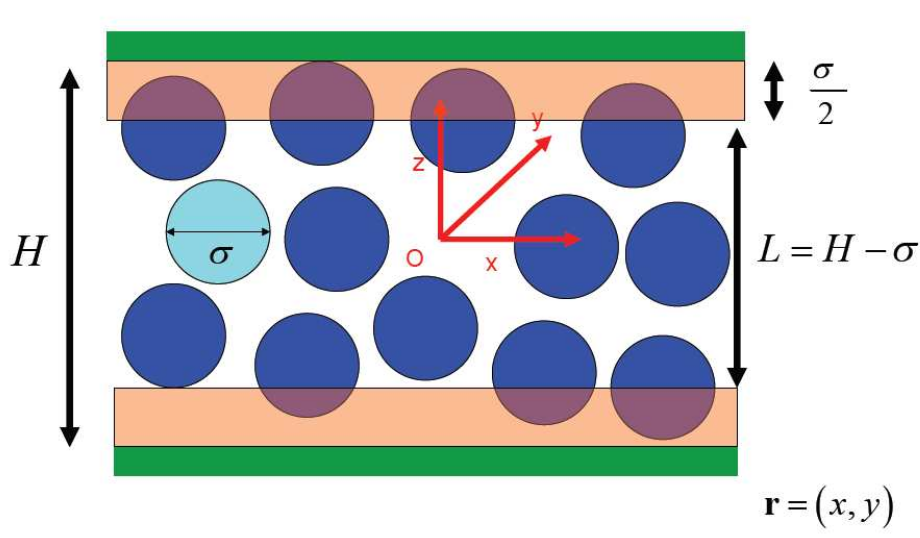}}
\caption[]{(colour online). Set up of confinement for a hard-sphere system. The accessible slit width $L$ is smaller than the physical wall separation $H$ by the hard-sphere diametre  $\sigma$. Planar coordinates $\vec{r} = (x,y)$ are unbounded, while lateral ones are confined $-L/2 \leq z \leq L/2$. }
\label{fig:wall_geometry}
\end{figure}

The basic setup consists of a liquid comprised of  $N$ structureless particles of mass $m$ confined by two flat, hard, and parallel walls with an accessible slit width  $L$. Hence for hard particles the separation of the walls will be $H= L+ \sigma$ where $\sigma$ is the diametre of the spheres, see Fig.~\ref{fig:wall_geometry} for an illustration. 
The thermodynamic limit is defined \delete{anticipated} such that the area density $n_0:= N/A$ remains fixed as the particle number $N\to\infty$ and the wall area $A\to \infty$ approach infinity  while the plate separation $L$ remains constant.  

The origin of a Cartesian coordinate system is  chosen in the centre between the boundaries with the $z$-axis perpendicular to the plates, while the $x$--$y$ plane is parallel to the walls. 
Lateral coordinates are abbreviated by two-dimensional vectors $\vec{r} = (x,y)$,  
the  time-dependent positions of the $N$ particles are denoted by $\{ \vec{x}_n(t) \} = \{ (\vec{r}_n(t), z_n(t)) \}$, 
%and similarly the momenta by  $\{ \vec{p}_n(t) \} = \{ (\vec{P}_n(t), P_n^z(t)) \}$, 
with $n=1,\ldots, N$. By the confinement all transverse positional coordinates fulfill $-L/2 \leq
 z_n(t) \leq L/2$. 

The most fundamental quantity is the microscopic density variable 
\begin{equation}
 \rho(\vec{r},z,t) = \sum_{n=1}^N \delta[\vec{r}-\vec{r}_n(t)] \delta[z-z_n(t)],
\end{equation}
 and its canonical expectation yields the density profile
\begin{equation}
 n(z) = \langle \rho(\vec{r},z,t) \rangle .
\end{equation}
Here $n(z)$  depends explicitly on the transverse coordinate but is uniform in the direction parallel to the walls by the  lateral symmetry. The modulation of the density profile $n(z)$ is expected to be significant if the slit width is comparable to the interaction range, i.e. if only a few monolayers fit into the slit. It is natural to decompose $n(z)$ into discrete Fourier modes 
\begin{equation}
 n(z) = \frac{1}{L} \sum_{\mu} n_\mu \exp(-i Q_\mu z) ,
\end{equation}
where the mode index $\mu \in \mathbb{Z}$ and the associated wavenumbers $Q_\mu = 2\pi \mu/L$ are discrete. Conversely, the Fourier coefficients are determined by
\begin{equation}
 n_\mu = \int_{-L/2}^{L/2} n(z) \exp(i Q_\mu z) \diff z.
\end{equation}
An analogous decomposition can be performed for the local volume per particle, $v(z) =1/n(z)$, and one verifies the relation between the Fourier coefficients
\begin{equation}
\frac{1}{L^2}  \sum_{\kappa} n_{\mu-\kappa} v_{\kappa-\nu} = \delta_{\mu\nu} .
\end{equation}

Similarly, the fluctuating part of the microscopic density $\delta \rho(\vec{r},z,t) = \rho(\vec{r},z,t) - n(z)$ is decomposed into a set of 
symmetry-adapted modes. For the lateral dependence  the conventional Fourier modes with in-plane wavevectors $\vec{q}$ are employed, whereas for the transverse direction the discrete set of wavenumbers $Q_\mu$ is appropriate. The proper choice of variables is therefore
\begin{equation}
 \delta \rho_\mu(\vec{q},t) = \sum_{n=1}^N \exp[i Q_\mu z_n(t) ] e^{i\vec{q}\cdot \vec{r}_n(t)} - A n_\mu \delta_{\vec{q},0}.
\end{equation}
The last term is relevant only for wave vectors $\vec{q}$ identical to zero and can be safely ignored in the following. 

The confinement implies that 
the  continuity equation for the density modes assumes the form 
\begin{equation}
 \delta \dot{\rho}_\mu(\vec{q},t) = i q j_\mu^\parallel(\vec{q},t) + i Q_\mu j_\mu^\perp(\vec{q},t) 
= i \sum_{\alpha =  \parallel, \perp} b^\alpha(q,Q_\mu) j_\mu^\alpha(\vec{q},t),
%\equiv \sum_{\alpha = \parallel, \perp} q_\mu^\alpha j_\mu^\alpha(\vec{q},t)
\end{equation}
with current densities $j_\mu^\alpha(\vec{q},t)$ to channel index $\alpha \in \{ \parallel , \perp \} $ parallel and perpendicular to the planes. 
%The short-hand notation $q_\mu^\alpha$ with $q_\mu^\parallel = q$, $q_\mu^\perp = Q_\mu$ is convenient for sums over  channel indices. 
The selector $b^\alpha(q,Q_\mu) = q \delta_{\alpha,\parallel} + Q_\mu \delta_{\alpha,\perp}$ permits a short-hand notation to account for both channel indices. 
The splitting of the currents has drastic consequences for the formulation of the theory and implies that the mathematical structure of the mode-coupling equations will differ from the conventional MCT of the glass transition.

The quantities of primary interest are then the time-dependent density-density correlation functions
\begin{equation}
 S_{\mu\nu}(q,t) = \frac{1}{N} \langle \delta \rho_{\mu}(\vec{q},t)^* \delta \rho_\nu(\vec{q}) \rangle ,
\end{equation}
 also referred to as generalized intermediate scattering function (ISF). 
Here we used the convention that suppressing the time indicates that the variable is to be evaluated at time $t=0$. By rotational symmetry around the $z$-axis, the ISF does not depend on the direction of the in-plane wave vector but only on its magnitude $q=|\vec{q}|$. The matrix-valued quantity $S_{\mu\nu}(q,t)$ constitutes the proper generalization of the intermediate scattering function in bulk systems, in particular, $S_{00}(q,t)$ measures the decay of density modulations within the plane only. The intermediate scattering function is directly measurable in neutron-scattering experiments for atomic systems or photon-correlation spectroscopy for colloidal suspensions~\cite{Hansen2006}. 
The initial value $S_{\mu\nu}(q) \equiv S_{\mu\nu}(q,t=0)$ encodes valuable information on the short-range order of the confined fluid and is referred to as generalized static structure factor. The corresponding  direct correlation function $c_{\mu\nu}(q)$ is then obtained by decomposing the Ornstein-Zernike equation into the symmetry-adapted modes~\cite{Lang2010,Lang2014a}
\begin{equation}
\boldsymbol{S}^{-1} = \frac{n_0}{L^2} [ \boldsymbol{v} - \boldsymbol{c} ] ,
\end{equation}
where a natural matrix notation has been employed.  Here, bold symbols indicate matrices in the mode indices, e.g. $[\mathbf{S}]_{\mu\nu} = S_{\mu\nu}$. Furthermore, the dependence on the wavenumber $q$ is suppressed if all quantities in the equation refer to the same $q$. The matrix corresponding to the local volume is $[\boldsymbol{v}]_{\mu\nu} = v_{\nu-\mu}$. 

Exact equations of motion for time correlation functions can be derived within the Zwanzig---Mori projection operator formalism~\cite{Lang2010,Goetze:Complex_Dynamics}. %To be specific we consider Newtonian dynamics of particles of mass $m$ at temperature $T$, i.e. the thermal velocity $v_\text{th} = \sqrt{k_B T/m}$. Then Newton's equations imply that observables $A(t) $are driven by the Liouville operator ${\cal L}$, $\partial_t A(t) = \{ A(t), H\} \equiv {\cal L} A$, where $H$ denotes the Hamilton function and $\{\cdot,\cdot\}$ the Poisson bracket. 
Then the intermediate scattering function satisfies the 
integro-differential equation
\begin{equation}\label{eq:eom_1}
 \dot{\mathbf{S}}(t) + \int_0^t \mathbf{K}(t-t') \mathbf{S}^{-1} \mathbf{S}(t') \diff t' = 0.
\end{equation}
 The initial condition for the ISF is merely the static structure factor $\mathbf{S}(t=0) = \mathbf{S}$. Explicit formal expressions for the matrix of the current kernel  $\mathbf{K}(t)$ are available. By the continuity equation, the current kernel also naturally splits
\begin{equation}
 K_{\mu\nu}(q,t) = \sum_{\alpha, \beta = \parallel,\perp} b^\alpha(q,Q_\mu) {\mathcal K}_{\mu\nu}^{\alpha \beta}(q,t) b^\beta(q,Q_\nu).
\end{equation}
Quantities associated with mode indices and channel indices are indicated by calligraphic symbols and again we use matrix notation, for example $[\boldsymbol{\mathcal K}(q,t) ]^{\alpha\beta}_{\mu\nu} = {\mathcal K}^{\alpha\beta}_{\mu\nu}(q,t)$. 

Within the Zwanzig---Mori formalism, a second equation of motion for the currents is derived within Newtonian dynamics
\begin{equation}\label{eq:eom_Newton}
 \boldsymbol{\mathcal J}^{-1} \dot{\boldsymbol{\mathcal K}} +  \boldsymbol{\mathcal D} \boldsymbol{\mathcal K}(t) + \int_0^t  \boldsymbol{\mathcal M}(t-t') \boldsymbol{\mathcal K}(t') \diff t' = 0 ,
\end{equation}
subject to the initial condition $\boldsymbol{\mathcal K}(t=0) = \boldsymbol{\mathcal J}$, where $\mathcal{J}_{\mu\nu}^{\alpha\beta}(q) = N^{-1} \langle j_\mu^\alpha(\vec{q})^* j_\nu^\beta(\vec{q}) \rangle = (k_B T/m) (n_{\mu-\nu}^*/n_0) \delta_{\alpha\beta}$ is the static current correlator matrix. An instantaneous damping term $\boldsymbol{\mathcal D}(q)$ has been split off to account for a regular short-time decay~\cite{Lang2013}.  The many-body dynamics is hidden in the force kernel $\boldsymbol{\mathcal M}(t)$.  

The mode-coupling ansatz provides a detailed prescription to approximate the force kernel $\boldsymbol{\mathcal M}(t)$ as bilinear functional local in time of the intermediate scattering functions
\begin{equation}\label{eq:MCT_ansatz}
 {\mathcal M}_{\mu\nu}^{\alpha\beta}(q,t) = {\mathcal F}_{\mu\nu}^{\alpha\beta}[\boldsymbol{S}(t), \boldsymbol{S}(t);q] .
\end{equation}
Explicitly the mode-coupling functional reads
\begin{eqnarray}\label{eq:MCT_functional}
 {\mathcal F}[ \boldsymbol{E},\boldsymbol{F};q] =& \frac{1}{4N} \sum_{\vec{q}_1,\vec{q}_2 = \vec{q}-\vec{q}_1}  \sum_{\mu_{1}\mu_{2} \atop \nu_{1} \nu_{2}} \mathcal{Y}^\alpha_{\mu,\mu_{1} \mu_{2}}(\vec{q},\vec{q}_{1}\vec{q}_{2}) \nonumber \\
&\times  \left[ E_{\mu_{1}\nu_{1}}(q_{1}) F_{\mu_{2}\nu_{2}}(q_{2}) + (1 \leftrightarrow 2) \right] \mathcal{Y}^\beta_{\nu,\nu_{1} \nu_{2}}(\vec{q},\vec{q}_{1}\vec{q}_{2})^*,
\end{eqnarray}
where the vertices $\mathcal{Y}^\alpha_{\mu,\mu_{1} \mu_{2}}(\vec{q},\vec{q}_{1}\vec{q}_{2})$ play the 
role of  coupling constants determined by static correlation functions only. Upon a convolution 
approximation to account for three-particle correlations~\cite{Lang2010,Lang2012} the vertex can be determined to
\begin{eqnarray}\label{eq:vertex}
& \mathcal{Y}^\alpha_{\mu,\mu_{1} \mu_{2}}(\vec{q},\vec{q}_{1}\vec{q}_{2}) = \nonumber \\
&= \frac{n_0^2}{L^4} \sum_\kappa v_{\mu-\kappa}^* \left[ b^\alpha(
\hat{\vec{q}} \cdot \vec{q}_1, Q_{\kappa-\mu}) c_{\kappa-\mu,\mu_1}(q_1) + (1\leftrightarrow 2) \right]. 
\end{eqnarray}

The two equations of motion for the intermediate scattering function and the current 
kernel, Eqs.~(\ref{eq:eom_1}),(\ref{eq:eom_Newton}), supplemented by their respective initial conditions, together 
with the MCT ansatz, Eqs.~(\ref{eq:MCT_ansatz}),(\ref{eq:MCT_functional}),(\ref{eq:vertex}) constitute a complete 
set of equations for the intermediate scattering function $S_{\mu\nu}(q,t)$. 

One can show that there are unique solutions~\cite{Lang2013}
 and moreover, that the solutions represent equilibrium correlation functions:
\begin{equation}
 \mathbf{S}(t) = \int_{\mathbb{R}} e^{-i \Omega t} \mathbf{R}(\diff \Omega).
\end{equation}
Here $\mathbf{R}(\Omega)$ is a self-adjoint symmetric matrix-valued measure, i.e. $R_{\mu\nu}(\Omega)$ is a complex finite Borel measure on the real line $\Omega \in \mathbb{R}$ and a positive-semidefinite matrix for fixed $\Omega$, symmetric means that $\mathbf{R}(\Omega)-\mathbf{R}(0) = \mathbf{R}(0) -\mathbf{R}(-\Omega)$. 
In particular, the measure fulfills $\mathbf{S} = \int_\mathbb{R} \mathbf{R}(\diff \Omega)$. 
The representation property states that, although the ${\bf S}(t)$ predicted within MCT  may differ from the experimentally measurable intermediate scattering function, there exists a different stochastic process [hopefully close to the real one] which yields ${\bf S}(t)$ as a result of correlating observables. Let us recall the most important consequences of the representation property. First, one finds that the ISF is hermitian $\mathbf{S}^\dagger(t) = \mathbf{S}(t)$, symmetric in time $\mathbf{S}(-t) = \mathbf{S}(t)$, and  bounded 
\begin{equation}
  \mathbf{S} \succeq \mathbf{S}(t) \succeq -\mathbf{S}  ,
\end{equation}
where $\mathbf{A}\succeq \mathbf{B}$ indicates that $\mathbf{B}(q)-\mathbf{A}(q)$ is a positive semidefinite matrix for each wavenumber $q$. More generally, for any finite set of times $t_i, i=1,\ldots , m$ and associated complex weighting factors $\lambda_i$, matrix-valued correlation functions are positive-semidefinite in the following sense
\begin{equation}
 \sum_{i=1}^m \sum_{j=1}^m \lambda_i^* \mathbf{S}(t_i-t_j) \lambda_j  = \int_{\mathbb{R}} \left| \sum_{j=1}^m \lambda_j e^{i\Omega t_j} \right|^2 \mathbf{R}(\diff \Omega) \succeq 0.
\end{equation}
By Bochner's theorem~\cite{Feller:Probability_2} the latter property is in fact equivalent to the representation property of correlation functions. 

It is of interest also to consider the one-sided Fourier transforms of the intermediate scattering functions
\begin{equation}
 \hat{\mathbf{S}}(z) = i \int_0^\infty e^{i z t} \mathbf{S}(t) \diff t ,
\end{equation}
for complex frequencies $z\in \mathbb{C}_+ = \{ z\in \mathbb{C}: \,\text{ Im}[z] > 0 \}$ in the upper complex half-plane. The representation property then yields directly a representation in terms of a Hilbert transform of the associated measure
\begin{equation}
 \hat{\mathbf{S}}(z) = \int_{\mathbb{R}} \frac{1}{\Omega-z} \mathbf{R}(\diff \Omega) .
\end{equation}
Then one verifies the following properties for frequencies $z\in\mathbb{C}_+$ 
\begin{itemize}
 \item[(1)] $\hat{\mathbf{S}}(z)$ is analytic 
\item[(2)] $\hat{\mathbf{S}}(-z^*) = - \hat\mathbf{S}^\dagger(z)$
\item[(3)] $\lim_{\eta\to \infty} \eta\, \text{Im}[\hat{\mathbf{S}}(z= i\eta) ]$ is finite
\item[(4)] $\text{Im}[\hat{\mathbf{S}}(z)] \succeq 0$
\end{itemize}
Here $\text{Im}[\hat{\mathbf{S}}(z)] := [\hat{\mathbf{S}}(z)- \hat{\mathbf{S}}^\dagger(z) ] /2i$ is the proper generalization of the imaginary part to matrices. In particular, property (4) implies that the measurable power spectrum associated with any linear combination of density modes $\rho_\mu(\vec{q},t)$ is non-negative in every frequency interval. The Riesz-Herglotz theorem~\cite{Feller:Probability_2,Gesztesy:2000} reveals that properties (1)-(4) are also sufficient for $\hat{\mathbf{S}}(z)$ being the one-sided Fourier transform of a matrix-valued correlation function $\mathbf{S}(t)$.

For the following discussion it is instructive to transform the equations of motion to the Fourier domain. 
By the convolution theorem, the equations of motion for the ISF become algebraic equations in the frequency domain
\begin{equation}\label{eq:isf_frequency}
 \hat{\mathbf{S}}(z) =  -\left[ z \mathbf{S}^{-1} + \mathbf{S}^{-1} \hat{\mathbf{K}}(z) \mathbf{S}^{-1} \right]^{-1}.
\end{equation}
The contraction over the channel indices readily transfers to the Fourier domain
\begin{equation}
 \hat{K}_{\mu\nu}(q,z) = \sum_{\alpha,\beta = \parallel, \perp} b^\alpha(q,Q_\alpha) \hat{\mathcal K}_{\mu\nu}^{\alpha\beta}(q,z) b^{\beta}(q,Q_\nu),
\end{equation}
and again the equation of motion for the current matrix $\mathcal{K}_{\mu\nu}^{\alpha\beta}(q,z)$ become  algebraic matrix equations
\begin{equation}\label{eq:currents_frequency}
 \hat{\boldsymbol{\mathcal{K}}}(z) = - \left[ z  \boldsymbol{\mathcal{J}} + \boldsymbol{\mathcal{D}}^{-1} + \hat{\boldsymbol{\mathcal{M}}}(z) \right]^{-1}.
\end{equation}
While the Zwanzig---Mori equations of motion become simple in the frequency domain since frequency is merely a parameter, the 
MCT kernel couples intermediate scattering functions to different frequencies, and one better sticks to the representation in time.

Glass states in the non-equilibrium state diagram are defined by a non-vanishing long-time limit of the intermediate scattering function
\begin{equation}
\mathbf{F} := \lim_{t\to\infty} \mathbf{S}(t) \neq 0,
\end{equation}
referred to as non-ergodicity parameter or glass form factor.
Conversely, ergodic liquid states are characterized by a trivial long-time limit $\mathbf{F} \equiv 0$. The representation 
property implies that the non-ergodicity parameter is positive-semidefinite and bounded by the structure factor
\begin{equation}
  \mathbf{S} \succeq \mathbf{F} \succeq 0 .
\end{equation}
For the case of the conventional mode-coupling theory of the glass transition, the existence of the long-time limit is guaranteed~\cite{Franosch:2014} 
and one anticipates that this result also holds for  the case of split currents. Hence, in principle the non-equilibrium state diagram can be constructed by solving the dynamic equations for given control parameters and classifying the solutions according to their respective long-time limits. Since the numerical solution is involved, one would like to circumvent the evaluation of the time-dependent solutions and derive simplified equations for the non-ergodicity parameters alone. Necessary conditions for $\mathbf{F}$ can be derived, by connecting it to the long-time limit of the force kernel.
A first relation is obtained by specializing  the mode-coupling functional to the limit of infinite times
\begin{equation}\label{eq:static_MCT}
 \boldsymbol{\mathcal{N}}(q) := \lim_{t\to \infty } \boldsymbol{\mathcal{M}}(q,t) = \boldsymbol{\mathcal{F}}[\mathbf{F},\mathbf{F};q] .
\end{equation}
A second relation follows by evaluating the equations of motion in the limit of small frequencies. A non-trivial long-time limit of the force kernel yields a simple pole in the frequency domain
\begin{equation}
 \hat{\boldsymbol{\mathcal{M}}} = - z^{-1} \boldsymbol{\mathcal{N}}  + o(z^{-1}) .
\end{equation}
Then the representation of the current correlator, Eq.~(\ref{eq:currents_frequency}), shows that for small frequencies $\hat{\boldsymbol{\mathcal{K}}}(z) = z \boldsymbol{\mathcal{N}}^{-1} + o(z)$. The contraction leads to leading order to $\hat{\mathbf{K}}(z) = z \mathbf{N}^{-1} + o(z)$ with
\begin{equation}\label{eq:static_contraction}
 [\mathbf{N}(q)^{-1}]_{\mu\nu} = \sum_{\alpha,\beta=\parallel,\perp} b^\alpha(q,Q_\mu) [\boldsymbol{\mathcal{N}}^{-1}(q) ]^{\alpha\beta}_{\mu\nu} b^\beta(q,Q_\nu) .
\end{equation}
One can show that $\mathbf{N} = \mathbf{N}[\mathbf{F}]$ considered as functional of the long-time limits $\mathbf{F}$ displays the  properties of an effective mode-coupling functional~\cite{Lang2012} in the space of matrices with mode indices $\mu,\nu$. Evaluating the equations of motion for the ISF, Eq.~(\ref{eq:isf_frequency}) shows that $\hat{\mathbf{S}}(z) = - z^{-1} \mathbf{F} + o(z^{-1})$ for $z\to 0$, where the glass form factor fulfills
\begin{equation}\label{eq:fixed_point}
 \mathbf{F} = \mathbf{S} - [\mathbf{S}^{-1} + \mathbf{N}[\mathbf{F}] ]^{-1} .
\end{equation}
The set of equations, Eqs. (\ref{eq:static_MCT}),(\ref{eq:static_contraction}),(\ref{eq:fixed_point}) is complete and called fixed-point 
equations. They  are necessarily fulfilled by the glass form factor $\mathbf{F}$, yet, in general the equations allow for many solutions, in particular, one checks that the trivial case $\mathbf{F}(q) = 0$ for 
all wavenumbers $q$ is always a solution. Furthermore, only solutions that correspond to long-time limits of correlation functions, i.e. 
positive-semidefinite ones, $\mathbf{F} \succeq 0$, are acceptable. Hence it appears as a non-trivial question, which solution
 actually represents the long-time limit of the full time-dependent solution of the MCT equations. 

The dynamic mode-coupling equations encode a property referred to as generalized covariance 
principle \cite{Lang2012,Lang2013,Goetze:Complex_Dynamics} which states that given a particular solution $\bar{\mathbf{F}} \succeq 0$ of the 
fixed-point equations, then the equations of motion for the remainder $\mathbf{S}(t)-\bar{\mathbf{F}}$ are of the same 
form as the original equations of motion. In particular, $\mathbf{S}(t)-\bar{\mathbf{F}}$ corresponds again to a correlation function, with 
associated spectral measure $\mathbf{R}(\Omega) - \bar{\mathbf{F}} \vartheta(\Omega)$. The generalized covariance principle turns out to be crucial to sort 
out the relevant solution  of the many solutions of the fixed-point equations.  

A particular solution can be constructed by iterating the set of equations, thereby generating a sequence $\mathbf{F}^{(n)}, n=0,1,\ldots$. The 
sequence is initialized with the static structure factor $\mathbf{F}^{(0)} = \mathbf{S}$, then, given $\mathbf{F}^{(n)}$, a 
new $\boldsymbol{\mathcal N}^{(n)} = \boldsymbol{\mathcal{F}}[\mathbf{F}^{(n)},\mathbf{F}^{(n)}]$ is calculated and contracted according to Eq.~(\ref{eq:static_contraction}) to yield the new $\mathbf{N}^{(n)}$. Last, the next element of the sequence is 
generated using Eq.~(\ref{eq:fixed_point}) by $\mathbf{F}^{(n+1)} = \mathbf{S} - [ \mathbf{S}^{-1} + \mathbf{N}^{(n)} ]^{-1}$. One can show~\cite{Lang2012}
that the sequence never leaves the space of positive-semidefinite elements $\mathbf{F}^{(n)} \succeq 0$, and is monotonically 
decreasing, hence convergence $\mathbf{F} = \lim_{n\to\infty} \mathbf{F}^{(n)}$ is guaranteed. 

The generalized covariance principle allows establishing the maximum principle, stating that the solution of the fixed-point equations 
assumed by the long-time limit $\mathbf{F}$ of the dynamic MCT equations is maximal, i.e all other positive-semidefinite 
solutions $\bar{\mathbf{F}}$ fulfill $\mathbf{F} \succeq \bar{\mathbf{F}}$. Furthermore, it has been demonstrated that the solution 
constructed via the monotonic sequence above is maximal~\cite{Lang2012}, hence it coincides with the long-time limit of the dynamical problem.  

The iteration scheme permits constructing the non-equilibrium phase diagram and separate ergodic from glassy states. The boundary between 
the two states is referred to as glass transition singularity and it corresponds to a singular dependence of the solution of the fixed-point 
equations as functions of the control parameters. Since the fixed-point equations are purely algebraic with smooth dependence on the static input, singular behaviour can only emerge as a result of bifurcations of the $A_\ell$-type according to the classification of Arnol{\textquoteright}d~\cite{Arnold:1975}. Then the 
mathematical properties of the singular behaviour are identical to the ones of multicomponent mixtures~\cite{Franosch:2002,Voigtmann:Dissertation}, 
provided that $\mathbf{N}[\mathbf{F}]$ is used as effective mode-coupling functional.
The simplest and also generic corresponds to 
the $A_2$ fold bifurcation. There the glass form factor assumes a nonzero value $\mathbf{F}^c \succeq 0$ directly at the transition. Hence, upon a smooth path in the state diagram starting from the liquid state, the long-time limit will be strictly zero until it jumps to $\mathbf{F}^c$ and 
will increase further upon penetrating the glass state. More precisely, one can introduce a separation parameter $\sigma\in \mathbb{R}$ with negative
values in the liquid and positive in the glass, such that it increases linearly along the path in the vicinity of the transition~\cite{Goetze:Complex_Dynamics}. Hence it is proportional to any generic distance measure, e.g. $\sigma = C (\varphi-\varphi^c)$ for the case of hard spheres with  packing fraction $\varphi$ and respective  critical value $\varphi^c$.   
The crossing of the glass transition singularity is referred to as critical point. 
Traversing the singularity into the glassy state, the non-ergodicity parameter increases according to
\begin{equation}\label{eq:glass_sqrt}
 \mathbf{F}(q) = \mathbf{F}^c(q) + \mathbf{H}(q) \sqrt{\frac{\sigma}{1-\lambda}}  + \mathcal{O}(\sigma).
\end{equation}
Here $\mathbf{H} \succeq 0$ is called the critical amplitude and $\lambda$ is known as  exponent parameter and for both quantities as well as $\sigma$ explicit expressions can be constructed~\cite{Voigtmann:Dissertation} starting from the effective mode-coupling functional $\mathbf{N}[\mathbf{F}]$. The square-root behaviour is characteristic of the fold bifurcations. The leading corrections $\mathcal{O}(\sigma)$ can also be worked out~\cite{Voigtmann:Dissertation} following the calculation of the MCT in bulk for monocomponent systems~\cite{Franosch_c:1997,Fuchs:1998}

It is anticipated that the splitting of the currents does not spoil the asymptotic behaviour  elaborated for time-dependent 
matrix-valued correlation function close to the transition. Directly at the critical point, $\sigma=0$, the critical glass form factor is 
approached as a power law relaxation \cite{Goetze:Complex_Dynamics,Voigtmann:Dissertation}
\begin{equation}\label{eq:critical_relaxation}
 \mathbf{S}(q,t) = \mathbf{F}^c(q) + \mathbf{H}(q) (t/t_0)^{-a} + \mathcal{O}(t^{-2a}),
\end{equation}
where the exponent $0<  a < 1/2$ is determined by the exponent parameter via 
\begin{equation}
 \lambda = \frac{\Gamma(1-a)^2}{\Gamma(1-2a)} = \frac{\Gamma(1+b)^2}{\Gamma(1+2b)},
\end{equation}
for the generic case $0< \lambda < 1$. We have included also the relation to the von-Schweidler exponent $b> 0$ that will become important below.
The time scale $t_0$ is of the order of microscopic times and is to be determined by matching the full dynamic solutions to the 
asymptotic critical  law, Eq.~(\ref{eq:critical_relaxation}). Close to the transition the solutions follow the critical law up to a time scale 
 $t_\sigma$
 diverging upon approaching the transition. For times larger than $t_\sigma$, the glass form factors $\mathbf{F}(q)$ are approached exponentially fast for glassy states, while for liquid states the plateau value $\mathbf{F}^c(q)$ is traversed to enter the terminal relaxation, also known as $\alpha$-relaxation. The dynamics in the vicinity of the plateau follows the first scaling law
\begin{equation}\label{eq:first_scaling_law}
 \mathbf{S}(q,t) = \mathbf{F}^c(q) + \mathbf{H}(q) \sqrt{|\sigma|} g_{\pm}( t/t_\sigma) + \mathcal{O}(\sigma),
\end{equation}
valid for intermediate times $t_0 \ll t \ll t_\sigma'$, where  $t_\sigma'< \infty$ denotes the $\alpha$-relaxation time for liquid states, while it becomes infinite for glassy states $t_\sigma' = \infty$. The scaling functions $g_\pm(\hat{t})$ fulfill the $\beta$-scaling equation of the MCT ~\cite{Goetze:Complex_Dynamics}
\begin{equation}
 \frac{\diff}{\diff \hat{t} } \int_0^{\hat{t}} g_\pm(\hat{t}') g_\pm(\hat{t}-\hat{t}') \diff \hat{t}' = \lambda g_\pm(\hat{t})^2 \pm 1 .
\end{equation}
Here the index $\pm$ refers to the glass $\sigma >0$ and liquid side $\sigma<0$. To match the critical law, Eq.~(\ref{eq:critical_relaxation}), at times $t_0 \ll t \ll t_\sigma$, the scaling functions have to fulfill $g_\pm(\hat{t}) \to \hat{t}^{-a}$ for $\hat{t}\to 0$ as short-time asymptote. Furthermore, this fixes the divergence of the  time scale 
\begin{equation}
t_\sigma = t_0 |\sigma|^{-1/2a},
\end{equation}
 upon approaching the glass transition singularity $\sigma \to 0$. 
For glass states, $g_+(\hat{t} \to \infty) = 1/\sqrt{1-\lambda}$, compatible with the result for the glass form factor, Eq.~(\ref{eq:glass_sqrt}).

For the liquid side, the scaling function diverges for long times as a power law $g_-(\hat{t} \gg 1) \simeq -B \hat{t}^b$ with the von-Schweidler exponent $b$. Hence, the first scaling law is valid only up to times 
\begin{equation}
 t_\sigma' = t_0 B^{-1/b} |\sigma|^{-\gamma}, \qquad \gamma = \frac{1}{2a} + \frac{1}{2b},
\end{equation}
and the MCT predicts for the early stage $t_\sigma \ll t \ll t_\sigma'$ of the relaxation process from the plateau 
\begin{equation}
 \mathbf{S}(q,t) = \mathbf{F}^c(q) - \mathbf{H}(q)  (t/t_\sigma')^b, 
\end{equation}
known as von-Schweidler law. 

The terminal relaxation from the plateau towards zero then fulfills the second scaling law~\cite{Goetze:Complex_Dynamics}
\begin{equation}\label{eq:second_scaling_law}
 \mathbf{S}(q,t) = \tilde{\mathbf{S}}(q, t/t_\sigma') ,
\end{equation}
for times $t \gtrsim t_\sigma'$. Hence, on a logarithmic time scale, the shape of the relaxation function decaying from its respective plateau value does not change upon approaching the glass transition singularity $\sigma \uparrow 0$, but is shifted merely to the right. Here, the relaxation functions are wavenumber-dependent, yet the shift in time scale is identical for all $q$.  The second scaling of the MCT is the theoretical foundation of the empirical time-temperature superposition principle or $\alpha$-scaling behaviour. 

The theory has been generalized also to capture the tagged-particle motion~\cite{Lang2014a}. Here a tracer particle is immersed into the 
confined fluid and its dynamics reflects aspects of the slow structural relaxation of the surrounding host fluid. Here we restrict the discussion
 to the most relevant case where 
 the tracer particle is identical to the constituent of the host fluid 
and  the tagged-particle motion can be measured in a computer simulation with high statistical accuracy. The simplest measurable quantities are the mean square displacement, e.g. in the case of confinement  parallel to the walls. More generally, the incoherent intermediate scattering function encodes all two-time correlation functions of tracer particle's dynamics. Again, generalized fluctuating density modes are considered
\begin{equation}
 \delta \rho^{(s)}_\mu(q,t) = \exp[ i Q_\mu z_s(t)] e^{i\vec{q}\cdot \vec{r}_s(t) }  - (n_\mu/n_0) \delta_{\vec{q},0}.
\end{equation}
Here $(\vec{r}_s(t),z_s(t))$ are the lateral and transverse coordinates of the tagged particle. The superscript $(s)$ indicates the single particle dynamics. The corresponding correlation function is the incoherent intermediate scattering function
\begin{equation}
 S_{\mu\nu}^{(s)}(q,t) = \langle \delta \rho_\mu(\vec{q},t)^* \delta \rho_\nu(\vec{q}) \rangle, 
\end{equation}
and, again $S_{00}^{(s)}(q,t)$ encodes  only the modulations along the plane. The mean square displacement parallel to the plates $\delta r^2_\parallel := \langle [\vec{r}_s(t)- \vec{r}_s(0)]^2 \rangle $ is contained in the long-wavelength expansion
\begin{equation}
 S_{00}^{(s)}(q,t) = 1 - q^2 \delta r^2_\parallel(t)/4 + \mathcal{O}(q^4).
\end{equation}
In particular the diffusion coefficient parallel to the confinement $D_\parallel$ describes the long-time behaviour of the mean square displacement in the liquid state
\begin{equation}
 \delta r^2_\parallel(t) \to 4 D_\parallel t \qquad \text{ for }\,\, t\to \infty.
\end{equation}

The equations of motion for $\mathbf{S}^{(s)}(t)$ are derived along the same lines as for the coherent motion and except for the decoration with the superscript $(s)$ are of the same form. The mode-coupling ansatz for the associated force kernel $\boldsymbol{\mathcal{M}}^{(s)}(t)$ yields a bilinear functional, coupling the single particle motion $\mathbf{S}^{(s)}(t)$ to the coherent dynamics $\mathbf{S}(t)$. Hence the slowing down of the coherent motion induces also the slow dynamics of the tracer particle. Again one anticipates that the splitting of the currents does not introduce qualitative new features with respect to the asymptotic scaling laws, such that Eqs. (\ref{eq:first_scaling_law}), (\ref{eq:second_scaling_law}) also hold for incoherent motion $\mathbf{S}^{(s)}(t)$.  The first scaling law translates for the mean square displacement~\cite{Goetze:Complex_Dynamics,Fuchs:1998} 
\begin{equation}
 \delta r^2_\parallel(t)/4 = \ell^2_{\parallel,c} - h_\parallel \sqrt{|\sigma|} g_\pm(t/t_\sigma) + \mathcal{O}(\sigma) ,
\end{equation}
in the $\beta$-scaling time window $t_0 \ll t \ll t_\sigma'$. Here $\ell^2_{\parallel,c} = \lim_{t\to \infty} \delta r^2_\parallel(t)/4$ corresponds to the localization length directly at the critical point, and $h_\parallel$ denotes the critical amplitude quantifying the coupling to the $\beta$-scaling function $g_{\pm}(\hat{t})$. Both quantities can be computed from the long-wavelength expansion of the first scaling law in terms of the critical glass form factor $F_{00,c}^{(s)}(q)$ and amplitude $H_{00}^{(s)}(q)$. 

The second scaling law implies for the mean square displacement
\begin{equation}
 \delta r^2_\parallel(t) = \delta \hat{r}^2_\parallel(t/t_\sigma'),
\end{equation}
for times $t\gtrsim t_\sigma'$, where $\delta \hat{r}^2_\parallel(\cdot)$ describes the shape of the terminal relaxation. In particular, one finds that the diffusion coefficient is expected to decrease rapidly as a power law
\begin{equation}
 D_\parallel \propto 1/t_\sigma' \propto |\sigma|^\gamma,
\label{eq:mct-power-law-D-sigma}
 \end{equation}
as the transition is approached $\sigma \uparrow 0$. 

It is also interesting to ask what happens with the confined fluid as the accessible slit width $L$ approaches zero or infinity. Physically, one anticipates that the equations of motion approach the limit of a purely two-dimensional system, respectively an ordinary unconfined bulk system. To prove convergence, one has to show that the matrix-valued correlation functions $S_{\mu\nu}(q,t)$ collapse to conventional 
intermediate scattering functions in the plane or in bulk. Furthermore, the Zwanzig---Mori matrix equations should yield scalar equations of motion, devoid of the mathematical subtleties of the splitting currents, and last the mode-coupling functional should reduce itself to the established expression for two-dimensional~\cite{Bayer:2007} and three-dimensional \cite{Goetze:Complex_Dynamics} systems. Since MCT requires the static structural properties as input, the question is intimately related to the convergence of the statics to their respective limits. For the three-dimensional case, it is not difficult to show~\cite{Lang2014b} that, indeed, both the static structure converges and the MCT equations of the glass transition in bulk are recovered. The limit of small wall separations turns out to be subtle~\cite{Lang2014c,Lang2014b}. 

Analytic progress is made observing, that for small wall separations, the lateral and transverse degrees of freedom decouple~\cite{Franosch:2012} allowing establishing a suitable perturbative scheme with the strongly interacting planar fluid as reference system. Convergence to the planar system can be shown only if the wall potential satisfies a certain smoothness criterion, in which case the corrections in the static can be calculated systematically in orders of the slit width $L$. The mode-coupling equations converge to the two-dimensional MCT equations for all finite times, yet the limits of vanishing slit width and infinite times do not commute~\cite{Lang2014b}. Hence a divergent time scale is anticipated separating the regime of decoupled lateral to transverse degrees of freedom and the strongly coupled one. The ramifications of this scenario still remain to be explored.

%%%%%%%%%%%%%%%%%%%%%%%%%%%%%%%%%%%%%%%%%%%%%%%%%%%%%%%%%%%%%%%%%%%%%%%%%
%%%%%%%%%%%%%%%%%%%%%%%%%%%%%%%%%%%%%%%%%%%%%%%%%%%%%%%%%%%%%%%%%%%%%%%%%
\subsection{Non-monotonic effect of confinement: MCT versus simulations}
\label{subsec:simulations}

The formalism described above allows to predict the state diagram for the ergodic-non-ergodic transition in  confined geometry. To this end, MCT requires as input the static structure factors, $S_{\mu\nu}(q)$, and the density profile, $n(z)$. The static structure factors are obtained from integral equation theory for inhomogeneous fluids within the  Percus---Yevick closure~\cite{Hansen2006}. The necessary density profiles, on the other hand, are calculated using the density functional theory with a fundamental measure functional~\cite{Roth2010}.

A test of these predictions for a monodisperse hard-sphere fluid via molecular dynamics simulations is not an easy task. The problem lies in the possibility of crystallization at high packing fractions~\cite{Palberg2014}. As shown by Hoover and coworkers roughly fifty years ago~\cite{Hoover1968}, the highest fluid density that remains thermodynamically stable in a monodisperse hard-sphere system corresponds to a packing fraction of $\phi_\mathrm{max, Fluid} \approx 0.49$. 
For packing fractions above this limiting value, the fluid develops a tendency towards crystallization, which grows with increasing $\varphi$. Since confinement does in general facilitate particle ordering and crystallization~\cite{Fortini2006b,Alba-Simionesco2006,Loewen2009,Loewen2011}, this problem is enhanced in the present case. Nevertheless, a standard way to circumvent this difficulty is to introduce size disparity among hard-sphere particles~\cite{Megen1994}. 

Interestingly, while at present MCT only deals with a monodisperse system, the fundamental measure theory (FMT)---whose task is to compute the local variation of density across the slit---can easily be generalized to account for the effect of polydispersity. The reliability of the FMT calculations in the high density regime of interest to MCT can thus be directly assessed via comparisons with simulations of densely packed polydisperse HS fluids. Once this is established, FMT is then applied to a monodisperse HS fluid in a narrow slit in order to provide the density profile necessary for MCT-calculations.

Within FMT, the polydispersity is accounted for by introducing $m$ components, $i=1,2,\ldots,m$, each with a diametre equal to $\sigma_i$. For a polydisperse HS fluid, the minimization of the functional for the grand potential then leads to the equation
\begin{equation}
 \label{eq:ELE}
 \ln n_i(z) = \beta\mu_i -\beta \frac{\delta F^{\rm ex}[n_i]}{\delta n_i(z)} - \beta V_i(z) \,,
\label{eq:nz}
 \end{equation}
where $n_i(z)$ is the partial number density of component $i$. Furthermore, $\mu_i$ and $V_i(z)$ are the corresponding chemical potential and the wall potential, respectively.

Equation~(\ref{eq:nz}) is solved using for the excess free energy functional, $F^{\rm ex}$, the fundamental measure functional, version White Bear II~\cite{Roth2010}. The bulk densities in the particle reservoir are taken from a Gaussian distribution as used in the simulation, and the chemical potentials $\mu_i$ correspond to these bulk densities. Interestingly, results obtained for $m=11$, 31 and 51 yield practically indistinguishable results for the density profile, meaning that a relatively small number of components is sufficient to emulate polydispersity.

\begin{figure}
\centering
\includegraphics*[width=0.6\linewidth]{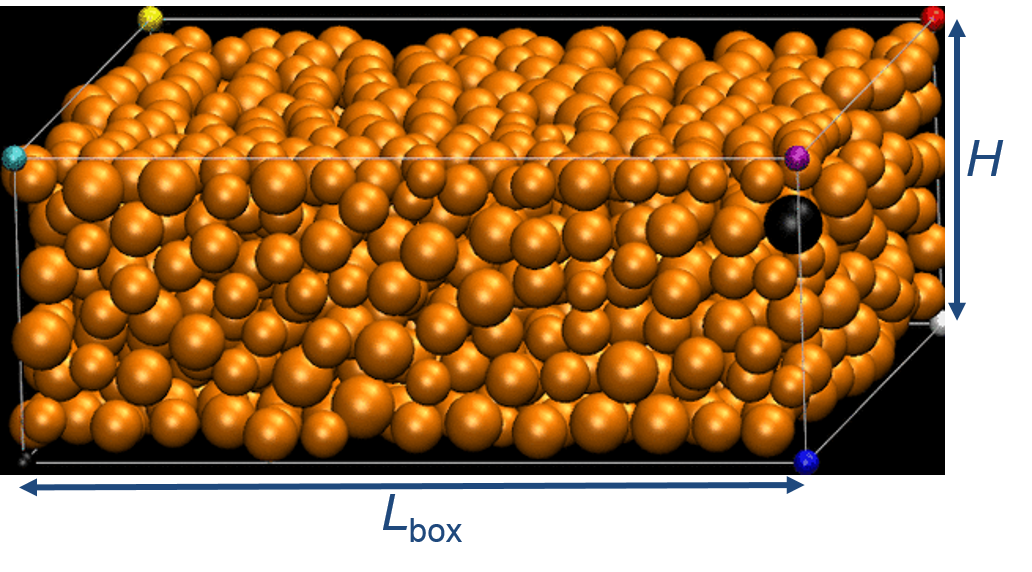}
\caption{(colour online).
The simulation setup. A polydisperse HS fluid is confined between two walls on the top and on the bottom. For a better visibility, the walls are not shown. The dimension of confinement is $H$. In order to keep the density at a desired value, the number of particles and the length $\Lbox$ are tuned. Following the motion of a single particle (e.g., the black sphere) allows one to determine the mean square displacements and thus the diffusion coefficient via the Einstein formula. Small particles on the corners are added for visualization purpose only.
They have no effect in the simulation.}
\label{fig:simulation-setup}
\end{figure}

As for simulations, event driven molecular dynamics method is used~\cite{Bannerman2011}. The particle sizes are drawn from a Gaussian around a mean diametre of $\bar \sigma$. The width of this Gaussian distribution relative to the mean particle diametre defines polydispersity. Two different polydispersities of 10\% and 15\% are investigated. The planar slit is defined by two planar hard walls placed in parallel at $\pm H/2$ (the width of the domain accessible to particle centers is approximately equal to $H-\bar\sigma$). Periodic boundary conditions are applied along the two independent directions parallel to the wall. Lengths is measured in units of the mean particle diametre, $\bar \sigma$, and  times in units of  $\tauHS=\sqrt{m \bar\sigma^2/\kB T}$ where $\kB$ is the Boltzmann constant, $T$ is temperature and  $m$ is the mass of a particle. This set of units is referred to as 'HS units'. The centre of particle $i$ with diametre $\sigma_i$ is confined 
to $-(H-\sigma_i)/2 \leq z_i \leq (H-\sigma_i)/2$. The volume of the simulation box is $V=\Lbox^2H$, where the lateral system size $\Lbox$ varies in the range from $60\bar\sigma$ to $75\bar\sigma$ (Fig.~\ref{fig:simulation-setup}). Depending on polydispersity, the packing fractions investigated lie in the range $\varphi \in [0.4,\; 0.49]$ (10\%) and  $\varphi \in [0.45\;\;\; 0.54]$ (15\%). Depending on $H, \Lbox$ and $\varphi$, the number of particles ranges between $8000$  and $30000$. Thermal equilibrium is ensured by sufficiently long simulations (extending up to 7 decades in time) and by explicitly testing the time-translation invariance of the properties of interest~\cite{Mandal2014}.

\begin{figure}
\centering
(a)\includegraphics*[width=0.6\linewidth]{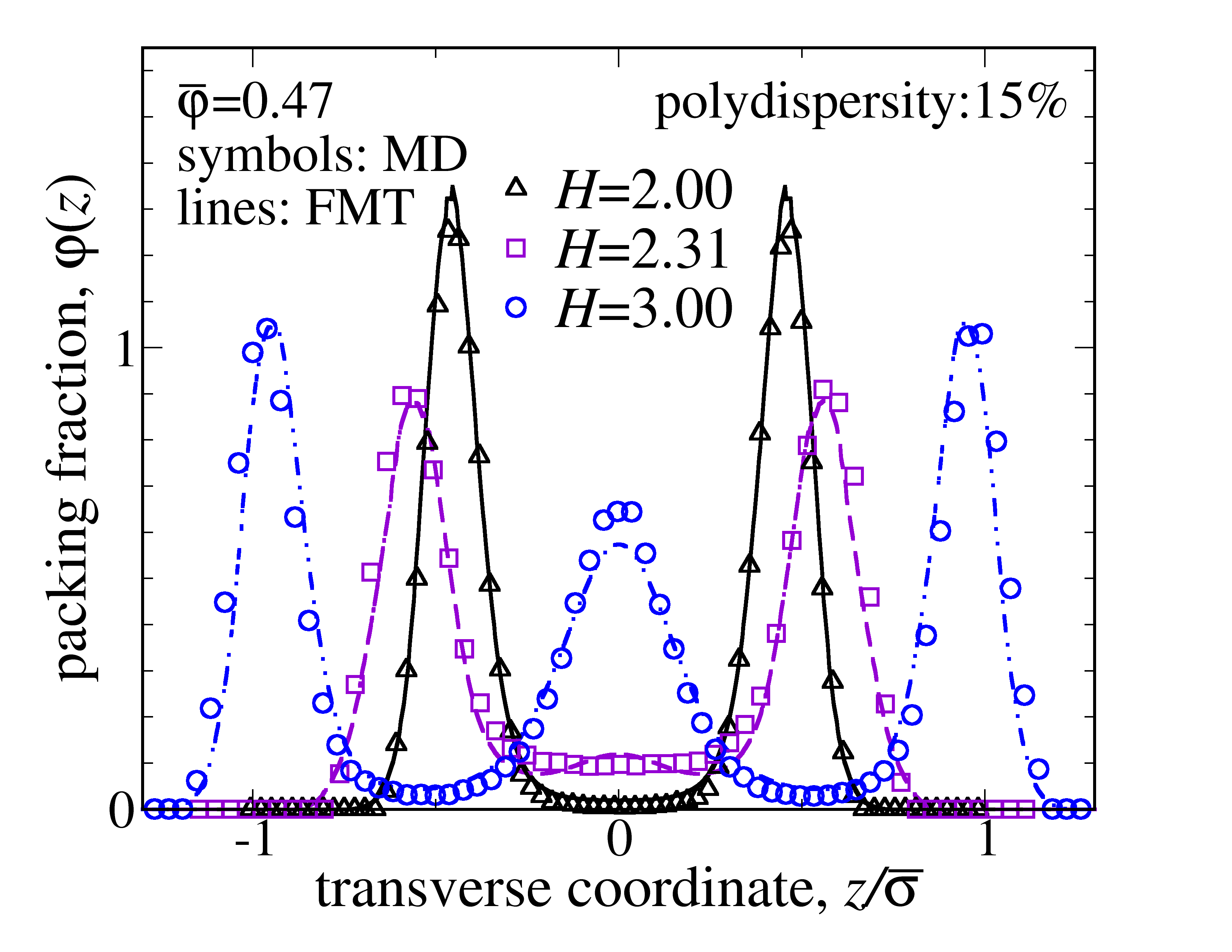}
(b)\includegraphics*[width=0.6\linewidth]{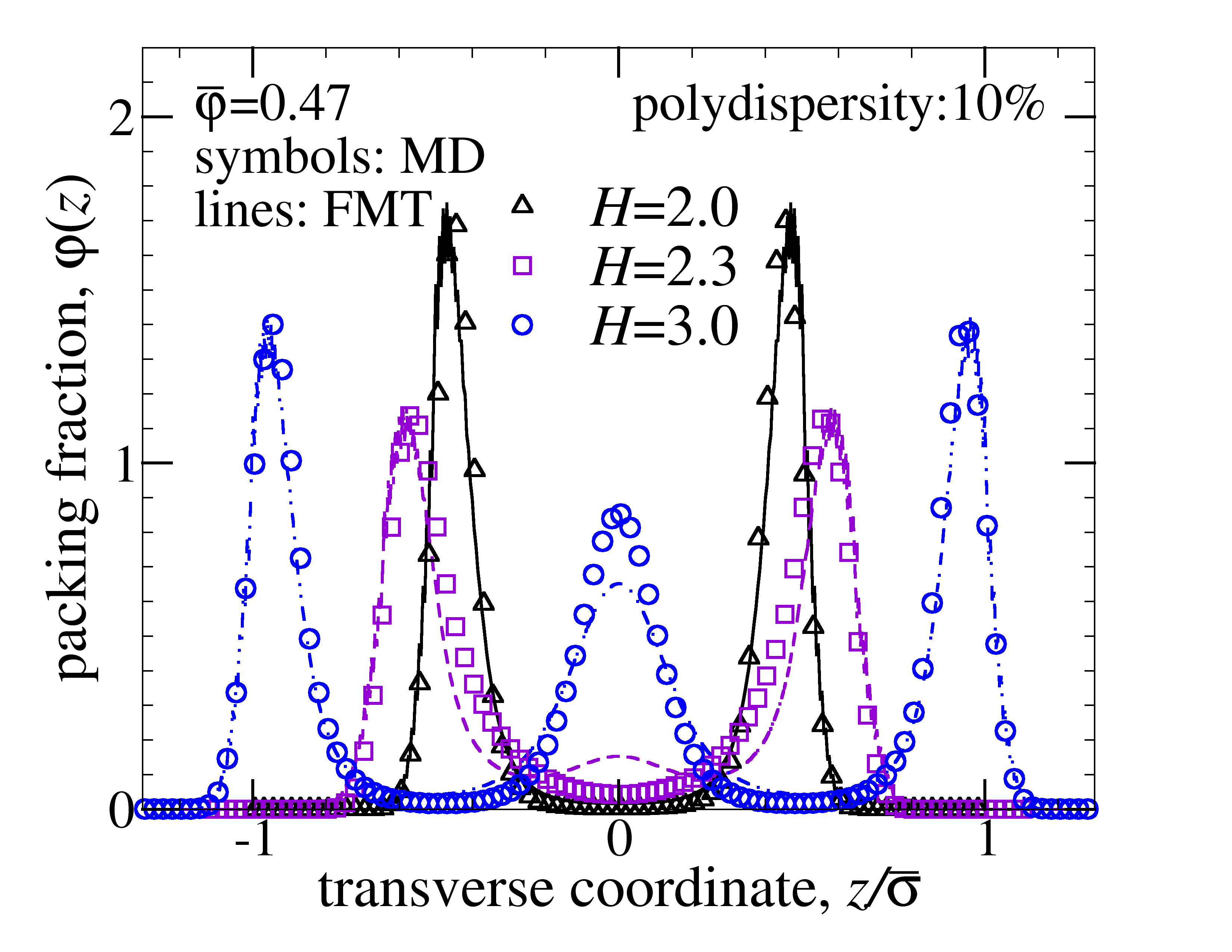}
(c)\includegraphics*[width=0.6\linewidth]{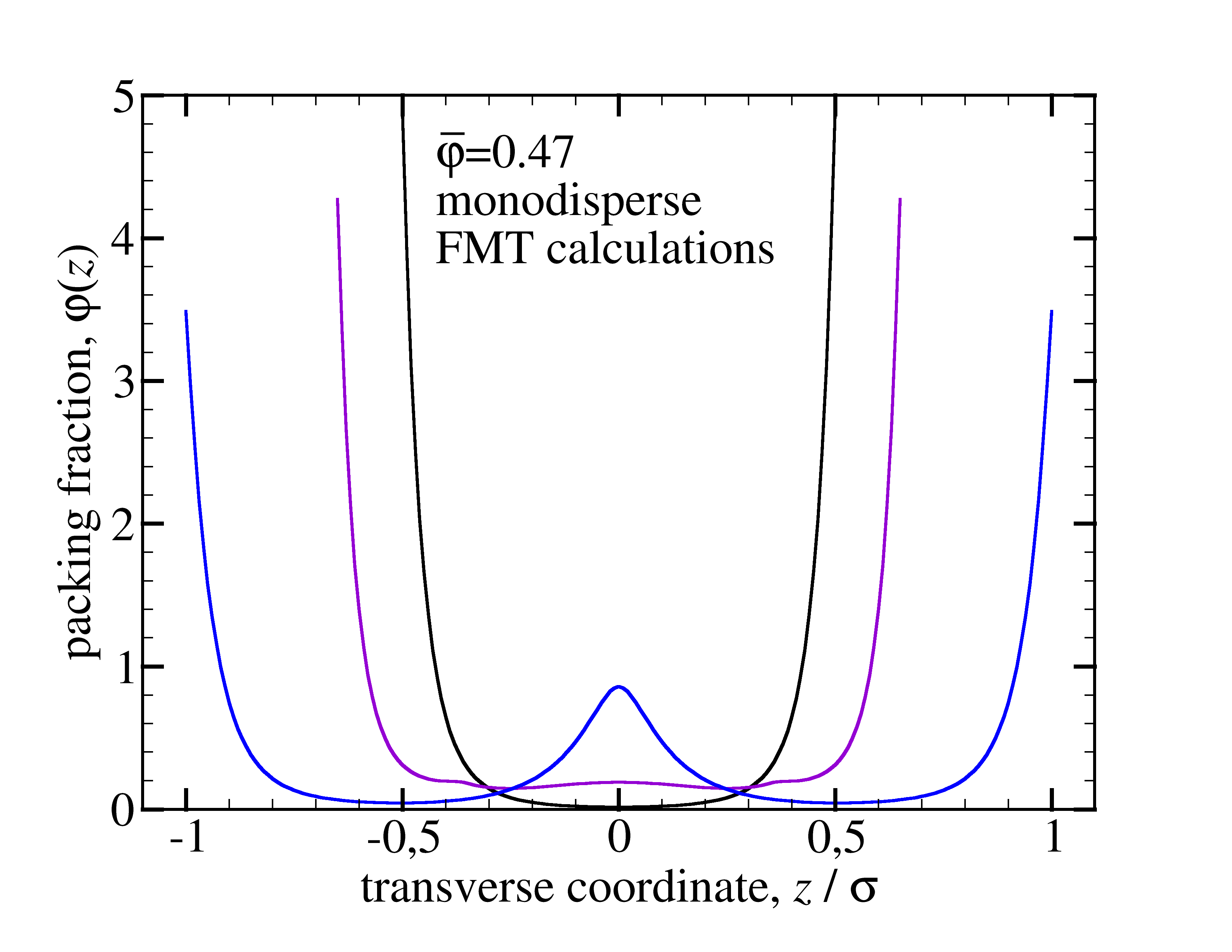}
\caption{
(colour online).
Local variation of the packing fraction, $\varphi(z)$, within a planar slit for a polydisperse hard-sphere fluid. Symbols in (a) and (b) indicate results obtained via event driven MD simulations and lines stand for density functional calculations using a fundamental measure functional at the same average packing fraction and polydispersity. Note the less pronounced peaks at the higher polydispersity. (c) FMT-results on density profile for the monodisperse case. These $\varphi(z)$ are part of the necessary input for the calculation of the system dynamics via MCT.}
\label{fig:phi_z_MD_vs_FMT}
\end{figure}

As can be inferred from  panels (a) and (b) of Fig.~\ref{fig:phi_z_MD_vs_FMT}, remarkable agreement is found between the results obtained from density functional calculations and MD simulations for the two choices of polydispersity. The deviations between the FMT and the MD simulations are probably due to the slight difference between slit and reservoir particle distributions. Panel (c) in Fig~\ref{fig:phi_z_MD_vs_FMT} illustrates the FMT results on $\varphi(z)$ for a monodisperse HS fluid ($n=1$) used as input to the dynamic MCT calculations.

\begin{figure}
\centering
(a)\includegraphics*[width=0.6\linewidth]{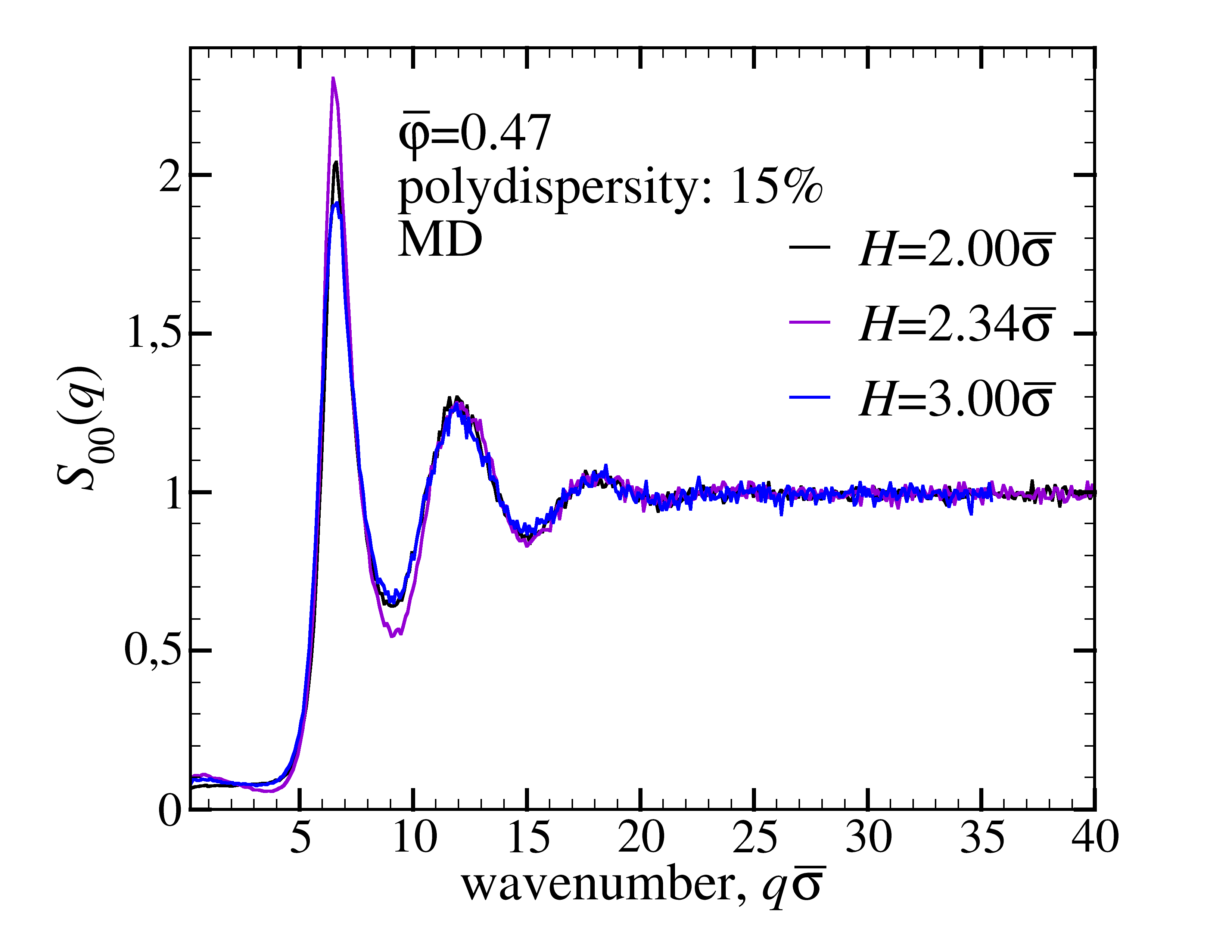}
(b)\includegraphics*[width=0.6\linewidth]{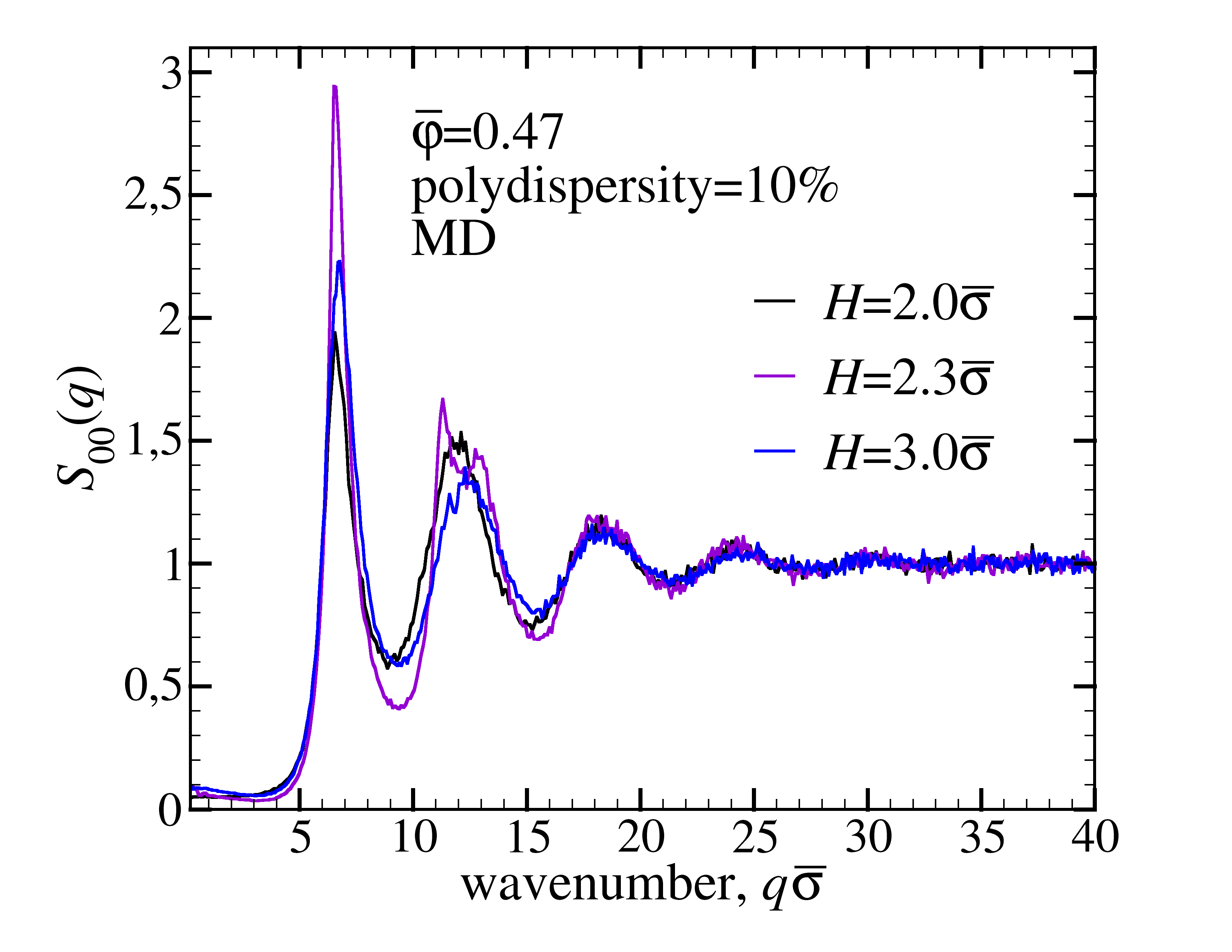}
(c)\includegraphics*[width=0.6\linewidth]{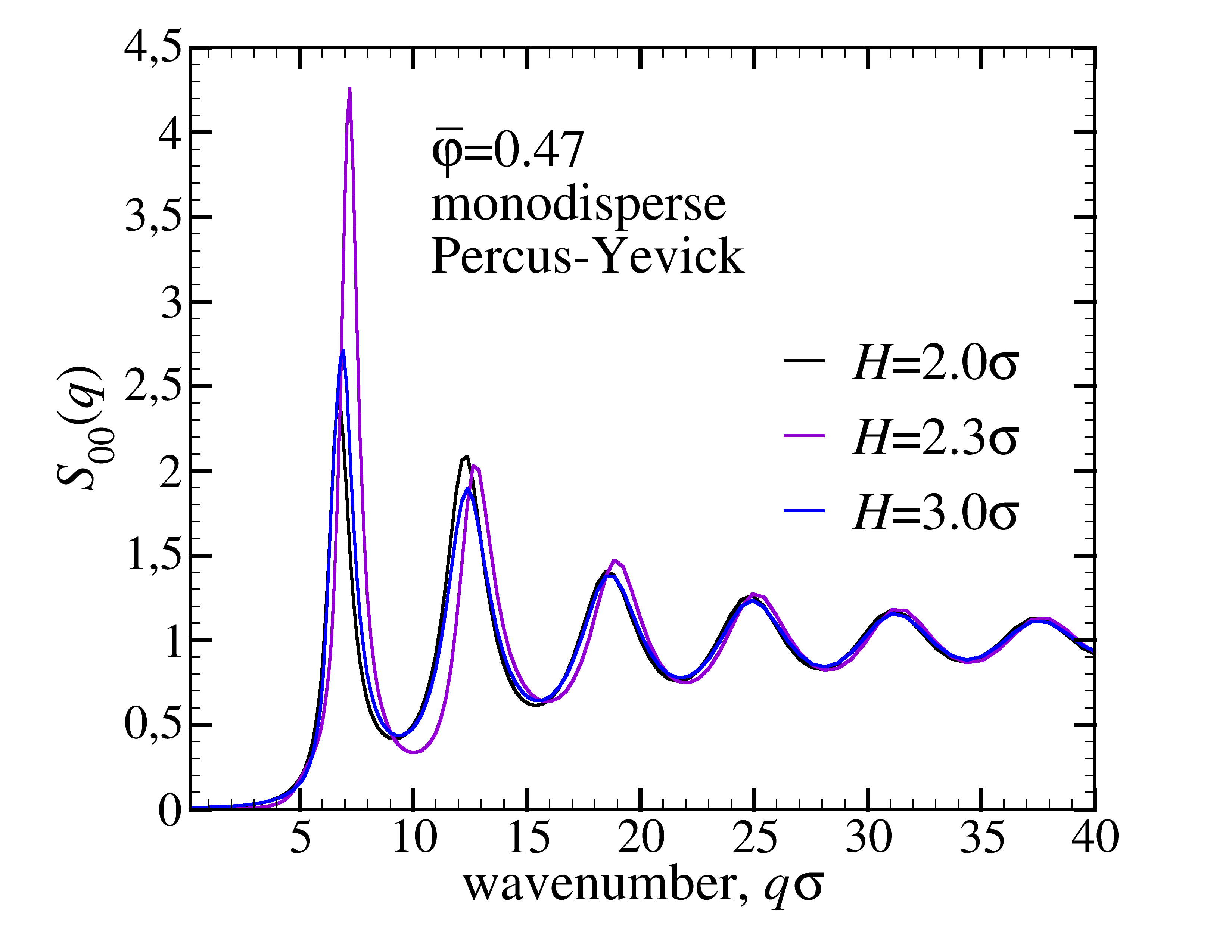}
\caption{(colour online).
The \revision{slit-averaged} static structure factor, $S_{00}(q)$, obtained from MD simulations (panels a,b) of polydisperse hard-sphere fluids, confined between two smooth hard plates, is compared to theoretical calculations using the Percus---Yevick closure (panel c). In theoretical calculations, the necessary density profiles are obtained from the fundamental measure theory at each investigated plate separation, $H$ (see, Fig.~\ref{fig:phi_z_MD_vs_FMT}c). Note that, both in simulations and theory, the first sharp diffraction peak varies in a non-monotonic 
way with plate separation. It is lowest for the weakest confinement, $H=3.0\bar\sigma$, increases to the highest value as $H$ raises to $H\approx 2.3\bar\sigma$ and decreases to an intermediate level for the strongest confinement investigates, $H=2.0\bar\sigma$.
Images (modified) from~\cite{Mandal2014} (reproduced with permission).
}
\label{fig:Sq_MD_vs_PY}
\end{figure}

The next important input to the dynamic MCT calculations is the static structure factor, $S(q)$. Figure~\ref{fig:Sq_MD_vs_PY} illustrates $S(q)$ obtained from event driven MD simulations of a confined HS fluid for the two polydispersities of 15\% (Fig.~\ref{fig:Sq_MD_vs_PY}a) and 10\% (Fig.~\ref{fig:Sq_MD_vs_PY}b) along with theoretical results~\cite{Lang2010,Lang2012,Lang2013} for the corresponding monodisperse system (Fig.~\ref{fig:Sq_MD_vs_PY}c). The quality of the Percus---Yevick closure, used in theoretical calculations of $S(q)$, has been corroborated recently for confined systems~\cite{Nygard2012}.

Despite the above mentioned difference in the theoretical setup with simulations, a comparison of the qualitative features in the respective static structure factors reveals itself as very instructive. Both in simulations and theoretical calculations, the structure factor exhibits a non-monotonic dependence on the plate separation, $H$. This is best seen by a survey of the first peak of $S_{00}(q)$. Starting from a plate separation of three (average) particle diametres and gradually enhancing the degree of confinement, the first maximum of $S_{00}(q)$, which has the smallest value at $H=3\bar\sigma$, jumps directly to the highest value  observed  at $H\approx 2.3\bar\sigma$ and then falls back to an intermediate level at $H=2.0\bar\sigma$. This non-monotonic effect 
is weak but discernible for the highest polydispersity investigated (Fig.~\ref{fig:Sq_MD_vs_PY}a), becomes stronger 
and clearly visible for a lower polydispersity of $10\%$ (Fig.~\ref{fig:Sq_MD_vs_PY}b) and is most prominent for the case of 
a monodisperse system (Fig.~\ref{fig:Sq_MD_vs_PY}c).

A qualitative interpretation of the observed behaviour of the density profiles and structure factors within planar slits can be given as resulting from a competition between the short-range local order and the imposed confinement. For an integer ratio of the slit width to the particle size, a number of identically filled layers can fit into the space between the slit walls, where the zones between adjacent layers are essentially devoid of particles' centre of mass. This case is referred to as a commensurate packing. In the incommensurate case, the slit width is not an integer multiple of the particle diametre so that identically filled layers cannot form. Moreover, the space between the adjacent layers is no longer empty but contains a finite fraction of particles. This dependence of the density profile on the slit width is most prominent for a monodisperse system, where two well-defined length scales compete with one another, the particle diametre and the slit width.

In a polydisperse system, on the other hand, the presence of particles of various sizes allows a more effective filling of the available space (the smallest particles playing to some extent the role of `interstitials`). In this case, even if the slit width is an integer multiple of the average particle diametre, some small particles may still reside in the intermediate space between the layers. As a result, a less heterogeneous density profile forms as compared to the equivalent monodisperse fluid. Accordingly and due to the same improved flexibility in space filling, a variation of the slit width has a less strong impact on the packing structure of a polydisperse HS fluid.

The agreement discussed above between the theory for monodisperse confined hard spheres and the simulation for a hard-sphere system with size dispersity on the static level is of central importance for a meaningful comparison of computer simulation and MCT for the dynamics in the vicinity of the glass transition. Indeed, the pioneering experiments~\cite{Megen1994} on hard spheres with 4\% polydispersity have been quantitatively rationalized within the one-component monodisperse MCT. Moreover, empirical studies of the MCT solutions for several components in bulk~\cite{Weysser2010} demonstrate only slight quantitative changes with respect to the 
monodisperse case. These observations motivate us to compare the present case of a confined polydisperse HS fluid with MCT predictions for the corresponding monodisperse system.

\begin{figure}
\centering
(a)\includegraphics*[width=0.6\linewidth]{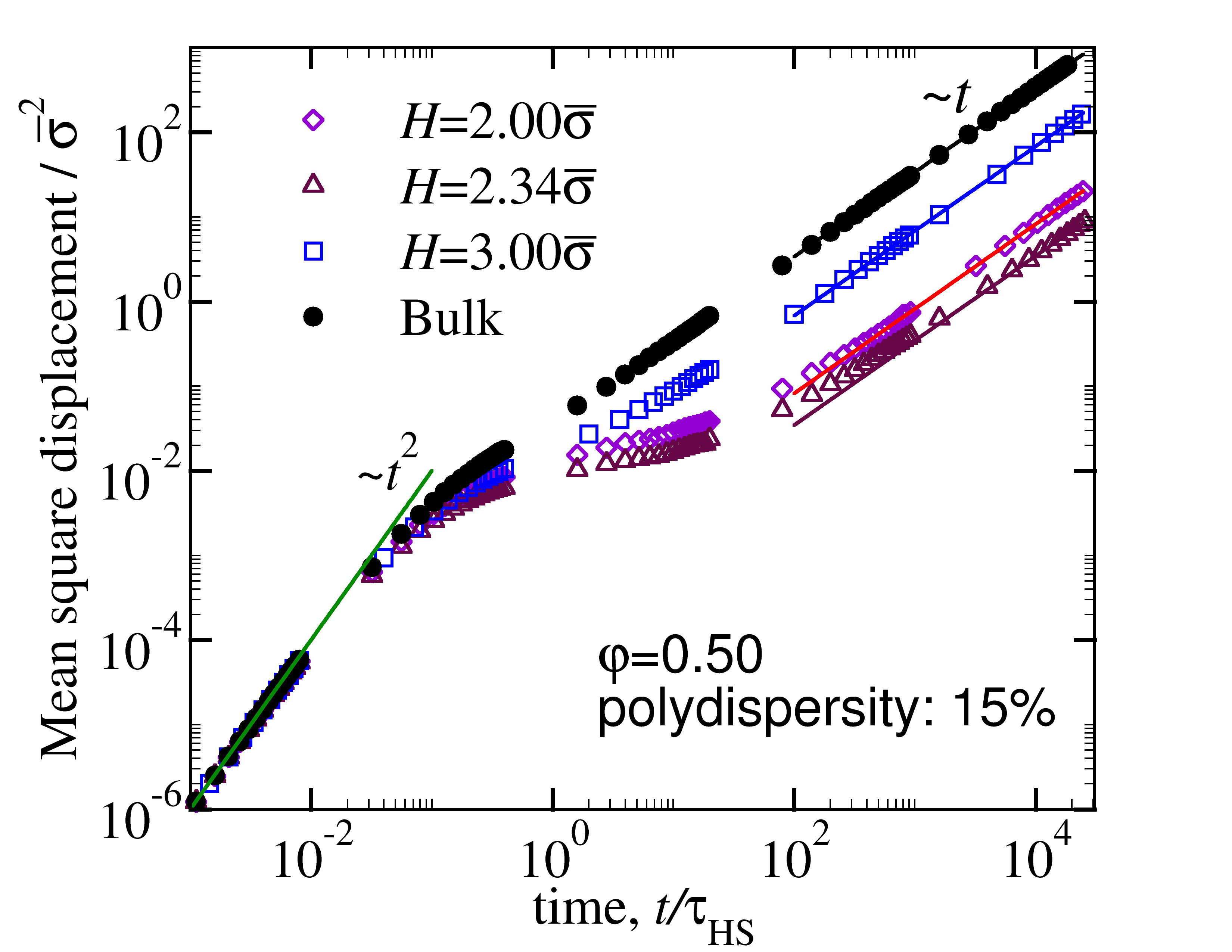}
(b)\includegraphics*[width=0.6\linewidth]{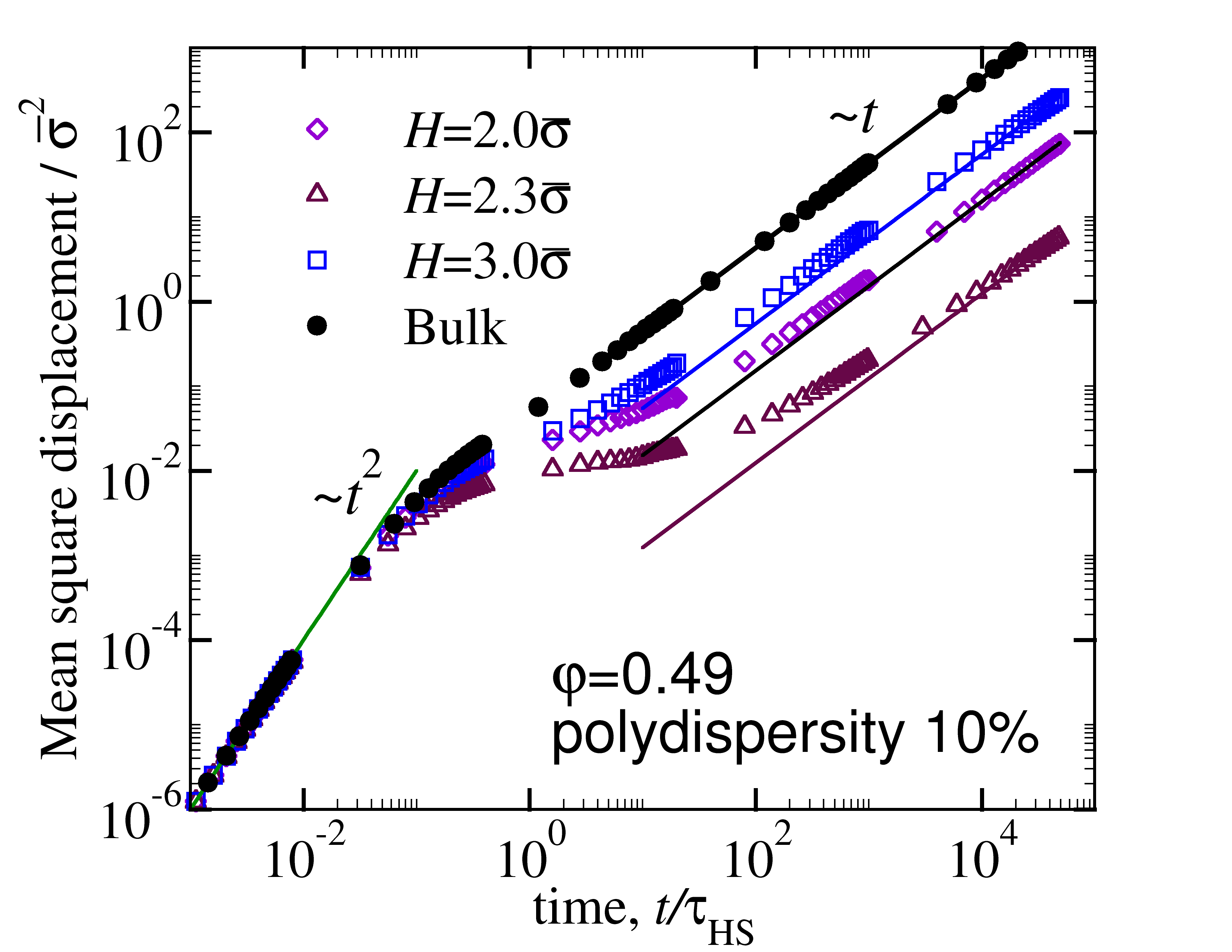}
\caption{(colour online).
The component of the \revision{slit-averaged}  mean square displacement in the direction parallel to the slit wall versus time for different slit widths, $H$, as indicated. The data correspond to the average over the entire slit. After a short-time ballistic motion (MSD$\sim t^2$), the single particle dynamics 
crosses gradually over to the long-time diffusive behaviour (MSD$\sim t$). The onset of the diffusive dynamics is associated with the structural relaxation. In qualitative agreement with the confinement effect on the static structure, the dynamics is fastest for the weakest confinement investigated, $H=3\bar\sigma$. It then slows down as $H$ decreases to $H\approx 2.3\bar\sigma$ but accelerates again as $H$ is reduced further 
to $H=2.0\bar\sigma$. For reference, the bulk data are also shown for the same packing fraction. The polydispersity is 15\% in (a) and  10\% in (b). The unit of time if $\tauHS$ and the unit of 
Images (modified) from~\cite{Mandal2014} (reproduced with permission).
}
\label{fig:msd-poly15+10}
\end{figure}

On the basis of the above discussed structure factors, one expects a corresponding non-monotonic effect on 
the system dynamics. This expectation is born out by a survey of single particle dynamics in MD simulations. As illustrated in Fig.~\ref{fig:msd-poly15+10}, the fastest mean square displacement (MSD) occurs for the reference bulk system, indicating 
that the overall effect of confinement is to slow down the system dynamics. However, in qualitative agreement with the 
confinement effect on the system structure, this effect turns out to be non-monotonic as the degree of confinement
 is increased gradually. Letting aside the bulk system, the dynamics is fastest for the 
weakest confinement investigated, then it slows down and accelerates again 
as $H$ decreases from $H=3.0\bar\sigma$ through $H\approx 2.3\bar\sigma$ to $H=2.0\bar\sigma$.

Interestingly, as compared to  bulk, the extent of the crossover region from ballistic to diffusive dynamics is significantly 
increased, signaling the onset of a two-step relaxation, a hallmark of the dynamics of supercooled liquids. Thus, at constant 
average density, the confinement plays here a role qualitatively similar to that of densification. As will be shown below, this 
observation is perfectly in line with the MCT calculations on the dynamics of confined HS fluids, which extrapolate this analogy to higher densities and predict that confinement even shifts the liquid-glass transition line.

Before exploring the dynamic behaviour of the system close to the glass transition line, it is important to first check whether the confinement induces any type of phase separation~\cite{Fortini2006} between small and large particles. For the present analysis, it is reassuring that no such effect occurs in the system. In particular, the distribution of particle sizes is homogeneous across the slit~\cite{Mandal2014}. Another issue worth worrying about in the context of glass transition is the possibility for confinement-induced long-range order. Indeed, as the packing fraction increases, long-range order becomes visible at some threshold $\varphi$, which depends on polydispersity and degree of confinement. The higher is polydispersity, the higher is this threshold and thus the larger is the range of packing fractions, which exhibit no long-range order and are thus suitable for dynamical studies.

\begin{figure}
\centering
(a)\includegraphics*[width=0.6\linewidth]{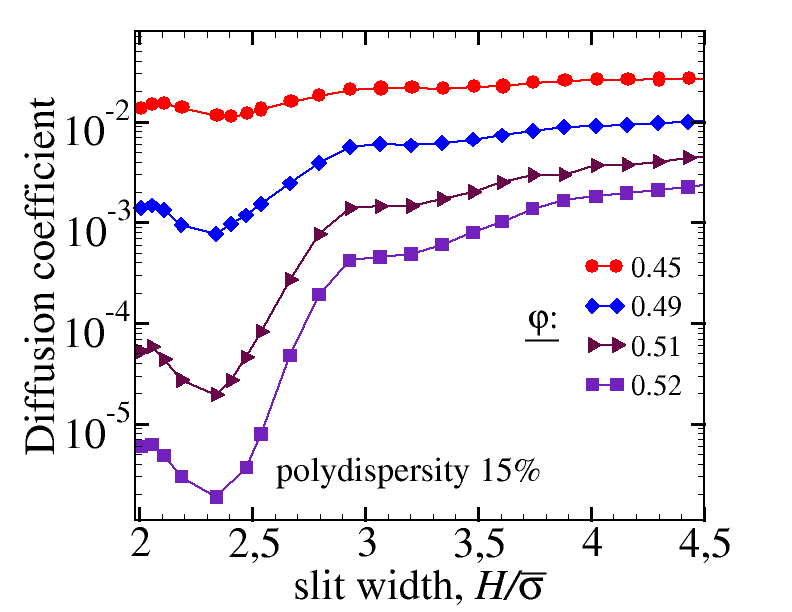}\quad
(b)\includegraphics*[width=0.6\linewidth]{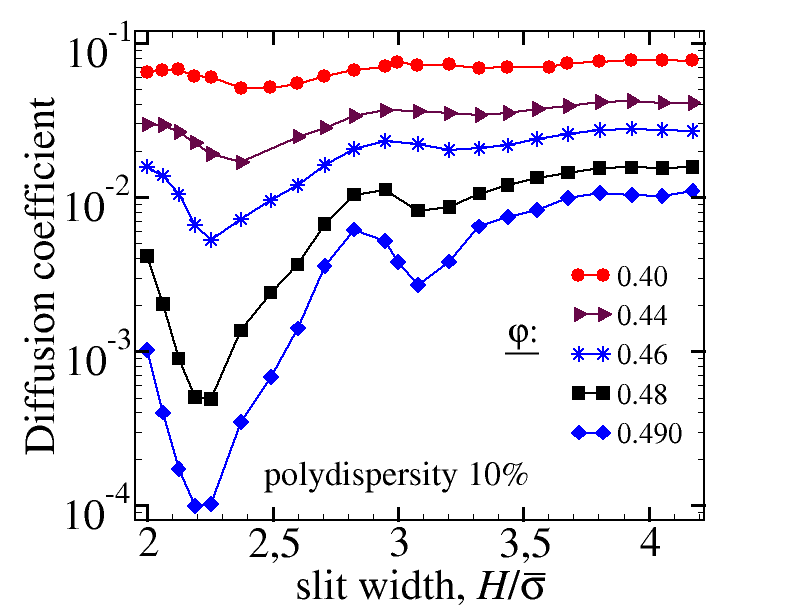}
\caption{(colour online).
Slit-averaged diffusion coefficient, $D$, as function of the plate separation, $H$, for different slit-averaged packing fractions, $\varphi$, as indicated. Note the weakening of the oscillations in $D(H)$ as polydispersity increases from \revision{10\% in panel (b) to 15\%} in panel (a). Images (modified) from~\cite{Mandal2014} (reproduced with permission).
}
\label{fig:diffusion_m4}
\end{figure}

The diffusion coefficient is obtained from the long-time behaviour of the single particle mean square displacement using the Einstein formula $D=\lim_{t \to \infty} \langle [x(t)-x(0)]^2 \rangle/2t$ (Fig.~\ref{fig:diffusion_m4}). A reliable criterion here to have reached the diffusive regime is that a particle moves a distance comparable to the average particle size, MSD$\approx \sigma^2$. As mentioned in section \ref{subsec:strong-confinement}, a non-monotonic dependence of $D$ on the plate separation had been observed for moderate densities~\cite{Mittal2008}. There, $D$ changed by about a factor of 2 for practically the same range of slit width as addressed in Fig.~\ref{fig:diffusion_m4}. Introducing polydispersity shifts the crystallization threshold to higher packing fractions and extends the accessible dynamic range considerably. For a polydispersity of 15\%, for example, the slit-averaged packing fraction can be set as high as $\varphi=0.52$, where the diffusion coefficient varies by a factor of 1000 upon a variation of $H$ (\revision{Fig.~\ref{fig:diffusion_m4}a}).

The strong non-monotonic variation of the diffusion coefficient with slit width is a further consequence of the above discussed competition of the two length scales of confinement and particle size. This view is corroborated by the qualitatively similar non-monotonic dependence on $H$ both for the diffusion coefficient (Fig.~\ref{fig:diffusion_m4}) and in the static structure factor (Fig.~\ref{fig:Sq_MD_vs_PY}). Consistently with this, polydispersity shows the same effect in both cases as it weakens the sensitivity to a variation of the degree of confinement.

Despite its oscillations with the slit width, the diffusion coefficient remains a monotonic function of the average packing fraction $\varphi$ for a fixed $H$. A question thus arises whether the simulation results on $D(\varphi)$, for each selected slit width, can be adequately described by the mct-power law, Eq.~(\ref{eq:mct-power-law-D-sigma}), which is asymptotically predicted by the (idealized) MCT in the bulk~\cite{Gotze2009} and has been also shown to persist under confinement~\cite{Gallo2000,Varnik2002e,Gallo2012}. Recalling that $\sigma\propto \varphic-\varphi$ for the case of a HS fluid, this equation reads

\begin{equation}
D(\varphi) \propto (\varphic-\varphi)^\gamma. 
\label{eq:mct-power-law-D-phi}
\end{equation}

\begin{figure}
\centering
\includegraphics*[width=0.6\linewidth]{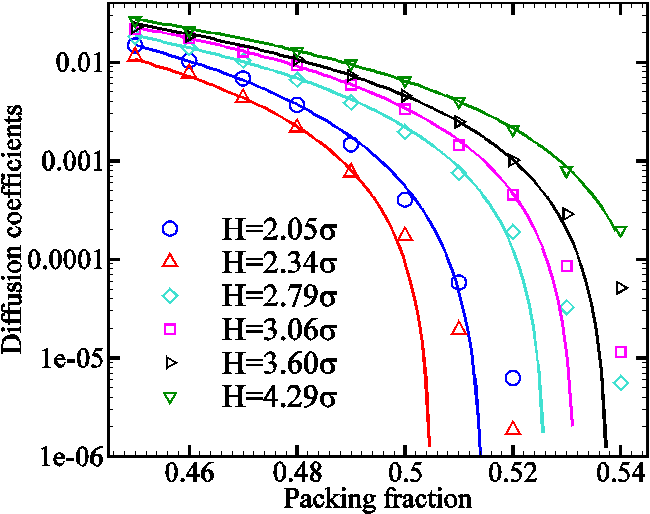}
\caption{(colour online).
A test of the applicability of Eq.~(\ref{eq:mct-power-law-D-phi}), which is a reformulation of the mct-prediction Eq.~(\ref{eq:mct-power-law-D-sigma}) for the case of a HS fluid, via event driven MD simulations. \revision{Slit-averaged} self diffusion coefficient (symbols) are plotted in log-linear scale versus the packing fraction $\varphi$. The solid lines are fits to the mct-prediction. Polydispersity is 15\%.
}
\label{fig:power-law_m5}
\end{figure}

Figure~\ref{fig:power-law_m5} shows that Eq.~(\ref{eq:mct-power-law-D-phi}) describes the variation of the diffusion coefficient with packing fraction over roughly two decades in time. \revision{Even though this range is rather limited, such an analysis is useful as it provides one, at least, with a rough estimate of the MCT glass transition line under confinement. Interestingly, the} exponent $\gamma=2.1 \pm 0.1$ is found to be rather robust, and depends only weakly on polydispersity and $H$. This means that the fit to Eq.~(\ref{eq:mct-power-law-D-phi}) probes essentially the location of the critical packing fraction, $\varphic$, for each $H$.

\begin{figure*}
\centering
\hspace*{3mm} (a) \hspace*{3mm}\includegraphics*[width=0.65\linewidth]{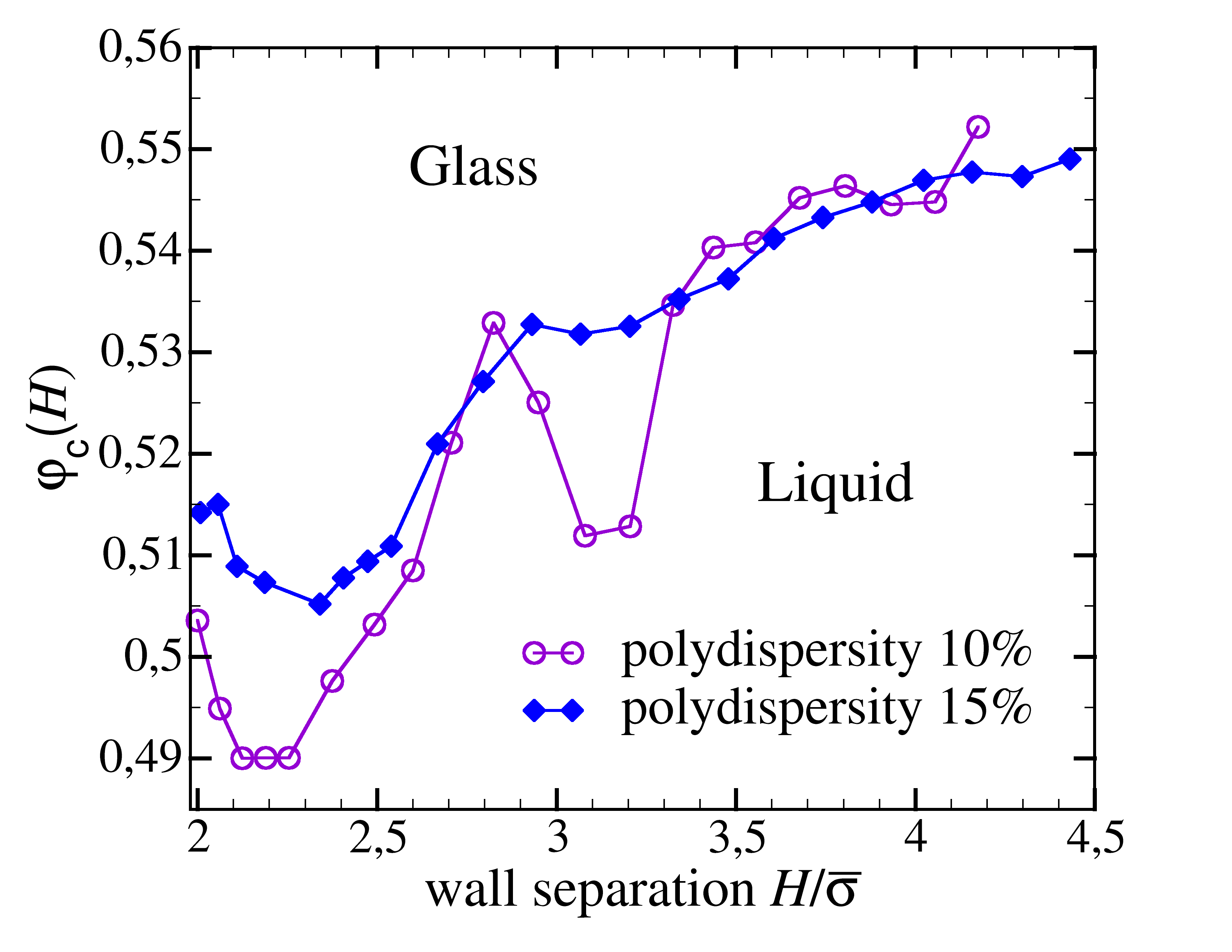}
(b)\includegraphics*[width=0.65\linewidth]{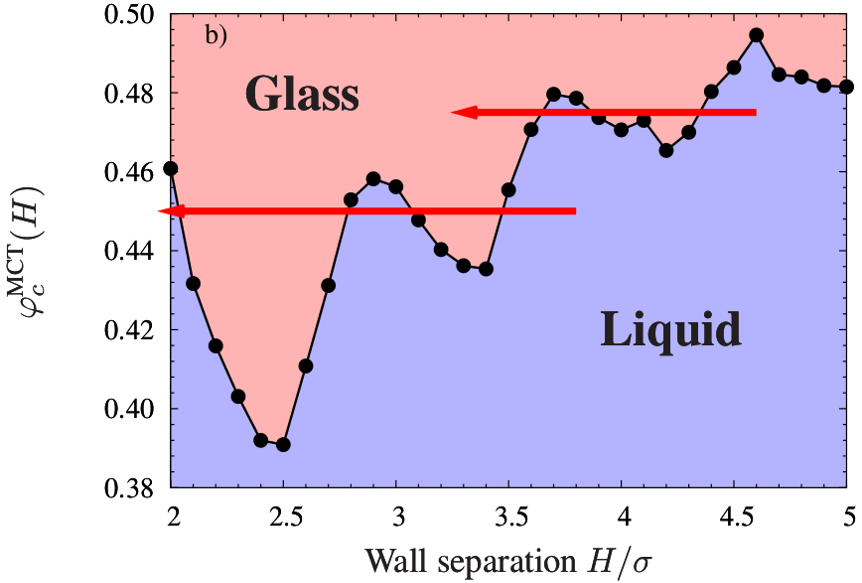}
\caption{(colour online).
Non-equilibrium state diagram of confined polydisperse hard spheres as obtained from (a) power law fits of the simulated diffusion coefficients to Eq.~(\ref{fig:power-law_m5}) and (b) numerical solution of the MCT equations for a hard-sphere fluid, confined between two parallel smooth hard walls. The arrows indicate paths of equal densities where (multiple) reentrant behaviour occurs. The simulated system is polydisperse, while the MCT result is obtained for a monodisperse HS fluid. A comparison of the state diagram for the two simulated polydispersities and theory (monodisperse case) is consistent with the idea that the main cause for the observed oscillations of $\varphic$ lies in the competition between packing effects and layering tendency, the latter being imposed by confinement. Since the layering tendency decreases upon increasing polydispersity (see Fig.~\ref{fig:phi_z_MD_vs_FMT}), the resulting 
non-monotonic effect is weakened.
Images (modified) from~\cite{Mandal2014} (reproduced with permission).
}
\label{fig:phase_diagram_m6}
\end{figure*}

The critical packing fraction obtained from fits to Eq.~(\ref{eq:mct-power-law-D-phi}) is an indicator for the glass-transition line. Figure~\ref{fig:phase_diagram_m6} compares the simulation results on $\varphic(H)$,  to the predictions of the MCT, developed in the last section. As seen from this figure, both the simulation and the theoretical results on the state diagram exhibit oscillations with a period comparable to the (average) hard-sphere diametre. Presumably, this is a manifestation of the competition between wall-induced layering and local packing. This view is in line with the observation in Fig.~\ref{fig:phase_diagram_m6} that the resulting non-monotonic effect is weakened at higher polydispersity, which exhibits a less strong layering tendency than the less polydisperse system (see Fig.~\ref{fig:phi_z_MD_vs_FMT}; note also the scale of the vertical axis).

This mechanism is of a generic character so that the resulting reentrant behaviour is expected to be a robust phenomenon, observable in a large variety of glass forming liquids. Arrows in Fig.~\ref{fig:phase_diagram_m6}b highlight also the possibility of multiple transitions from a liquid state into a non-ergodic glass state and back to the liquid state upon a variation of the slit width along lines of constant packing fraction.

%%%%%%%%%%%%%%%%%%%%%%%%%%%%%%%%%%%%%%%%%%%%%%%%%%%%%%%%%%%%%%%%%%%%%%%%%
%%%%%%%%%%%%%%%%%%%%%%%%%%%%%%%%%%%%%%%%%%%%%%%%%%%%%%%%%%%%%%%%%%%%%%%%%
\subsection{Multiple liquid-glass domains at constant chemical potential}

\begin{figure}
\centering
\includegraphics*[width=0.8\linewidth]{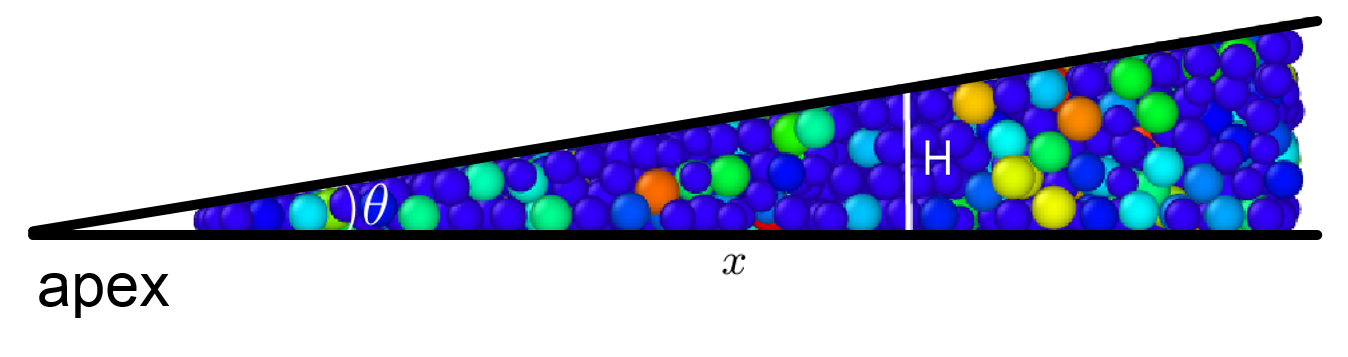}
\caption{(colour online).
A snapshot of MD simulations of a wedge-shaped channel containing a polydisperse hard-sphere fluid. The local height $H$ obeys $H= x \tan(\theta)$, where $x$ is the distance from the apex. In this example, the tilt angle is $\theta\approx 9^\circ$. In real experiments, one usually has $\theta\approx 1^\circ$.
Image (modified) from~\cite{Mandal2014} (reproduced with permission).
}
\label{fig:wedge_m7}
\end{figure}

While it is rather easy in computer simulations to fix the average packing fraction within a slit at any given slit width, such a path is quite difficult to follow in real experiments, where the chemical potential would be the more appropriate control parameter. An example is a wedge-shaped channel, where the chemical potential is spatially uniform but the local packing fraction/density at a given distance from the apex (see Fig.~\ref{fig:wedge_m7}) is not necessarily position independent. Indeed, in such a geometry, density variations may occur and it is expected that this effect will give rise to the formation of multiple liquid-glass domains at sufficiently high packing fractions. A way to study this issue is to transfer the results obtained for a variable slit width at fixed packing fraction to the case of a constant chemical potential. The results obtained from this procedure support the above conjecture that regions of low density may coexist with more densely packed domains in an alternating arrangement. Further evidence is also provided by direct computer simulations at constant chemical potential for moderate densities.

The procedure which allows to transfer the results obtained at constant density to the case of a fixed chemical potential shall now be described. A liquid of $N$ particles confined between two parallel flat walls of surface area $A$ separated by a distance $H$ (assumed to be comparable to the bulk correlation length) is characterized by the grand potential $\Omega= \Omega(T,A,H,\mu)$ (with the temperature $T$).

The grand potential fulfills the fundamental thermodynamic relation
\begin{equation}
 \diff \Omega = -S \diff T - p_L H \diff A - p_N A \diff H -N \diff \mu,
\label{eq_fund}
\end{equation}
where $S$ is the entropy, $\mu$ is the chemical potential and $p_L$ and $p_N$ are the lateral~\footnote{It does not really matter if one refers the lateral pressure $p_L$ to the physical plate separation $H$ or the accessible slit width $L$. The measurable quantity is the product $(p_N-p_L) H$ which plays the role of a surface tension. In the following equations only this product enters.} and normal pressures, respectively. It is clear that for macroscopic plate separations $H \gg \sigma$, only the total volume $V = A H$ enters the grand potential. Then $\diff V = H \diff A + A \diff H$ and one concludes that both pressures $p_L$ and $p_N$ become identical in this limit and coincide with the conventional bulk pressure $p$. In contrast for small plate separations these pressures can differ drastically and reflect the resistance to compressing the system laterally, respectively normally.

Considering the grand potential per area $\omega = \Omega /A$, extensivity implies that $\omega = \omega(T,H,\mu)$ depends only on the intensive thermodynamic control parameters $T$ and $\mu$, as well as the plate distance $H$. Taking the total differential of $\Omega(T,A,H,\mu) = A \,\omega(T,H,\mu)$ reveals that $\omega = - p_L H$ which is the proper generalization of the well-known relation $\Omega = - p V$ for homogeneous bulk systems.  Furthermore, this yields the fundamental relation 
\begin{equation}\label{eq:omega_diff}
 \diff \omega = - \sigma \diff T - p_N \diff H - n \diff \mu,
\end{equation}
with the entropy per area $\sigma = S/A$ and the area density $n= N/A$. 

The thermodynamics of wedges has been studied extensively, mostly in the context of density functional theory~\cite{Henderson2004}. Here a much simpler approach is followed. For small tilt angles, the plates are almost parallel and the fluid can be viewed as being locally confined with a wall separation $H$. Since particles are free to move along the wedge, each section is in chemical and thermal contact with its 
neighbours. Hence, the chemical potential and the temperature along the wedge are spatially constant, whereas the particle density adjusts locally to this constraint. The thermodynamic coefficient characterizing the density variation along the wedge is therefore
\begin{equation}
 \left( \frac{\diff n}{\diff H} \right)_{\text{wedge}} = \left( \frac{\partial n}{\partial H} \right)_{T,\mu} = - \left( \frac{\partial^2 \omega}{\partial H \partial \mu} \right)_T = \left( \frac{\partial p_N}{\partial \mu } \right)_{T,H},
\end{equation}
where for the last equality a Maxwell relation following from Eq.~(\ref{eq:omega_diff}) has been employed. Last the thermodynamic derivative of the normal pressure can be simplified using the thermodynamic calculus (temperature is constant throughout)
\begin{eqnarray}
 \left( \frac{\partial p_N}{\partial \mu} \right)_{T,H} &= \left. \frac{\partial (p_N,H)}{\partial (\mu,H)} \right|_T= \left. \frac{\partial (p_N,H)}{\partial (\omega,H)}
\frac{\partial (\omega,H)}{\partial (\mu,H)} \right|_T \nonumber \\
 &= \left(\frac{\partial p_N}{\partial \omega} \right)_{T,H}  \left( \frac{\partial \omega}{\partial \mu} \right)_{T,H}.
\end{eqnarray}
With Eq.~(\ref{eq:omega_diff}) and recalling that $\omega=-p_L H$,  the desired thermodynamic coefficient is expressed as
\begin{eqnarray}
 \left( \frac{\diff n}{\diff H} \right)_{\text{wedge}}  = - n  \left( \frac{\partial p_N}{\partial \omega} \right)_{T,H} =
\frac{n}{H} \left( \frac{\partial p_N}{\partial p_L} \right)_{T,H} .
\label{eq:du_dH}
\end{eqnarray}

\begin{figure}
\centering
(a)\includegraphics*[width=0.6\linewidth]{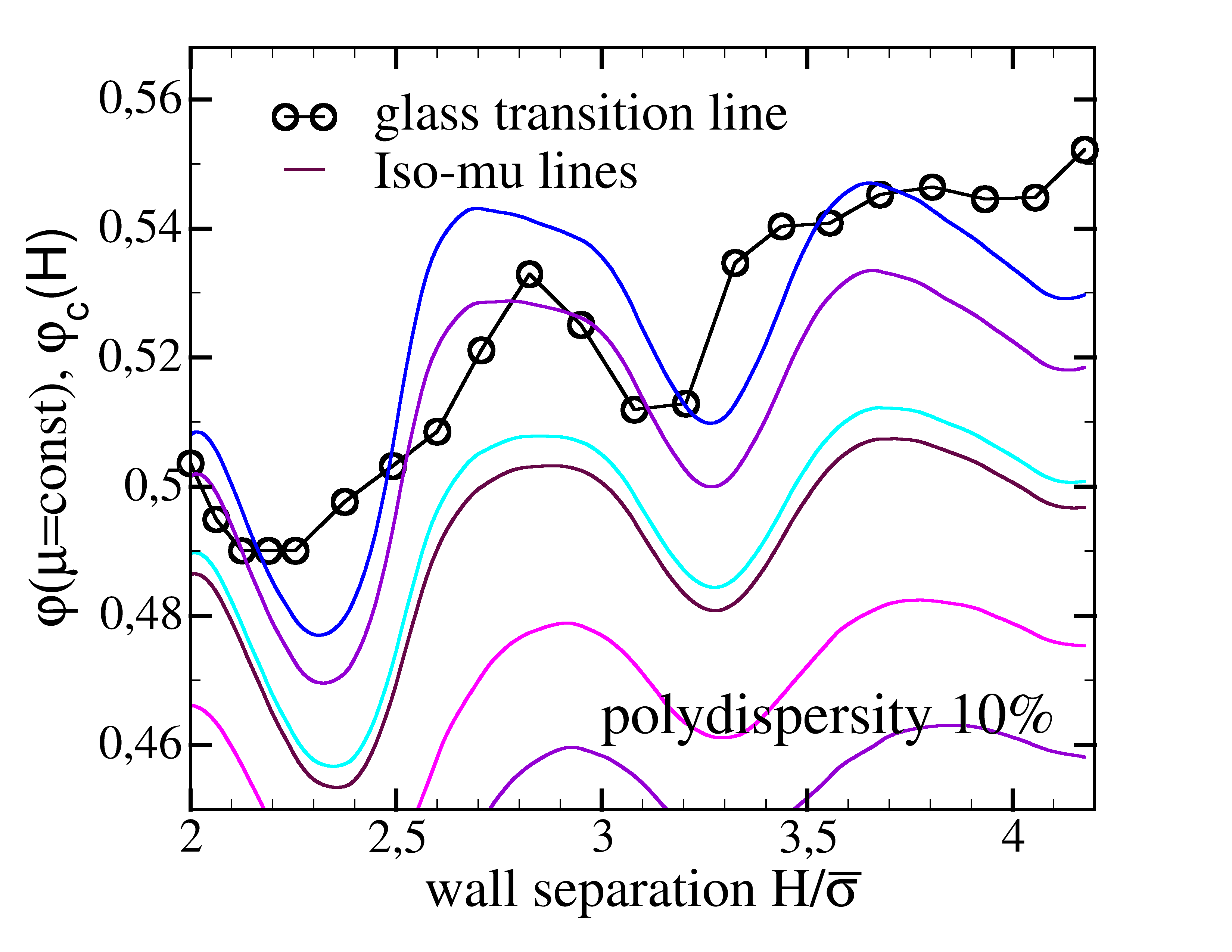}
(b)\includegraphics*[width=0.6\linewidth]{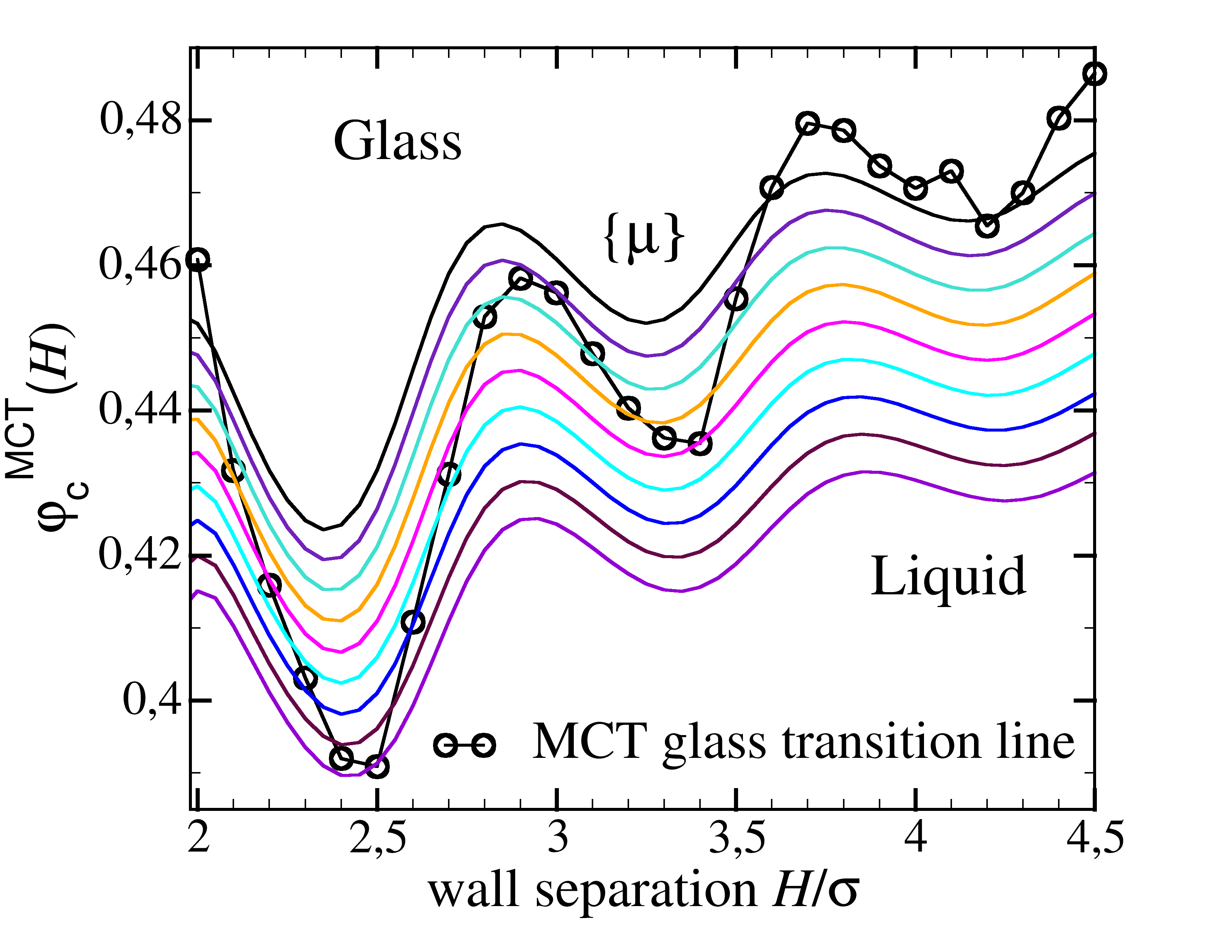}
\caption{(colour online).
(a) Results of the thermodynamic mapping for the local packing fraction at constant chemical potential (`iso-$\mu$ lines'). 
Simulation data for the planar slit are converted to the case of a wedge-shaped channel for a polydispersity of 10\%. Panel (b) shows the results obtained from the fundamental measure theory for the same type of situation.
Images from~\cite{Mandal2014} (reproduced with permission).
}
\label{fig:isomu_m8}
\end{figure}

Using Eq.~(\ref{eq:du_dH}) allows to obtain the dependence of the packing fraction, $\varphi$, on $H$ at constant chemical potential (iso-$\mu$ lines) from the simulated normal and lateral pressures. The thus obtained results are depicted in Fig.~\ref{fig:isomu_m8}a. These data are equivalent to the experimentally more accessible situation of a wedge-shaped geometry, which has been used in a similar context already~\cite{Neser1997,Nugent2007}. Increasing $H$  in  Fig.~\ref{fig:isomu_m8}a corresponds to moving away from the apex in a corresponding wedge-shaped channel.

The plot also contains the simulation results on the critical packing fraction, $\varphic(H)$. Moving along one of the iso-$\mu$ lines leads to multiple crossing points with this glass transition line. For a wedge-shaped channel, this means that one would hit multiple glass-liquid and glass-liquid transitions as one moves along the wedge at a properly chosen chemical potential. In other words, multiple glassy and liquid-like zones may 'co-exists' along the wedge. This expectation is corroborated by Fig.~\ref{fig:isomu_m8}b, where DFT results on iso-$\mu$ lines are shown together with the glass-transition line predicted by MCT, underlying the existence of multiple glass-liquid-glass transitions.

\begin{figure}
\centering
(a)\includegraphics*[width=0.6\linewidth]{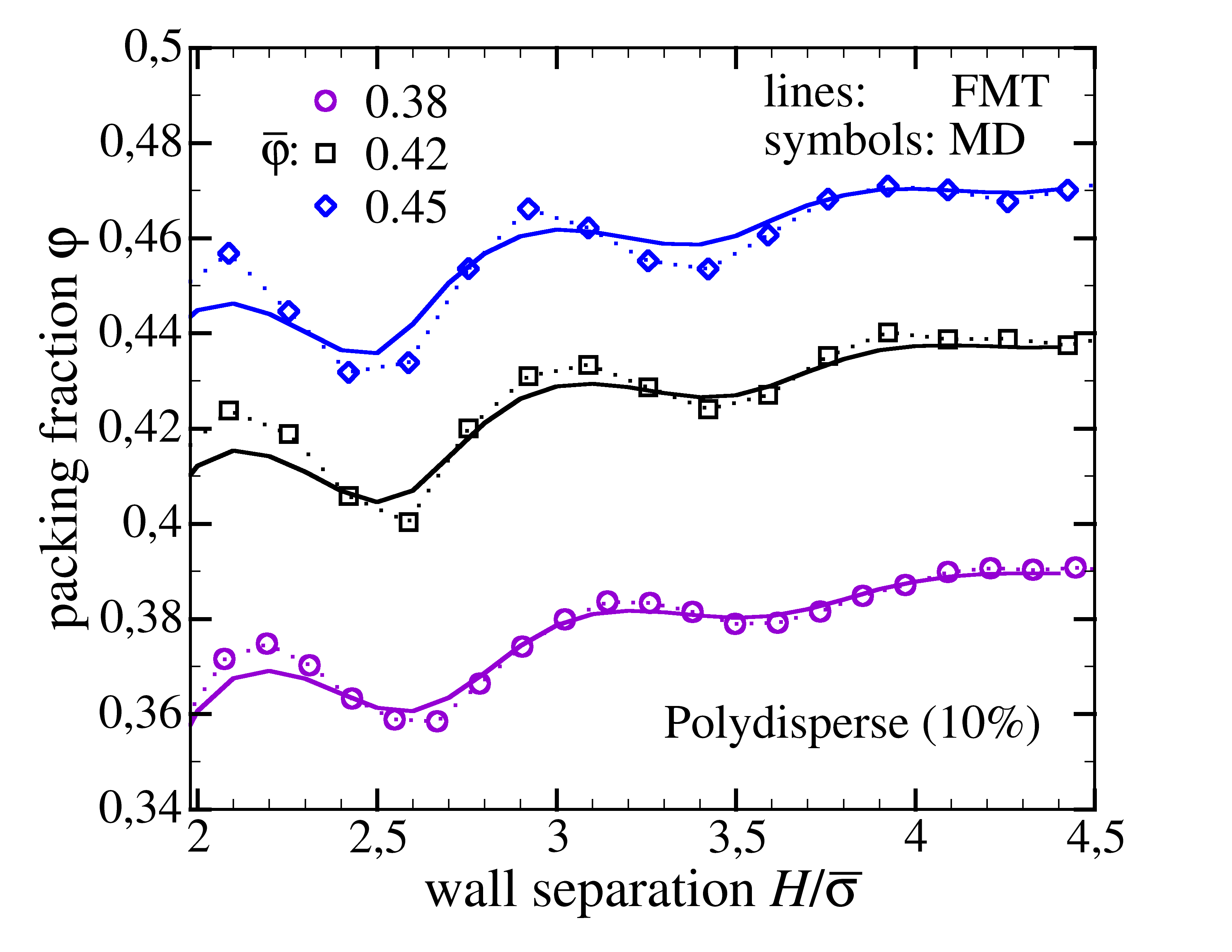}
(b)\includegraphics*[width=0.6\linewidth]{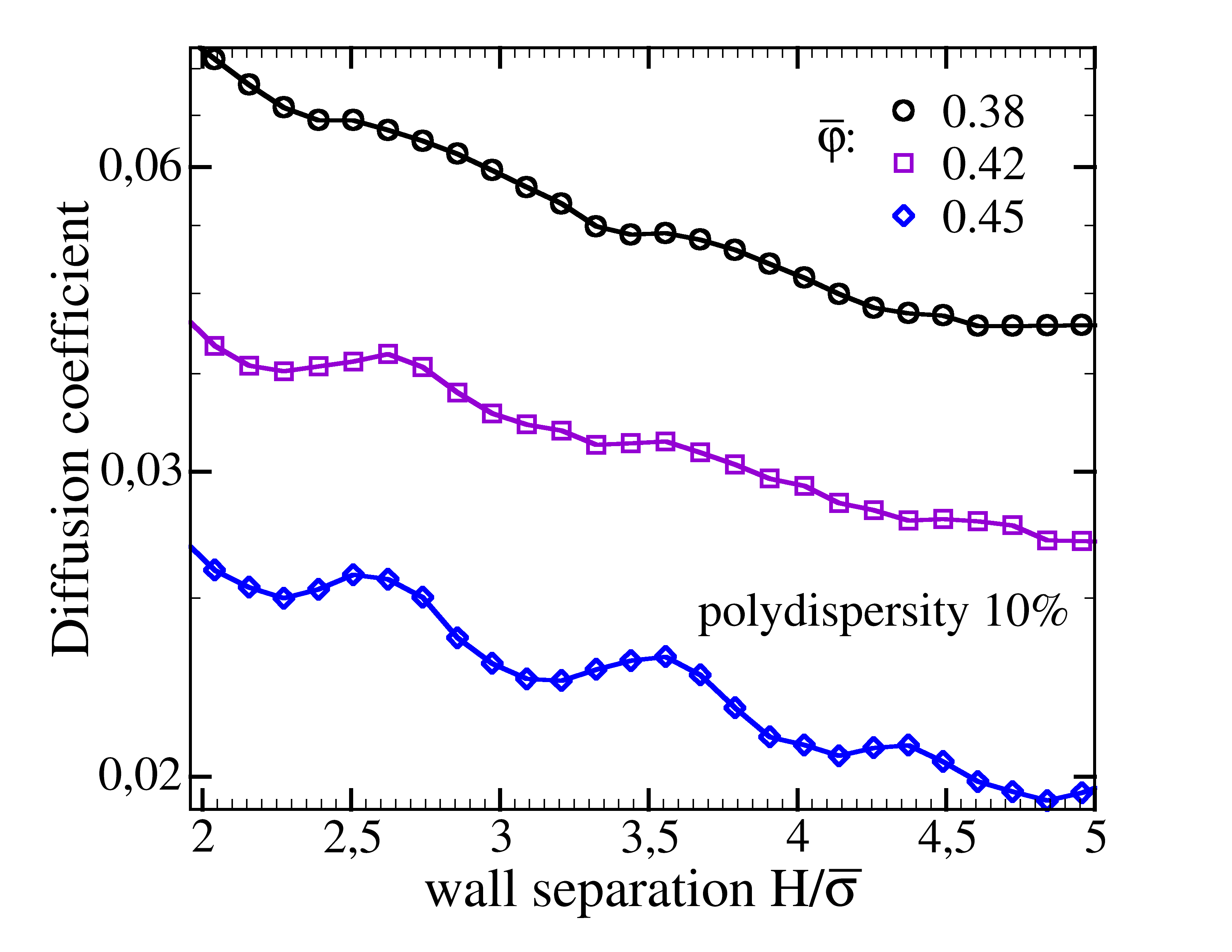}
\caption{(colour online).
Spatial variations of (a) the packing fraction and (b) the diffusion coefficient of a polydisperse (10\%) HS fluid in a wedge-shaped channel (Fig.~\ref{fig:wedge_m7}). \revision{Note that the data are evaluated within a wedge at various distances, $x$, from the apex but are plotted versus the equivalent wall separation at that position, obtained via $H=x\tan(\theta)$}. The tilt angle is $\theta\approx 9^{\circ}$. Symbols correspond to direct simulations and lines gives the FMT results. In simulations, the chemical potential is tuned via a variation of the wedge-averaged packing fraction, $\bar\varphi$, defined as the total volume occupied by all the particles within the wedge divided by the volume of the wedge. 
Images from~\cite{Mandal2014} (reproduced with permission).
}
\label{fig:wedge-MD+FMT}
\end{figure}

Let us close this section by providing a further piece of evidence for the non-monotonic scenario. Figure~\ref{fig:wedge-MD+FMT} displays the local packing fraction and diffusion coefficient, obtained from direct MD simulations of a polydisperse HS fluid in a wedge-shaped channel (see Fig.~\ref{fig:wedge_m7} for a snapshot). Results of FMT calculations are also shown.  Both simulation and theory show oscillations of the local packing fraction and diffusion constant along the wedge and an enhancement of these oscillations upon increasing the total wedge-averaged density or, equivalently, chemical potential. The FMT results are obtained making the assumption that, at a given distance $x$ from the apex, a portion $dx$ of the wedge can be approximated as being confined by two (upper and lower) parallel plates. Due to the finite tilt angle of $\theta\approx 9^\circ$, this assumption is not perfectly valid and may be the cause of the slight differences between simulation and theory.

%%%%%%%%%%%%%%%%%%%%%%%%%%%%%%%%%%%%%%%%%%%%%%%%%%%%%%%%%%%%%%%%%%%%%%%%%
%%%%%%%%%%%%%%%%%%%%%%%%%%%%%%%%%%%%%%%%%%%%%%%%%%%%%%%%%%%%%%%%%%%%%%%%%
%%%%%%%%%%%%%%%%%%%%%%%%%%%%%%%%%%%%%%%%%%%%%%%%%%%%%%%%%%%%%%%%%%%%%%%%%
%%%%%%%%%%%%%%%%%%%%%%%%%%%%%%%%%%%%%%%%%%%%%%%%%%%%%%%%%%%%%%%%%%%%%%%%%
\section{Summary and outlook}
\label{sec:summary}

Starting from early studies in the 1990s~\cite{MansfieldEtal:Macro1991,KeddieEtal:EPL1994,JacksonMcKenna1996,KajiyamaEtal:Polymer1998}, it has now become well established that geometric confinement may have strong influence on the glass-transition phenomenon. 
A non-negligible fraction of investigations in this field were motivated by the desire to provide at least an indirect evidence for the existence of a growing length scale upon approaching the glass transition~\cite{Adam1965}. In the past ten years, however, the focus has been gradually shifted 
towards a new type of confinement effects, which arise as the scale of confinement becomes comparable to the particle scale. 
Restricting, for simplicity, the discussion to a planar slit made of two planar hard and perfectly smooth walls, one of the most 
salient observations here is that the single particle dynamics becomes a non-monotonic function of the slit width along lines of constant density.

A number of authors have proposed arguments to rationalize this intriguing observation. Among these models, we address in this review ideas based on an extension of the excess entropy approach~\cite{Rosenfeld1977,Rosenfeld1999,Dzugutov1996} to confined~\cite{Mittal2006,Mittal2007,Mittal2007b,Mittal2008,Goel2008,Goel2009} and periodic inhomogeneous~\cite{Bollinger2014} fluids. 
One of the features, making excess free energy based arguments very attractive, is the simplicity of the underlying 
physical principles. Central to all variants of the excess entropy based approaches to the relaxation dynamics is the 
assumption that a transition from a given starting state to a new one (realized, e.g., either by a single particle displacement or a collective rearrangement event) is proportional to the number of available non-forbidden states, hence the link to entropy. The usefulness of 
these ideas is highlighted in a number of recent works where a one-to-one correspondence between the excess entropy and 
single particle diffusion has been established both in  bulk and for confined fluids at low and intermediate densities.

The research along this direction has been quite rapid in the past years but there is still room for new fundamental work. An example 
is the linear proportionality between local diffusion and local packing fraction in a planar slit, which was shown to provide a 
reasonable approximation to the simulated data at low and intermediate densities~\cite{Mittal2008} but have recently 
been found to be reversed at higher densities~\cite{Bollinger2015}. This observation underlines the need for a 
deeper analysis of the problem. In particular, the empirical argument used in deriving the 
linearity relation between the local density and the local single particle dynamics, which assumes that the single particle dynamics 
is proportional to the Widom's insertion probability, shall be revisited to adequately account for the dynamics at high densities. 

A further interesting issue here regards the anisotropy of diffusion in the proximity of 
a confining wall~\cite{Mittal2006,Mittal2008} shown in Fig.~\ref{fig:Mittal}. One often observes 
that the dynamics in the direction parallel to the wall is faster than that in the perpendicular direction but it remains 
still unclear how this relates to the number of available states and the closely related excess entropy. In this context, it might also 
be interesting to use the concept of the free volume and consider its directional dependence close to a planar boundary.

Another path to rationalize confinement effects on the dynamics of glass forming liquids is provided by the mode-coupling theory of
 the glass transition, generalized in the recent years to 
include effects of a geometric confinement~\cite{Lang2010,Lang2012,Lang2013,Franosch:2014}. The mode-coupling theory has the  
appealing feature that it deals, from the outset, with the limit of highly dense liquids and thus lends
 itself naturally to study confinement effects on the glass-transition phenomenon. 
An important part of this review is, 
therefore, devoted to this theoretical development and a test of its predictions via computer simulations. A first prediction of the theory is that confinement has a dramatic impact on the dynamics of structural relaxation, thereby leading to a shift of the glass-transition line. Another
 prediction of the MCT is that, when increasing the confinement along lines of constant density, this effect 
is non-monotonic and can give rise to a (multiple) reentrant glass-transition scenario. These predictions are found 
to be in agreement with the results obtained from MD simulations.

We provide an interpretation of these predictions and observations in terms of a competition between confinement-induced layering and local particle packing structure, which may lead, depending on the ratio of the slit width to particle diametre, to subtle incommensurability effects. This effect is most prominent in the case of monodisperse HS fluids but becomes drastically enhanced near the glass-transition line and can then be fairly strong also for moderately polydisperse systems. As a task for future work, one expects that these competing trends should manifest themselves also in the glass form factors as function of wavenumber and mode index. Similarly, a study of the shape of the 
structural relaxation dynamics should contain valuable information on how commensurability controls the glass transition.

A very promising topic for future research in this field is the study of the dynamics and structure of dense glass forming 
fluids in a wedge-shaped channel (as it occurs, e.g., in a cone-plate chamber). Simulations and density functional calculations at intermediate densities reveal density oscillations as function of the distance from the apex. Thermodynamic mapping of the planar slit-data to the wedge-geometry even suggest the coexistence of alternating liquid-glass-liquid domains along the wedge. This setup is particularly interesting as it simplifies experimental investigations considerably. There is no need to keep the density constant. Rather, by tuning the chemical potential (via, e.g., a variation of the external pressure), it is possible to establish the coexistence of alternating liquid-glass regions.

The study of densely packed glass forming fluid mixtures in narrow slits also bears great potential for future work. Binary fluid mixtures exhibit interesting mixing effects upon variation of the composition and the size ratio of the constituents~\cite{Gotze2003,Voigtmann2011}. Confinement introduces a new length scale into this problem and can strongly shift, and even qualitatively modify, these mixing effects. Tuning the confinement effects may thus provide a genuinely new material design strategy in order to, depending on the ones requirements, enhance or suppress oscillations in the non-equilibrium state diagram of fluid mixtures.

\section{Acknowledgments}

This work has benefited from interaction with a number of colleagues. We would like to acknowledge valuable discussions with (in alphabetical order) 
Wolfgang G\"otze, Simon Lang, Suvendu Mandal, Martin Oettel, Dierk Raabe, Rolf Schilling, Nima Siboni, and Matthias Sperl. 
We are also grateful to all those colleagues who provided us with valuable results/plots. In particular, we are indebted 
to Thomas Eckert and Eckhard Bartsch for Figure 1a, Thomas Markland, Kunimasa Miyazaki and David Reichman (Figure 1b), Jaeetain Mittal, 
Gaurav Goel and Tom Truskett (Figures 3,6,7-9), Dillip Satapathy and Johannes Friso van der Veen (Figure 4), and  Jos\'e Rafael Bordin and Marcia Barbosa (Figure 5). Funding by the Deutsche Forschungsgemeinschaft (DFG) via the research unit  FOR1394 'Nonlinear Response to probe Vitrification' and the Max-Planck Society are also acknowledged. FV thanks, on behalf of the ICAMS the state NRW, the European Union and all its industrial sponsors.

\section*{References}

\bibliographystyle{prsty_with_title}
\bibliography{references-2016-01-06}

\begin{thebibliography}{100}

\bibitem{Anderson1995}
In a short note to Science (volume 267, page 1615, (1995)), Philip W.\ Anderson
  wrote "The deepest and most interesting unsolved problem in solid state
  theory is probably the theory of the nature of the glass and the glass
  transition".

\bibitem{Stillinger1995}
F.~H. Stillinger, A Topographic View of Supercooled Liquids and Glass
  Formation, Science {\bf 267},  1935  (1995).

\bibitem{Angell1995}
C.~A. Angell, Formation of Glasses from Liquids and Biopolymers, Science {\bf
  267},  1924  (1995).

\bibitem{Binder1999b}
K. Binder, Understanding the glass transition and the amorphous state of
  matter: can computer simulation solve the challenge?, Computer Physics
  Communications {\bf 121-122},  168  (1999), proceedings of the Europhysics
  Conference on Computational Physics \{CCP\} 1998.

\bibitem{Pham2002}
K.~N. Pham, A.~M. Puertas, J. Bergenholtz, S.~U. Egelhaaf, A. Moussaid, P.~N.
  Pusey, A.~B. Schofield, M.~E. Cates, M. Fuchs, and W.~C.~K. Poon, Multiple
  glassy states in a simple model system, Science {\bf 296},  104  (2002).

\bibitem{Tant1999}
{\em Structure and Properties of Glassy Polymers}, Vol.~710 of {\em ACS
  Symposium Series} (Oxford University Press, Oxford, 1999).

\bibitem{Loeffler2003}
J.~F. L\"offler, Bulk metallic glasses, Intermetallics {\bf 11},  529  (2003).

\bibitem{Telford2004}
M. Telford, The case for bulk metallic glass, Materials Today {\bf 7},  36
  (2004).

\bibitem{LeBourhis2007}
E.~L. Bourhis, {\em Glass: Mechanics and Technology} (Wiley-VCH, Heidelberg,
  2007).

\bibitem{Greer2009}
A.~L. Greer, Metallic glasses…on the threshold, Materials Today {\bf 12},  14
    (2009).

\bibitem{Sherrington1975}
D. Sherrington and S. Kirkpatrick, Solvable Model of a Spin-Glass, Phys. Rev.
  Lett. {\bf 35},  1792  (1975).

\bibitem{Binder1986}
K. Binder and A.~P. Young, Spin glasses: Experimental facts, theoretical
  concepts, and open questions, Rev. Mod. Phys. {\bf 58},  801  (1986).

\bibitem{Mezard1987}
M. Mezard, G. Parisi, and M.~A. Virasoro, {\em Spin glass theory and beyond}
  (World Scientific, Singapore, 1987).

\bibitem{Nisoli2013}
C. Nisoli, R. Moessner, and P. Schiffer, \textit{Colloquium} : Artificial spin
  ice: Designing and imaging magnetic frustration, Rev. Mod. Phys. {\bf 85},
  1473  (2013).

\bibitem{Pusey1991}
P.~N. Pusey,  in {\em Liquids, Freezing and the Glass Transition}, edited by
  D.~L. J-P~Hansen and J. Zinn-Justin (Elsevier, ADDRESS, 1991), Chap.~10, p.\
  7769.

\bibitem{Dawson2000}
K. Dawson, G. Foffi, M. Fuchs, W. G\"otze, F. Sciortino, M. Sperl, P.
  Tartaglia, T. Voigtmann, and E. Zaccarelli, Higher-order glass-transition
  singularities in colloidal systems with attractive interactions, Phys. Rev. E
  {\bf 63},  011401  (2000).

\bibitem{Eckert2002}
T. Eckert and E. Bartsch, Re-entrant Glass Transition in a Colloid-Polymer
  Mixture with Depletion Attractions, Phys. Rev. Lett. {\bf 89},  125701
  (2002).

\bibitem{Poon_TOP2002}
W.~C.~K. Poon, The physics of a model colloid-polymer mixture, J. Phys.:
  Condens. Matter {\bf 14},  R859  (2002).

\bibitem{Poon2003}
W.~C.~K. Poon, K.~N. Pham, S.~U. Egelhaaf, and P.~N. Pusey, 'Unsticking' a
  colloidal glass, and sticking it again, Journal of Physics: Condensed Matter
  {\bf 15},  S269  (2003).

\bibitem{Pham2004}
K.~N. Pham, S.~U. Egelhaaf, P.~N. Pusey, and W.~C.~K. Poon, Glasses in hard
  spheres with short-range attraction, Phys. Rev. E {\bf 69},  011503  (2004).

\bibitem{Charbonneau2007}
P. Charbonneau and D.~R. Reichman, Phase behavior and far-from-equilibrium
  gelation in charged attractive colloids, Phys. Rev. E {\bf 75},  050401
  (2007).

\bibitem{Zaccarelli2009}
E. Zaccarelli, C. Valeriani, E. Sanz, W.~C.~K. Poon, M.~E. Cates, and P.~N.
  Pusey, Crystallization of Hard-Sphere Glasses, Phys. Rev. Lett. {\bf 103},
  135704  (2009).

\bibitem{Markland2011}
T.~E. Markland, J.~A. Morrone, B.~J. Berne, K. Miyazaki, E. Rabani, and D.~R.
  Reichman, Quantum fluctuations can promote or inhibit glass formation, Nature
  Physics {\bf 7},  134  (2011).

\bibitem{Foffi2003}
G. Foffi, W. G\"otze, F. Sciortino, P. Tartaglia, and T. Voigtmann, Mixing
  Effects for the Structural Relaxation in Binary Hard-Sphere Liquids, Phys.
  Rev. Lett. {\bf 91},  085701  (2003).

\bibitem{Zaccarelli2004}
E. Zaccarelli, H. L\"owen, P.~P.~F. Wessels, F. Sciortino, P. Tartaglia, and
  C.~N. Likos, Is There a Reentrant Glass in Binary Mixtures?, Phys. Rev. Lett.
  {\bf 92},  225703  (2004).

\bibitem{Krakoviack2005}
V. Krakoviack, Liquid-Glass Transition of a Fluid Confined in a Disordered
  Porous Matrix: A Mode-Coupling Theory, Phys. Rev. Lett. {\bf 94},  065703
  (2005).

\bibitem{Krakoviack2007}
V. Krakoviack, Mode-coupling theory for the slow collective dynamics of fluids
  adsorbed in disordered porous media, Phys. Rev. E {\bf 75},  031503  (2007).

\bibitem{Krakoviack2011}
V. Krakoviack, Mode-coupling theory predictions for the dynamical transitions
  of partly pinned fluid systems, Phys. Rev. E {\bf 84},  050501  (2011).

\bibitem{Kurzidim2009}
J. Kurzidim, D. Coslovich, and G. Kahl, Single-Particle and Collective Slow
  Dynamics of Colloids in Porous Confinement, Phys. Rev. Lett. {\bf 103},
  138303  (2009).

\bibitem{Kim2009}
K. Kim, K. Miyazaki, and S. Saito, Slow dynamics in random media: Crossover
  from glass to localization transition, EPL {\bf 88},  36002  (2009).

\bibitem{Kob1994}
W. Kob and H.~C. Andersen, , Phys. Rev. Lett. {\bf 73},  1376  (1994).

\bibitem{Boninsegni2006}
M. Boninsegni, N. Prokof'ev, and B. Svistunov, Superglass Phase of
  $^{4}\mathrm{He}$, Phys. Rev. Lett. {\bf 96},  105301  (2006).

\bibitem{Biroli2008}
G. Biroli, C. Chamon, and F. Zamponi, Theory of the superglass phase, Phys.
  Rev. B {\bf 78},  224306  (2008).

\bibitem{Hunt2009}
B. Hunt, E. Pratt, V. Gadagkar, M. Yamashita, A.~V. Balatsky, and J.~C. Davis,
  Evidence for a Superglass State in Solid $^4$He, Science {\bf 324},  632
  (2009).

\bibitem{Lang2010}
S. Lang, V. Bo\ifmmode~\mbox{\c{t}}\else \c{t}\fi{}an, M. Oettel, D. Hajnal, T.
  Franosch, and R. Schilling, Glass Transition in Confined Geometry, Phys. Rev.
  Lett. {\bf 105},  125701  (2010).

\bibitem{Mandal2014}
S. Mandal, S. Lang, M. Gross, M. Oettel, D. Raabe, T. Franosch, and F. Varnik,
  Multiple reentrant glass transitions in confined hard-sphere glasses, Nature
  Communications {\bf 5},  4435  (2014).

\bibitem{Nugent2007}
C.~R. Nugent, K.~V. Edmond, H.~N. Patel, and E.~R. Weeks, Colloidal Glass
  Transition Observed in Confinement, Phys. Rev. Lett. {\bf 99},  025702
  (2007).

\bibitem{Lang2012}
S. Lang, R. Schilling, V. Krakoviack, and T. Franosch, Mode-coupling theory of
  the glass transition for confined fluids, Phys. Rev. E {\bf 86},  021502
  (2012).

\bibitem{Lang2013}
S. Lang, R. Schilling, and T. Franosch, Mode-coupling theory for multiple decay
  channels, J. Stat. Mech.: Theor. and Exp. {\bf 2013},  P12007  (2013).

\bibitem{Mittal2008}
J. Mittal, T.~M. Truskett, J.~R. Errington, and G. Hummer, Layering and
  Position-Dependent Diffusive Dynamics of Confined Fluids, Phys. Rev. Lett.
  {\bf 100},  145901  (2008).

\bibitem{Rosenfeld1977}
Y. Rosenfeld, Relation between the transport coefficients and the internal
  entropy of simple systems, Phys. Rev. A {\bf 15},  2545  (1977).

\bibitem{Rosenfeld1999}
Y. Rosenfeld, A quasi-universal scaling law for atomic transport in simple
  fluids, Journal of Physics: Condensed Matter {\bf 11},  5415  (1999).

\bibitem{Dzugutov1996}
M. Dzugutov, A universal scaling law for atomic diffusion in condensed matter,
  Nature {\bf 381},  137  (1996).

\bibitem{Widom1963}
B. Widom, Some Topics in the Theory of Fluids, J. Chem. Phys. {\bf 39},  2808
  (1963).

\bibitem{Bollinger2015}
J.~A. Bollinger, A. Jain, J. Carmer, and T.~M. Truskett, Communication: Local
  structure-mobility relationships of confined fluids reverse upon
  supercooling, The Journal of Chemical Physics {\bf 142},  161102  (2015).

\bibitem{Franosch:2014}
T. Franosch, Long-time limit of correlation functions, Journal of Physics A:
  Mathematical and Theoretical {\bf 47},  325004  (2014).

\bibitem{Varnik2002e}
F. Varnik, J. Baschnagel, and K. Binder, Static and dynamic properties of
  supercooled thin polymer films, Eur. Phys. J. E {\bf 8},  175  (2002).

\bibitem{Asakura1954}
S. Asakura and F. Oosawa, On Interaction between Two Bodies Immersed in a
  Solution of Macromolecules, The Journal of Chemical Physics {\bf 22},  1255
  (1954).

\bibitem{Tuinier2000}
R. Tuinier, G.~A. Vliegenthart, and H.~N.~W. Lekkerkerker, Depletion
  interaction between spheres immersed in a solution of ideal polymer chains,
  The Journal of Chemical Physics {\bf 113},  10768  (2000).

\bibitem{Fuchs2000}
M. Fuchs and K.~S. Schweizer, Structure and thermodynamics of colloid-polymer
  mixtures: A macromolecular approach, Europhys. Lett. {\bf 51},  621  (2000).

\bibitem{Fuchs2001}
M. Fuchs and K.~S. Schweizer, Macromolecular theory of solvation and structure
  in mixtures of colloids and polymers, Phys. Rev. E {\bf 64},  021514  (2001).

\bibitem{Fischer1992}
E.~W. Fischer, E. Donth, and W. Steffen, Temperature dependence of
  characteristic length for glass transition, Phys. Rev. Lett. {\bf 68},  2344
  (1992).

\bibitem{Donth2001}
E. Donth, {\em The Glass Transition} (Springer, Berlin--Heidelberg, 2001).

\bibitem{Biroli2006}
G. Biroli, J.-P. Bouchaud, K. Miyazaki, and D.~R. Reichman, Inhomogeneous
  Mode-Coupling Theory and Growing Dynamic Length in Supercooled Liquids, Phys.
  Rev. Lett. {\bf 97},  195701  (2006).

\bibitem{Adam1965}
G. Adam and J.~H. Gibbs, On the Temperature Dependence of Cooperative
  Relaxation Properties in Glass Forming Liquids, J. Chem. Phys. {\bf 43},  139
   (1965).

\bibitem{Fryer2001}
D.~S. Fryer, R.~D. Peters, E.~J. Kim, J.~E. Tomaszewski, J.~J. de~Pablo, P.~F.
  Nealey, C.~C. White, and W.~L. Wu, Dependence of the glass transition
  temperature of polymer films on interfacial energy and thickness,
  Macromolecules {\bf 34},  5627  (2001).

\bibitem{MorineauEtal:JCP2003}
D. Morineau and C. Alba-Simionesco, Liquids in confined geometry: How to
  connect changes in the structure factor to modifications of local order, J.
  Chem. Phys. {\bf 118},  9389  (2003).

\bibitem{AlbaEtal:EPJE2003}
C. Alba-Simionesco, G. Dosseh, E. Dumont, B. Frick, B. Geil, D. Morineau, V.
  Teboul, and Y. Xia, Confinement of molecular liquids: Consequences on
  thermodynamic, static and dynamical properties of benzene and toluene, Eur.
  Phys. J. E {\bf 12},  19  (2003).

\bibitem{JacksonMcKenna1996}
C.~L. Jackson and G.~B. McKenna, Vitrification and Crystallization of Organic
  Liquids Confined to Nanoscale Pores, Chem. Mater. {\bf 8},  2128  (1996).

\bibitem{WendtRichert1999}
H. Wendt and R. Richert, Cooperativity and heterogeneity of the dynamics in
  nano-confined liquids, J. Phys.: Condens. Matter {\bf 11},  A199  (1999).

\bibitem{PissisEtal:JPCM1998}
P. Pissis, A. Kyritsis, D. Daoukaki, G. Barut, R. Pelster, and G. Nimtz,
  Dielectric studies of glass transition in confined propylene glycol, J.
  Phys.: Condens. Matter {\bf 10},  6205  (1998).

\bibitem{ArndtEtal:PRL1997}
M. Arndt, R. Stannarius, H. Groothues, E. Hempel, and F. Kremer, Length Scale
  of Cooperativity in the Dynamic Glass Transition, Phys. Rev. Lett. {\bf 79},
  2077  (1997).

\bibitem{DemirelGranick:JCP2001}
A. Levent~Demirel and S. Granick, Origins of solidification when a simple
  molecular fluid is confined between two plates, J. Chem. Phys. {\bf 115},
  1498  (2001).

\bibitem{KeddieEtal:EPL1994}
J.~L. Keddie, R.~A.~L. Jones, and R.~A. Cory, Size-Dependent Depression of the
  Glass Transition Temperature in Polymer Films, Europhys. Lett. {\bf 27},  59
  (1994).

\bibitem{Jones:Current1999}
R.~A.~L. Jones, The dynamics of thin polymer films, Curr. Opinion Coll. \&
  Interf. Sci. {\bf 4},  153  (1999).

\bibitem{ForrestDalnoki:AdvColSci2001}
J.~A. Forrest and K. Dalnoki-Veress, The glass transition in thin polymer
  films, Adv. Coll. Interf. Sci. {\bf 94},  167  (2001).

\bibitem{Forrest:EPJE2002}
J.~A. Forrest, A decade of dynamics in thin films of polystyrene: Where are we
  now?, Eur. Phys. J. E {\bf 8},  261  (2002).

\bibitem{Herminghaus:EPJE2002}
S. Herminghaus, Polymer thin films and surfaces: Possible effects of capillary
  waves, Eur. Phys. J. E {\bf 8},  237  (2002).

\bibitem{HerminghausEtal:PRL2004}
S. Herminghaus, R. Seemann, and K. Landfester, Polymer Surface Melting Mediated
  by Capillary Waves, Phys. Rev. Lett. {\bf 93},  017801  (2004).

\bibitem{Johannsmann:EPJE2002}
D. Johannsmann, The glass transition and contact mechanical experiments on
  polymer surfaces, Eur. Phys. J. E {\bf 8},  257  (2002).

\bibitem{Reiter:EPJE2002}
G. Reiter, Are changes in morphology clear indicators for the glass transition
  in thin polymer films? Tentative ideas, Eur. Phys. J. E {\bf 8},  251
  (2002).

\bibitem{KajiyamaEtal:Polymer1998}
T. Kajiyama, K. Tanaka, and A. Takahara, Study of the surface glass transition
  behaviour of amorphous polymer film by scanning-force microscopy and surface
  spectroscopy, Polymer {\bf 39},  4665  (1998).

\bibitem{FukaoEtal:PRE2000}
K. Fukao and Y. Miyamoto, Glass transitions and dynamics in thin polymer films:
  Dielectric relaxation of thin films of polystyrene, Phys. Rev. E {\bf 61},
  1743  (2000).

\bibitem{SchoenhalsStauga:JCP1998}
A. Sch\"onhals and A. Stauga, Broadband dielectric study of anomalous diffusion
  in a poly(propylene glycol) melt confined to nanopores, J. Chem. Phys. {\bf
  108},  5130  (1998).

\bibitem{EllisonTorkelson:NatureMaterials2003}
C.~J. Ellison and J.~M. Torkelson, The distribution of glass-transition
  temperatures in nanoscopically confined glass formers, Nature Mat. {\bf 2},
  695  (2003).

\bibitem{SinghEtal:SolidFilms2004}
L. Singh, P.~J. Ludovice, and C.~L. Henderson, Influence of molecular weight
  and film thickness on the glass transition temperature and coefficient of
  thermal expansion of supported ultrathin polymer films, Thin Solid Films {\bf
  449},  231  (2004).

\bibitem{Pye2011}
J.~E. Pye and C.~B. Roth, Two Simultaneous Mechanisms Causing Glass Transition
  Temperature Reductions in High Molecular Weight Freestanding Polymer Films as
  Measured by Transmission Ellipsometry, Phys. Rev. Lett. {\bf 107},  235701
  (2011).

\bibitem{Yin2013}
H. Yin, D. Cangialosi, and A. Sch\"onhals, Glass transition and segmental
  dynamics in thin supported polystyrene films: The role of molecular weight
  and annealing, Thermochimica Acta {\bf 566},  186   (2013).

\bibitem{Yin2013b}
H. Yin and A. Sch\"onhals, Calorimetric glass transition of ultrathin
  poly(vinyl methyl ether) films, Polymer {\bf 54},  2067   (2013).

\bibitem{VarnikEtal:PRE2002}
F. Varnik, J. Baschnagel, and K. Binder, Reduction of the glass transition
  temperature in polymer films: A molecular-dynamics study, Phys. Rev. E {\bf
  65},  021507  (2002).

\bibitem{VarnikEtal:EPJE2002}
F. Varnik, J. Baschnagel, and K. Binder, Static and dynamic properties of
  supercooled thin polymer films, Eur. Phys. J. E {\bf 8},  175  (2002).

\bibitem{VarnikEtal:EPJE2003}
F. Varnik, J. Baschnagel, K. Binder, and M. Mareschal, Confinement effects on
  the slow dynamics of a supercooled polymer melt: Rouse modes and the
  incoherent scattering function, Eur. Phys. J. E {\bf 12},  167  (2003).

\bibitem{VarnikBinder:JCP2002}
F. Varnik and K. Binder, Shear viscosity of a supercooled polymer melt via
  nonequilibrium molecular dynamics simulations, J. Chem. Phys. {\bf 117},
  6336  (2002).

\bibitem{XuMattice:Macro2003}
G. Xu and W.~L. Mattice, Monte Carlo simulation on the glass transition of
  free-standing atactic polypropylene thin films on a high coordination
  lattice, J. Chem. Phys. {\bf 118},  5241  (2003).

\bibitem{StarrEtal:Macro2002}
F.~W. Starr, T.~B. Schr{\o}der, and S.~C. Glotzer, Molecular Dynamics
  Simulation of a Polymer Melt with a Nanoscopic Particle, Macromolecules {\bf
  35},  4481  (2002).

\bibitem{YoshimotoEtal:JCP2005}
K. Yoshimoto, T.~S. Jain, P.~F. Nealey, and J.~J. de~Pablo, Local dynamic
  mechanical properties in model free-standing polymer thin films, J. Chem.
  Phys. {\bf 122},  144712  (2005).

\bibitem{JaindePablo:PRL2004}
T.~S. Jain and J.~J. de~Pablo, Investigation of Transition States in Bulk and
  Freestanding Film Polymer Glasses, Phys. Rev. Lett. {\bf 92},  155505
  (2004).

\bibitem{Torres2000}
J.~A. Torres, P.~F. Nealey, and J.~J. de~Pablo, Molecular Simulation of
  Ultra-Thin Polymeric Films Near the Glass Transition, Phys. Rev. Lett. {\bf
  85},  3221  (2000).

\bibitem{MansfieldEtal:Macro1991}
K.~F. Mansfield and D.~N. Theodorou, Molecular Dynamics Simulation of a Glassy
  Polymer Surface, Macromolecules {\bf 24},  6283  (1991).

\bibitem{ManiasEtal:ColSurf2001}
E. Manias, V. Kuppa, D.-K. Yang, and D.~B. Zax, Relaxation of polymers in 2 nm
  slit-pores: confinement induced segmental dynamics and suppression of the
  glass transition, Colloids Surf. A {\bf 187--188},  509  (2001).

\bibitem{BaljonEtal:Macro2005}
A.~R.~C. Baljon, M.~H.~M. Van~Weert, R. Barber~DeGraaff, and R. Khare, Glass
  Transition Behavior of Polymer Films of Nanoscopic Dimensions, Macromolecules
  {\bf 38},  2391  (2005).

\bibitem{Roder:JCP2001}
A. Roder, W. Kob, and K. Binder, Structure and dynamics of amorphous silica
  surfaces, J. Chem. Phys. {\bf 114},  7602  (2001).

\bibitem{Scheidler2000}
P. Scheidler, W. Kob, and K. Binder, Static and dynamical properties of a
  supercooled liquid confined in a pore, J. Phys. IV {\bf 10},  33  (2000).

\bibitem{Scheidler2002}
P. Scheidler, W. Kob, and K. Binder, Cooperative motion and growing length
  scales in supercooled confined liquids, Europhys. Lett. {\bf 59},  701
  (2002).

\bibitem{Scheidler2004}
P. Scheidler, W. Kob, and K. Binder, The Relaxation Dynamics of a Supercooled
  Liquid Confined by Rough Walls, J. Phys. Chem. B {\bf 108},  6673  (2004).

\bibitem{GalloEtal:EPL2002}
P. Gallo, R. Pellarin, and M. Rovere, Mode Coupling relaxation scenario in a
  confined glass former, Europhys. Lett. {\bf 57},  212  (2002).

\bibitem{FehrLoewen:PRE1995}
T. Fehr and H. L\"owen, Glass transition in confined geometry, Phys. Rev. E
  {\bf 52},  4016  (1995).

\bibitem{BoddekerTeichler:PRE1999}
B. B\"oddeker and H. Teichler, Dynamics near free surfaces of molecular
  dynamics simulated $\mathrm{{Ni}_{0.5}{Zr}_{0.5}}$ metallic glass films,
  Phys. Rev. E {\bf 59},  1948  (1999).

\bibitem{Xia2013}
W. Xia, S. Mishra, and S. Keten, Substrate vs. free surface: Competing effects
  on the glass transition of polymer thin films, Polymer {\bf 54},  5942
  (2013).

\bibitem{Krishnan2012}
S.~H. Krishnan and K.~G. Ayappa, Glassy dynamics in a confined monatomic fluid,
  Phys. Rev. E {\bf 86},  011504  (2012).

\bibitem{Ingebrigtsen2013}
T.~S. Ingebrigtsen, J.~R. Errington, T.~M. Truskett, and J.~C. Dyre, Predicting
  How Nanoconfinement Changes the Relaxation Time of a Supercooled Liquid,
  Phys. Rev. Lett. {\bf 111},  235901  (2013).

\bibitem{Williams2013}
I. Williams, E.~C. O\u{g}uz, P. Bartlett, H. L\"owen, and C.~P. Royall, Direct
  measurement of osmotic pressure via adaptive confinement of quasi hard disc
  colloids, Nat. Commun. {\bf 4},  2555  (2013).

\bibitem{Rodriguez-Fris2014}
J.~A.~R. Fris, M.~A. Frechero, and G.~A. Appignanesi, Relaxation pathway
  confinement in glassy dynamics, The Journal of Chemical Physics {\bf 141},
  114905  (2014).

\bibitem{PGG:EPJE2000}
P.~G. de~Gennes, Glass transitions in thin polymer films, Eur. Phys. J. E {\bf
  2},  201  (2000).

\bibitem{McCoyCurro:JCP2002}
J.~D. McCoy and J.~G. Curro, Conjectures on the glass transition of polymers in
  confined geometries, J. Chem. Phys. {\bf 116},  9154  (2002).

\bibitem{Truskett2003}
T.~M. Truskett and V. Ganesan, Ideal glass transitions in thin films: An energy
  lanscape perspective, J. Chem. Phys. {\bf 119},  1897  (2003).

\bibitem{Mittal2004}
J. Mittal, P. Shah, and T.~M. Truskett, Using Energy Landscapes To Predict the
  Properties of Thin Films, J. Phys. Chem. B {\bf 108},  19769  (2004).

\bibitem{Chow:JPCM2002}
T.~S. Chow, The glass transition of nanoscale polymeric films, J. Phys.:
  Condens. Matter {\bf 14},  L333  (2002).

\bibitem{LongLequeux:EPJE2001}
D. Long and F. Lequeux, Heterogeneous dynamics at the glass transition in van
  der Waals liquids, in the bulk and in thin films, Eur. Phys. J. E {\bf 4},
  371  (2001).

\bibitem{MerabiaEtal:EPJE2004}
S. Merabia, P. Sotta, and D. Long, Heterogeneous nature of the dynamics and
  glass transition in thin polymer films, Eur. Phys. J. E {\bf 15},  189
  (2004).

\bibitem{Mayes:Macro1994}
A.~M. Mayes, Glass Transition of Amorphous Polymer Surfaces, Macromolecules
  {\bf 27},  3114  (1994).

\bibitem{Ngai:EPJE2003}
K.~L. Ngai, The effects of changes of intermolecular coupling on glass
  transition dynamics in polymer thin films and glass-formers confined in
  nanometer pores, Eur.\ Phys.\ J. E {\bf 12},  93  (2003).

\bibitem{Mirigian2014}
S. Mirigian and K.~S. Schweizer, Communication: Slow relaxation, spatial
  mobility gradients, and vitrification in confined films, The Journal of
  Chemical Physics {\bf 141},  161103  (2014).

\bibitem{Scheidler2000b}
P. Scheidler, W. Kob, and K. Binder, The relaxation dynamics of a simple glass
  former confined in a pore, Europhys. Lett. {\bf 52},  277  (2000).

\bibitem{Baschnagel2005}
J. Baschnagel and F. Varnik, Computer simulation of supercooled polymer melts
  in the bulk and in confined geometry, J.Phys.: Condens. Matter {\bf 17},
  R851  (2005).

\bibitem{Varnik2009}
F. Varnik and K. Binder, Multiscale modelling of polymers at interfaces, Int.
  J. Mater. Res. {\bf 100},  1494  (2009).

\bibitem{Mittal2006}
J. Mittal, J.~R. Errington, and T.~M. Truskett, Thermodynamics Predicts How
  Confinement Modifies the Dynamics of the Equilibrium Hard-Sphere Fluid, Phys.
  Rev. Lett. {\bf 96},  177804  (2006).

\bibitem{Satapathy2008}
D.~K. Satapathy, O. Bunk, K. Jefimovs, K. Nyg\aa{}rd, H. Guo, A. Diaz, E.
  Perret, F. Pfeiffer, C. David, G.~H. Wegdam, and J.~F. van~der Veen,
  Colloidal Monolayer Trapped near a Charged Wall: A Synchrotron X-Ray
  Diffraction Study, Phys. Rev. Lett. {\bf 101},  136103  (2008).

\bibitem{Satapathy2009}
D.~K. Satapathy, K. Nyg\r{a}rd, O. Bunk, K. Jefimov, E. Perret, A. Diaz, F.
  Pfeiffer, C. David, and J.~F. van~der Veen, Buckling and layering transitions
  in confined colloids, EPL {\bf 87},  34001  (2009).

\bibitem{Bordin2012}
J.~R. Bordin, A.~B. de~Oliveira, A. Diehl, and M.~C. Barbosa, Diffusion
  enhancement in core-softened fluid confined in nanotubes, The Journal of
  Chemical Physics {\bf 137},  084504  (2012).

\bibitem{Nygard2014}
K. Nyg{\aa}rd, S. Sarman, and R. Kjellander, Packing frustration in dense
  confined fluids, The Journal of Chemical Physics {\bf 141},  094501  (2014).

\bibitem{Bastea2003}
S. Bastea, Transport properties of dense fluid argon, Phys. Rev. E {\bf 68},
  031204  (2003).

\bibitem{Hoyt2000}
J.~J. Hoyt, M. Asta, and B. Sadigh, Test of the Universal Scaling Law for the
  Diffusion Coefficient in Liquid Metals, Phys. Rev. Lett. {\bf 85},  594
  (2000).

\bibitem{Bretonnet2002}
J.-L. Bretonnet, Self-diffusion coefficient of dense fluids from the pair
  correlation function, The Journal of Chemical Physics {\bf 117},  9370
  (2002).

\bibitem{Samanta2004}
A. Samanta, S.~M. Ali, and S.~K. Ghosh, New Universal Scaling Laws of Diffusion
  and Kolmogorov-Sinai Entropy in Simple Liquids, Phys. Rev. Lett. {\bf 92},
  145901  (2004).

\bibitem{Goel2008}
G. Goel, W.~P. Krekelberg, J.~R. Errington, and T.~M. Truskett, Tuning Density
  Profiles and Mobility of Inhomogeneous Fluids, Phys. Rev. Lett. {\bf 100},
  106001  (2008).

\bibitem{Goel2009}
G. Goel, W.~P. Krekelberg, M.~J. Pond, J. Mittal, V.~K. Shen, J.~R. Errington,
  and T.~M. Truskett, Available states and available space: static properties
  that predict self-diffusivity of confined fluids, Journal of Statistical
  Mechanics: Theory and Experiment {\bf 2009},  P04006  (2009).

\bibitem{Mittal2007}
J. Mittal, V.~K. Shen, J.~R. Errington, and T.~M. Truskett, Confinement,
  entropy, and single-particle dynamics of equilibrium hard-sphere mixtures, J.
  Chem. Phys. {\bf 127},  154513  (2007).

\bibitem{Mittal2007b}
J. Mittal, J.~R. Errington, and T.~M. Truskett, Relationships between
  Self-Diffusivity, Packing Fraction, and Excess Entropy in Simple Bulk and
  Confined Fluids, The Journal of Physical Chemistry B {\bf 111},  10054
  (2007), pMID: 17629320.

\bibitem{Hansen2006}
H. Hansen-Goos and R. Roth, Density functional theory for hard-sphere mixtures:
  the White Bear version mark II, J. Phys.: Condens. Matter {\bf 18},  8413
  (2006).

\bibitem{Goetze:Complex_Dynamics}
W. G\"otze, {\em Complex Dynamics of Glass-Forming Liquids -- A Mode-Coupling
  Theory} (Oxford, Oxford, 2009).

\bibitem{Lang2014a}
S. Lang and T. Franosch, Tagged-particle motion in a dense confined liquid,
  Phys. Rev. E {\bf 89},  062122  (2014).

\bibitem{Feller:Probability_2}
W. Feller, {\em An Introduction to Probability Theory and Its Applications}
  (John Wiley \& Sons, ADDRESS, 1966), Vol.~II.

\bibitem{Gesztesy:2000}
F. Gesztesy and E. Tsekanovskii, On Matrix–Valued Herglotz Functions,
  Mathematische Nachrichten {\bf 218},  61  (2000).

\bibitem{Arnold:1975}
V.~I. Arnol{\textquoteright}d, , Russian Math. Surveys {\bf 30},    (1975).

\bibitem{Franosch:2002}
T. Franosch and $\text{Th.}$ Voigtmann, {Completely monotone solutions of the
  mode-coupling theory for mixtures}, {J. Stat. Phys.} {\bf {109}},  {237}
  (2002).

\bibitem{Voigtmann:Dissertation}
\relax{Th}. Voigtmann, Doctoral thesis, Technische Universit\"at M\"unchen,
  2003.

\bibitem{Franosch_c:1997}
T. Franosch, M. Fuchs, W. G\"otze, M.~R. Mayr, and A.~P. Singh, Asymptotic laws
  and preasymptotic correction formulas for the relaxation near
  glass-transition singularities, Phys. Rev. E {\bf 55},  7153  (1997).

\bibitem{Fuchs:1998}
M. Fuchs, W. G\"otze, and M.~R. Mayr, Asymptotic laws for tagged-particle
  motion in glassy systems, Phys. Rev. E {\bf 58},  3384  (1998).

\bibitem{Bayer:2007}
M. Bayer, J.~M. Brader, F. Ebert, M. Fuchs, E. Lange, G. Maret, R. Schilling,
  M. Sperl, and J.~P. Wittmer, Dynamic glass transition in two dimensions,
  Phys. Rev. E {\bf 76},  011508  (2007).

\bibitem{Lang2014b}
S. Lang, R. Schilling, and T. Franosch, Glassy dynamics in confinement: Planar
  and bulk limits of the mode-coupling theory, Phys. Rev. E {\bf 90},  062126
  (2014).

\bibitem{Lang2014c}
S. Lang, T. Franosch, and R. Schilling, Structural quantities of
  quasi-two-dimensional fluids, The Journal of Chemical Physics {\bf 140},
  104506  (2014).

\bibitem{Franosch:2012}
T. Franosch, S. Lang, and R. Schilling, Fluids in Extreme Confinement, Phys.
  Rev. Lett. {\bf 109},  240601  (2012), erratum: {\bf 110}, 059901(E) (2013).

\bibitem{Roth2010}
R. Roth, Fundamental measure theory for hard-sphere mixtures: a review, J.
  Phys.: Condens. Matter {\bf 22},  063102  (2010).

\bibitem{Palberg2014}
T. Palberg, Crystallization kinetics of colloidal model suspensions: recent
  achievements and new perspectives, Journal of Physics: Condensed Matter {\bf
  26},  333101  (2014).

\bibitem{Hoover1968}
W.~G. Hoover and F.~H. Ree, Melting Transition and Communal Entropy for Hard
  Spheres, The Journal of Chemical Physics {\bf 49},  3609  (1968).

\bibitem{Fortini2006b}
A. Fortini and M. Dijkstra, Phase behaviour of hard spheres confined between
  parallel hard plates: manipulation of colloidal crystal structures by
  confinement, J. Phys. Condens. Matter {\bf 18},  L371  (2006).

\bibitem{Alba-Simionesco2006}
C. Alba-Simionesco, B. Coasne, G. Dosseh, G. Dudziak, K.~E. Gubbins, R.
  Radhakrishnan, and M. Sliwinska-Bartkowiak, Effects of conﬁnement on
  freezing and melting, J. Phys.: Condens. Matter {\bf 18},  R15  (2006).

\bibitem{Loewen2009}
H. L\"owen, Twenty years of confined colloids: from confinement-induced
  freezing to giant breathing, Journal of Physics: Condensed Matter {\bf 21},
  474203  (2009).

\bibitem{Loewen2011}
H. L\"owen, E.~C. O\v{g}uz, L. Assoud, and R. Messina,  in {\em Advances in
  Chemical Physics} (John Wiley \& Sons, Inc., ADDRESS, 2011), pp.\ 225--249.

\bibitem{Megen1994}
W. van Megen and S.~M. Underwood, Glass transition in colloidal hard spheres:
  Measurement and mode-coupling-theory analysis of the coherent intermediate
  scattering function, Phys. Rev. E {\bf 49},  4206  (1994).

\bibitem{Bannerman2011}
M. Bannerman, R. Sargant, and L. Lue, DynamO: A free ${\cal{O}}(N)$ general
  event-driven molecular dynamics simulator, J. Comput. Chem. {\bf 32},  3329
  (2011).

\bibitem{Nygard2012}
K. Nyg\aa{}rd, R. Kjellander, S. Sarman, S. Chodankar, E. Perret, J.
  Buitenhuis, and J.~F. van~der Veen, Anisotropic Pair Correlations and
  Structure Factors of Confined Hard-Sphere Fluids: An Experimental and
  Theoretical Study, Phys. Rev. Lett. {\bf 108},  037802  (2012).

\bibitem{Weysser2010}
F. Weysser, A.~M. Puertas, M. Fuchs, and T. Voigtmann, Structural relaxation of
  polydisperse hard spheres: Comparison of the mode-coupling theory to a
  Langevin dynamics simulation, Phys. Rev. E {\bf 82},  011504  (2010).

\bibitem{Fortini2006}
A. Fortini, M. Schmidt, and M. Dijkstra, Phase behavior and structure of model
  colloid-polymer mixtures confined between two parallel planar walls, Phys.
  Rev. E {\bf 73},  051502  (2006).

\bibitem{Gotze2009}
W. G\"otze, {\em Complex Dynamics of Glass-Forming Liquids-A Mode-Coupling
  Theory} (Oxford University, Oxford, 2009).

\bibitem{Gallo2000}
P. Gallo, M. Rovere, and E. Spohr, Supercooled Confined Water and the Mode
  Coupling Crossover Temperature, Phys. Rev. Lett. {\bf 85},  4317  (2000).

\bibitem{Gallo2012}
P. Gallo, M. Rovere, and S.-H. Chen, Water confined in MCM-41: a mode coupling
  theory analysis, J. Phys.: Condens. Matter {\bf 24},  064109  (2012).

\bibitem{Henderson2004}
J.~R. Henderson, Interfacial statistical geometry: Fluids adsorbed in wedges
  and at edges, The Journal of Chemical Physics {\bf 120},  1535  (2004).

\bibitem{Neser1997}
S. Neser, C. Bechinger, P. Leiderer, and T. Palberg, Finite-Size Effects on the
  Closest Packing of Hard Spheres, Phys. Rev. Lett. {\bf 79},  2348  (1997).

\bibitem{Bollinger2014}
J.~A. Bollinger, A. Jain, and T.~M. Truskett, Structure, Thermodynamics, and
  Position-Dependent Diffusivity in Fluids with Sinusoidal Density Variations,
  Langmuir {\bf 30},  8247  (2014), pMID: 24984592.

\bibitem{Gotze2003}
W. G\"otze and T. Voigtmann, Effect of composition changes on the structural
  relaxation of a binary mixture, Phys. Rev. E {\bf 67},  021502  (2003).

\bibitem{Voigtmann2011}
T. Voigtmann, Multiple glasses in asymmetric binary hard spheres, EPL
  (Europhysics Letters) {\bf 96},  36006  (2011).

\end{thebibliography}

\end{document}